\documentclass[a4paper]{JHEP3}
\usepackage{amsfonts,amsmath,pifont}
\pdfoutput=1
\usepackage{graphicx,float,cite}
\newcommand{\eref}[1]{(\ref{#1})}

\newcommand{\fref}[1]{Figure~\ref{#1}}
\newcommand{\cref}[1]{Chapter~\ref{#1}}
\newcommand{\beq}{\begin{equation}}
\newcommand{\eeq}{\end{equation}}
\newcommand{\ba}{\begin{array}}
\newcommand{\ea}{\end{array}}
\newcommand{\bcenter}{\begin{center}}
\newcommand{\ecenter}{\end{center}}

\def\IB{\relax\hbox{$\inbar\kern-.3em{\rm B}$}}
\def\ID{\relax\hbox{$\inbar\kern-.3em{\rm D}$}}
\def\IE{\relax\hbox{$\inbar\kern-.3em{\rm E}$}}
\def\IF{\relax\hbox{$\inbar\kern-.3em{\rm F}$}}
\def\IG{\relax\hbox{$\inbar\kern-.3em{\rm G}$}}
\def\IGa{\relax\hbox{${\rm I}\kern-.18em\Gamma$}}
\def\IH{\relax{\rm I\kern-.18em H}}
\def\IK{\relax{\rm I\kern-.18em K}}
\def\IL{\relax{\rm I\kern-.18em L}}
\def\IP{\relax{\rm I\kern-.18em P}}
\def\II{\relax{\rm I\kern-.18em I}}

\def\IC{\mathbb{C}}
\def\IR{\mathbb{R}}
\def\IZ{\mathbb{Z}}
\def\Q{\mathbb{Q}}
\def\T{\mathbb{T}}

\def\sCC{{\kern 0.27em\vrule height1.45ex width0.03em depth0em
          \kern-0.30em\rm C}}
\def\C{{\mathchoice
  {\sCC}
  {\sCC}
  {\kern 0.225em \vrule height1.05ex width0.025em depth0em \kern-0.25em \rm C}
  {\kern 0.180em \vrule height0.78ex width0.02em depth0em \kern-0.2em \rm C}
        }}
\def\sHH{{\rm I\kern-.16em{}H}}
\def\H{{\mathchoice
  {\sHH}
  {\sHH}
  {\rm I\kern-.13em{}H}
  {\rm I\kern-.13em{}H} }}
\def\sNN{{\rm I\kern-.16em{}N}}
\def\N{{\mathchoice
  {\sNN}
  {\sNN}
  {\rm I\kern-.12em{}N}
  {\rm I\kern-.10em{}N} }}
\def\sPP{{\rm I\kern-.16em{}P}}
\def\P{{\mathchoice
  {\sPP}
  {\sPP}
  {\rm I\kern-.12em{}P}
  {\rm I\kern-.10em{}P} }}
\def\sQQ{{\kern 0.27em \vrule height1.45ex width0.03em depth0em
          \kern-0.30em \rm Q}}
\def\sRR{{\rm I\kern-0.16em{}R}}
\def\R{{\mathchoice
  {\sRR}
  {\sRR}
  {\rm I\kern-0.12em{}R}
  {\rm I\kern-0.10em{}R} }}
\def\sZZ{{\rm Z\kern-0.32em{}Z}}
\def\Z{{\mathchoice
  {\sZZ}
  {\sZZ} 
  {\rm Z\kern-0.3em{}Z}     %.3
  {\rm Z\kern-0.25em{}Z} }}  %.25
\def\ZZZ{{\rm Z\kern-0.24em{}Z}}
\def\sII{{\rm I\kern-0.16em{}I}}
\def\I{{\mathchoice
  {\sII}
  {\sII}
  {\rm I\kern-0.12em{}I}
  {\rm I\kern-0.10em{}I} }}

\def\inbar{\,\vrule height1.5ex width.4pt depth0pt}

\def\smiley{\hbox{\large$\bigcirc$\hspace{-0.80em}\raise.2ex
\hbox{$\cdot\cdot$}\kern-.61em\lower.2ex\hbox{\scriptsize$\smile$}}\ }
\def\frowny{\hbox{\large$\bigcirc$\hspace{-0.80em}\raise.2ex
\hbox{$\cdot\cdot$}\kern-.635em\lower.2ex\hbox{\scriptsize$\frown$}}\ }

\def\I{{\rlap{1} \hskip 1.6pt \hbox{1}}}

\makeatletter
\let\hangafter\@hangfrom
\makeatother

\newcommand{\be}{\begin{equation}}
\newcommand{\ee}{\end{equation}}
\newcommand{\bea}{\begin{eqnarray}}
\newcommand{\eea}{\end{eqnarray}}
\newcommand{\bean}{\begin{eqnarray*}}
\newcommand{\eean}{\end{eqnarray*}}
\newcommand{\nn}{\nonumber}

\newcommand{\dd}{\mathrm{d}}

\newcommand{\tr}{\mbox{tr}}
\def\T{\relax\hbox{$\inbar\kern-.23em{\rm T}$}}

\newcommand{\ex}{\mathrm{e}}
\newcommand{\CM}{\mathcal{M}}
\newcommand{\f}{\mathcal{F}^{\flat}}
\newcommand{\comment}[1]{}

\newcommand{\me}{\mathrm{e}}
\newcommand{\ii}{\mathrm{i}}

\preprint{}

\title{(Un)Higgsing the M2-brane}
\author{\parbox{14cm}{Nessi Benishti$^1$, Yang-Hui He$^{1,2,3}$ and
James Sparks$^3$ \vspace{1cm}}\\
~\\
\begin{tabular}{rl}
$^1$ & Rudolf Peierls Centre for Theoretical Physics, Oxford University, 
   1 Keble Road, Oxford, \\ 
&  OX1 3NP, U.K.\\
$^2$ & Collegium Mertonese in Academia Oxoniensi, Oxford, OX1 4JD, U.K.\\
$^3$ & Mathematical Institute, University of Oxford,
  24-29 St.\ Giles', Oxford, OX1 3LB, U.K.
\end{tabular}

\vspace{1cm}

\email{nessibenishti@gmail.com, hey@maths.ox.ac.uk, sparks@maths.ox.ac.uk}

\vspace{2.5cm}
}

\abstract{We study various aspects of $\mathcal{N}=2$ quiver-Chern-Simons theories, conjectured to be dual to M2-branes at toric Calabi-Yau four-fold singularities, under Higgsing. In particular we study in detail the orbifold $\IC^4/\IZ_2^3$, obtaining a number of different quiver-Chern-Simons phases for this model, and all 18 toric partial resolutions thereof. In the process we develop a general un-Higgsing algorithm that allows one to construct quiver-Chern-Simons theories by blowing up, thus obtaining a plethora of new models. In addition we explain how turning on torsion $G$-flux non-trivially affects the supergravity dual of Higgsing, showing that the supergravity and field theory analyses precisely match in an example based on the Sasaki-Einstein manifold $Y^{1,2}(\mathbb{CP}^2)$.

\vspace{2.5cm}

}


\begin{document}

%%%%%%%%%%%%%%%%%%%%%%%%%%%%%%%%%%%%%%%%%%%%%%%%%%%%%%

%=====================================================

\section{Introduction and overview}\label{sec:intro}

%=====================================================

There has been considerable interest recently in supersymmetric Chern-Simons (CS) matter theories as candidate AdS$_4$/CFT$_3$ duals to M2-branes at various conical singularities. A key breakthrough was made in \cite{Aharony:2008ug}, following work by \cite{BL, gust}, in which Aharony-Bergman-Jafferis-Maldacena (ABJM) constructed a $U(N)_k\times U(N)_{-k}$ quiver-Chern-Simons (QCS) theory, with $\mathcal{N}=6$ superconformal symmetry, and conjectured this to be the low-energy theory on $N$ M2-branes probing a $\IC^4/\IZ_k$ singularity. (Here the generator of $\IZ_k$ acts with equal charge on each coordinate of $\IC^4$.) In the field theory the Chern-Simons level $k\in\IZ$ plays the role of a coupling constant, with the $k=1$ theory on a stack of $N$ parallel M2-branes in flat spacetime being strongly coupled. For $k=1,2$ the theory has enhanced $\mathcal{N}=8$ superconformal symmetry, as expected for the theories on M2-branes transverse to $\IC^4$, $\IC^4/\IZ_2$, respectively. In the field theory this is a quantum enhancement of supersymmetry, involving monopole operators which create quantized magnetic flux in the diagonal $U(1)$ gauge group \cite{Aharony:2008ug}.

This work was soon generalized to QCS theories with less supersymmetry  \cite{Benna:2008zy, Hosomichi:2008jd, Hosomichi:2008jb, Schnabl:2008wj, Imamura:2008nn, Terashima:2008ba, Jafferis:2008qz, Martelli:2008si, Hanany:2008cd, Hanany:2008fj, Imamura:2008dt, Franco:2008um, Hanany:2008gx, Franco:2009sp, Davey:2009sr, Ooguri:2008dk, Davey:2009qx}, which have been conjectured to be dual to M2-branes probing other geometries admitting parallel spinors. These include orbifolds of $\IC^4$, preserving various fractions of supersymmetry, as well as non-trivial hyperK\"ahler, Calabi-Yau and $Spin(7)$ holonomy cones $C(Y_7)$, where $Y_7$ is a compact tri-Sasakian, Sasaki-Einstein, or weak $G_2$ manifold, respectively. Here we focus on QCS theories with $\mathcal{N}=2$ supersymmetry that conjecturally describe M2-branes on Calabi-Yau four-fold cones. This is the fewest number of supercharges for which supersymmetry still provides a useful constraint on the infra-red (IR) dynamics. For example, the scaling dimensions of chiral primary operators are given exactly by their R-charges under the $U(1)_R$ symmetry.

In this paper we study various aspects of $\mathcal{N}=2$ QCS theories under Higgsing. As for the more well-studied case of D3-branes at Calabi-Yau three-fold singularities \cite{Beasley:1999uz, Feng:2000mi, Feng:2001xr}, the Higgs mechanism is a useful way to construct new QCS theories from old. A necessary condition to interpret a QCS theory as a worldvolume theory on an M2-brane probing a Calabi-Yau four-fold singularity $X$ is that its vacuum moduli space (VMS) is, or at least contains, $X$. At the level of the VMS, the Higgsing leads to a partial resolution $\pi:\hat{X}\rightarrow X$ of $X$ induced by turning on Fayet-Iliopoulos (FI) parameters, and the IR limit is then a near-horizon limit in $\hat{X}$. Indeed, this process of partial resolution is the basis for the \emph{inverse algorithm} of \cite{Feng:2000mi}, by which one can in principle obtain a D3-brane quiver gauge theory for any toric Calabi-Yau three-fold singularity by partial resolution of an appropriate Abelian orbifold of $\IC^3$. The latter gauge theory may be constructed straightforwardly as a Douglas-Moore (DM) projection of $\mathcal{N}=4$ super-Yang-Mills theory \cite{Douglas:1996sw}.

Motivated by this early work of \cite{Beasley:1999uz, Feng:2000mi} on partial resolutions of $\IC^3/\IZ_2^2$ and $\IC^3/\IZ_3^2$, here we study the Abelian orbifold $\IC^4/\IZ_2^3$. At present there is no known general method for constructing QCS theories for orbifolds $\IC^4/\Gamma$ as a projection of the ABJM theory -- for certain choices of $\Gamma$ one can use a DM projection, but for the $\IC^4/\IZ_2^3$ singularity of interest this is \emph{not} the case. (For very recent work on orbifolds of the ABJM theory, see \cite{Berenstein:2009ay}.) This leads us to construct an \emph{un-Higgsing} algorithm where one starts with a QCS theory for a singularity $X$, and then enlarges the quiver in a specific way, corresponding to ``blowing up'' $X$. Via this method, and others, we are able to construct a number of different QCS theories, starting from the ABJM theory, whose Abelian VMSs are the orbifold $\IC^4/\IZ_2^3$. We then systematically study the Higgsings of these theories, thus obtaining QCS theories for all 18 inequivalent toric partial resolutions of the singularity. This leads to a wealth of new models, many of which are new to the literature.

Another important difference between the M2-brane and D3-brane cases is that typically for the background AdS$_4\times Y_7$ one is allowed to turn on \emph{torsion} $G$-flux in $H^{4}(Y_7,\IZ)$; whereas for AdS$_5\times Y_5$ backgrounds, with $Y_5$ a toric Sasaki-Einstein five-manifold, there is never torsion in $H^{3}(Y_5,\IZ)$. Indeed, typically $H^{4}_{\mathrm{tor}}(Y_7,\IZ)$ is non-zero, and each different choice of flux should give a physically distinct theory. This was first discussed in this context by \cite{Aharony:2008gk}, who considered the ABJM model with $Y_7=S^{7}/\IZ_k$. In this case $H^{4}(Y_7,\IZ)\cong\IZ_k$, so there are $k$ distinct M-theory backgrounds corresponding to the $k$ choices of torsion $G$-flux. The authors of \cite{Aharony:2008gk} argued this corresponds to changing the \emph{ranks} of the ABJM theory from $U(N)_k\times U(N)_{-k}$ to $U(N+l)_k\times U(N)_{-k}$, where $0\leq l <k$. As we explain quite generally, theories with non-zero torsion $G$-flux have a richer behaviour under Higgsing than those without any flux. As for the D3-brane case, when there is no flux one can argue from the supergravity dual that one expects to obtain field theories for \emph{all} partial resolutions of a given singularity by Higgsing the original theory. However, once one turns on torsion flux the story is more complicated. The essential idea is that in the supergravity dual of the RG flow induced by the Higgsing one must extend the $G$-flux over the whole spacetime, satisfying the appropriate equations of motion. 
This can lead to interesting predictions for the expected patterns of Higgsings observed in the dual field theory. 
We examine this in detail in the example where $Y_7=Y^{1,2}(\mathbb{CP}^2)$ is a certain non-trivial Sasaki-Einstein seven-manifold, finding precise agreement between the supergravity analysis and field theory analysis for a new QCS theory we construct by un-Higgsing. A different QCS theory for this Calabi-Yau geometry, with a Type IIA construction, has already appeared in the literature, and we point out several puzzles encountered when trying to similarly interpret this as an M2-brane theory.

The organization of the paper is as follows. We begin in Section 2 with a brief review of quiver-Chern-Simons theories in $(2 + 1)$ dimensions and explain how to compute their moduli spaces; this is simply the generalization \cite{Hanany:2008gx} of the forward algorithm of \cite{Feng:2000mi}. In Section 3 we consider $\IC^2/\IZ_n \times \IC^2/\IZ_n$ theories, reviewing how these theories can be obtained by orbifold projection of the ABJM theory, and studying their general behaviour under Higgsing. In Section 4 we introduce the un-Higgsing algorithm and utilize it to produce a $\IC^4/\IZ_2^3$ phase, together with several sets of other phases. We examine the rules for transformations between certain types of dual theories, and study in detail the Higgsing behaviour of the  $\IC^4/\IZ_2^3$ phases.  In Section 5 we study partial resolutions of $C(Y_7)$ spaces with different configurations of torsion $G$-flux, examining in detail the example where $Y_7= Y^{1,2}(\mathbb{CP}^2)$. We conclude with some discussions and future prospects in Section 6. In two appendices we present the details of some orbifold projections, and also list additional QCS theories that do not appear in the main text.

%=====================================================
\section{$\mathcal{N}=2$ quiver-Chern-Simons theories and toric geometry}
%=====================================================

In this section we briefly review the $\mathcal{N}=2$ supersymmetric 
QCS theories of interest, focusing in particular on their vacuum moduli spaces. 
For further details the reader is referred to \cite{Gaiotto:2007qi, Martelli:2008si, Hanany:2008cd} and references therein. 
We shall make extensive use of toric geometry throughout the paper,  
so include a brief summary for completeness (a standard reference is \cite{Fulton}). We also state 
a necessary and sufficient condition on the toric diagram for the corresponding Calabi-Yau four-fold singularity 
to be isolated. 

\subsection{$\mathcal{N}=2$ QCS theories}\label{sec:CSreview}

Our starting point is an $\mathcal{N}=2$ gauge theory in $(2+1)$ dimensions with 
product gauge group $\prod_{i=1}^G U(N_i)$. The matter content will be specified by a quiver diagram with
$G$ nodes. To each arrow in the quiver going from node $i$ to node $j$ we associate 
a chiral superfield $X_{i,j}$ in the bifundamental representation of 
the corresponding two gauge groups. More precisely, we take the convention that $X_{i,j}$ transforms in the $({N}_i,\bar{N}_j)$ representation of the gauge groups at nodes $i$ and $j$, respectively.
When $i=j$ this is understood to be 
the adjoint representation, and we shall often denote such an adjoint field by $\phi_i$. The Lagrangian, in $\mathcal{N}=2$ superspace notation, is then
\bea\label{action}
\mathcal{L}= &&\int \dd^4 \theta\, \mathrm{Tr}\left[ \sum\limits_{X_{i,j}} X_{i,j}^\dagger \, \ex^{-V_{i}} X_{i,j} \, \ex^{V_{j}}
+ \sum\limits_{i=1}^G \frac{k_i}{2\pi} \int\limits_0^1 \dd t V_i \bar{D}^{\alpha}(\ex^{t V_i} D_{\alpha} \ex^{-tV_i})
\right] \nonumber \\
&& + 
\int \dd^2 \theta \, W(X_{i,j}) \, + \, \mathrm{c.c.}~.
\eea
Here $i=1,\ldots,G$ labels the nodes in the quiver, or equivalently factors in the gauge group, 
$V_i$ are the corresponding gauge multiplets,  
$D_\alpha$ denotes the superspace 
derivative, and $W$ is the superpotential. The latter is taken to be a gauge invariant polynomial 
in the chiral superfields formed from traces of closed loops in the quiver. 
The first and third terms in \eqref{action} are the same as the kinetic and 
superpotential terms in $(3+1)$-dimensional $\mathcal{N}=1$ field theories, respectively. 
The second term is special to $(2+1)$ dimensions and is the supersymmetric completion of the Chern-Simons interaction. 
The integers $k_i\in\IZ$ are the CS levels. We may denote these in the quiver 
diagram by attaching an integer label to each node, as shown for the quivers of the $\IC^4$ phases in Figure~\ref{f:c4}. In general we take the following two constraints on these CS levels
\begin{equation}\label{k-con}
\sum_{i=1}^{G} k_i = 0, \qquad \gcd(\{k_i\}) = 1~.
\end{equation}
The first ensures that the string theory dual has zero Romans mass \cite{Gaiotto:2009mv}, and thus 
has an M-theory lift, while for the second if $\gcd(\{k_i\})=h\in\mathbb{N}$, then 
the vaccum moduli space will simply be a $\IZ_h$ quotient of the moduli space 
with CS levels $\{k_i/h\}$. 

The classical VMS $\CM$ is determined by the following equations 
\cite{Martelli:2008si, Hanany:2008cd}
\begin{eqnarray}\label{VMSeqns}
\nn \partial_{X_{i,j}} W &=& 0~,\\
\nn \mu_i := -\sum\limits_{j=1}^G X_{j,i}^{\dagger} {X_{j,i}} + 
\sum\limits_{k=1}^G  {X_{i,k}} X_{i,k}^{\dagger} &=& 
\frac{k_i\sigma_i}{2\pi}~, \\
\label{DF} \sigma_i X_{i,j} - X_{i,j} \sigma_j &=& 0~,
\end{eqnarray}
where $\sigma_i$ is the scalar component of $V_i$. The first two equations are precisely analogous to the F-term and D-term equations of $\mathcal{N}=1$ 
gauge theories in $(3+1)$ dimensions, while the third equation is a new addition. 
To form $\CM$ one should identify vacuum 
solutions to these equations that are related by the  gauge symmetries 
of the theory. This is slightly more subtle than in $(3+1)$ dimensions due to the 
Chern-Simons interactions.

In this paper we will be particularly interested in Abelian theories, 
where the gauge group is $U(1)^G$ and where $\CM$ is a toric 
Calabi-Yau four-fold variety. For a stack of $N$ coincident M2-branes transverse to a Calabi-Yau four-fold singularity, one expects 
the moduli space to be the $N$th symmetric product of the four-fold. 
In \cite{Martelli:2008si} it was shown quite generally that the moduli space 
of the $U(N)^G$ theory is (or, more precisely, \emph{contains}) the 
$N$th symmetric product of the moduli space of the Abelian $N=1$ theory.
It is then natural to try to interpret 
such a QCS theory as the effective worldvolume theory on M2-branes transverse to the 
Calabi-Yau four-fold. 

In the Abelian case the 
moduli space $\CM$ is straightforward to describe. The third equation 
of \eref{VMSeqns} sets all $\sigma_i$ equal $\sigma_1=\cdots = \sigma_G=s$ to a single value
$s$ on the coherent component of the moduli space. The first equation 
describes the space of F-term solutions, which is by construction an affine 
algebraic set. For the theories we study in this paper, this is itself a toric variety, of dimension 
$4+(G-2) = G+2$. This is the so-called \emph{master space} $\f_{G+2}$, studied 
in detail in \cite{Forcella:2008bb}, and is the same as that in 
$(3+1)$-dimensional $\mathcal{N}=1$ theories. 
Finally, the combination of imposing the 
second equation in \eqref{VMSeqns} and identifying by the gauge symmetries
may be described as a K\"ahler quotient of $\f_{G+2}$ by a subgroup 
$U(1)^{G-2}\subset U(1)^G$. This subgroup is specified \cite{Martelli:2008si, Hanany:2008fj} 
by the integer kernel of the matrix
\begin{equation}\label{C}
C =\left(\begin{matrix}
1 & 1 & 1 & \ldots & 1 \\ k_1 & k_2 & k_3 & \ldots & k_G
\end{matrix}\right)~.
\end{equation}
In particular, this K\"ahler quotient precisely sets the $\mu_i$ in \eqref{VMSeqns}
equal to $k_is/2\pi$, where $s$, which may take any real value, is interpreted as a coordinate 
on the VMS. As discussed in \cite{Martelli:2008si}, 
this picks out a particular baryonic branch of $\f_{G+2}$ determined 
by the vector of CS levels. The four-fold  moduli space 
$\CM_4$ of the $(2+1)$-dimensional QCS theory then fibres 
over the three-fold moduli space of the corresponding $(3+1)$-dimensional 
$\mathcal{N}=1$ theory obtained by replacing the CS interaction 
by standard kinetic terms. The four-fold and three-fold are related by a $U(1)$ K\"ahler quotient
where $s$ is precisely the moment map level.
To summarize, 
\begin{equation}\label{symp}
\CM_4 \cong \f_{G+2} \, //\, U(1)^{G-2}~,
\end{equation}
where the K\"ahler quotient is taken at level zero, implying that $\CM_4$ is a 
K\"ahler cone.

\subsection{Toric Calabi-Yau four-folds}\label{sec:toric}

An affine toric four-fold variety $X=X_4$ is specified by a \emph{strictly convex 
rational polyhedral cone} $\mathcal{C}_4\subset\IR^4$. More invariantly, 
$\IR^4$ here is the Lie algebra of a torus $\mathbb{T}^4\cong U(1)^4$ of rank four.
By definition,
$\mathcal{C}_4$ takes the form
\bea
\mathcal{C}_4= \left\{\sum_{a=1}^D \lambda_a v_a\mid\lambda_a\in\IR_{\geq 0}\right\}
\eea
where the set of vectors $v_a\in\IR^4$, $a=1,\ldots,D$, are the generating 
rays of the cone. The condition of being rational means that 
$v_a\in\Q^4$, and without loss of generality we normalize these 
to be primitive vectors $v_a\in\IZ^4$. 
The condition of 
strict convexity is equivalent to saying that $\mathcal{C}_4$ 
is a cone over a compact convex polytope. 

For an affine toric Calabi-Yau four-fold the $v_a$ all have their endpoints in a single hyperplane, 
where the hyperplane is at unit distance from the origin/apex of the cone. By an appropriate 
choice of basis, we may therefore write $v_a=(1,w_a)$ where the $w_a\in\IZ^3$ 
are the vertices of the \emph{toric diagram} $\Delta$. The toric diagram is simply the 
convex hull of these lattice points, and so is a compact convex lattice 
polytope in $\IR^3$. Any affine toric Calabi-Yau four-fold is specified 
uniquely by $\Delta$, up to shifts of the origin and 
$SL(3,\IZ)$ transformations, which amount to $SL(4,\IZ)$ 
transformations of the original torus $\mathbb{T}^4\cong U(1)^4$. 
Much of the geometry of affine toric Calabi-Yau four-folds 
reduces to studying these lattice polytopes. The toric diagram 
for $X_4=\IC^4$ is shown as an example in Figure \ref{f:c4} (B).

Given a toric diagram $\Delta$, one can recover the corresponding 
Calabi-Yau four-fold via \emph{Delzant's construction}. In physics terms, this would be 
called a gauged linear sigma model (GLSM) description of the four-fold. A minimal 
presentation of the variety is as follows. 
One takes the external vertices $w_a\in\IZ^3$, $a=1,\ldots,D$, of the toric diagram $\Delta$
(the smallest set of points whose convex hull is $\Delta$), 
and constructs the linear map
\bea\label{globe}
A&:&\IR^D\rightarrow \IR^4\nonumber\\
&;&e_a\mapsto v_a~.
\eea
Here $\{e_a\}$ denotes the standard orthonormal basis of $\IR^D$. 
The fact that we started with a strictly convex cone implies that 
the map (\ref{globe}) is surjective.
Since $A$ maps lattice points in $\IR^D$ to lattice points in $\IR^4$, 
there is an induced map of tori
\bea
\mathbb{T}^D=\IR^D/\IZ^D \rightarrow \mathbb{T}^4=\IR^4/\IZ^4~.
\eea
The kernel is $\mathcal{G}\cong U(1)^{D-4}\times \Gamma$, where
$\Gamma\cong \IZ^4/\mathrm{span}_{\IZ}\{v_a\}$ is a finite Abelian group. 
The toric variety $X_4$ is then the K\"ahler quotient
\bea
X_4 \cong \IC^D\, //\, \mathcal{G}
\eea
at moment map level zero, so that it is a K\"ahler cone. In GLSM terms, the 
coordinates $p_1,\ldots,p_D$ on $\IC^D$ are identified with vacuum expectation values 
of the chiral fields; we shall thus generally refer to these as \emph{p-fields}. 
The moment map equation then arises as a D-term equation, while quotienting by 
$\mathcal{G}$ identifies gauge-equivalent vacua.
There is an induced action of $\mathbb{T}^4\cong U(1)^4\cong U(1)^D/\mathcal{G}$ 
on the K\"ahler variety $X_4$, and the image of the moment map 
is a polyhedral cone $\mathcal{C}_4^*$ which is the \emph{dual cone} to 
the polyhedral cone $\mathcal{C}_4$ with which we began.

With the exception of $\IC^4$, the apex of the cone always 
corresponds to a \emph{singular point} $p$ in the toric variety. 
An important question is whether this is an \emph{isolated} singular 
point, or whether there are other singular loci that intersect it. 
In the former case, $X_4\setminus\{p\}\cong \IR_+\times Y_7$ 
where $Y_7$ is a \emph{smooth} Sasakian seven-manifold. 
The condition for the singular point $p$ to be isolated 
is precisely the condition that the moment map cone 
$\mathcal{C}^*_4$ is \emph{good}, in the sense of \cite{Lerman}. 
This condition may be stated as follows. Let $F$ be a face of the cone, 
and let $\{v_{a_1},\ldots,v_{a_m}\}$ be the normals to the set of 
supporting hyperplanes meeting at the face $F$. Then 
the singularity is isolated if and only if for every face 
$F$ the $\{v_{a_1},\ldots,v_{a_m}\}$ may be extended to a 
$\IZ$-basis for $\IZ^4$. In particular, this means that necessarily $m=\mathrm{codim}\, F$. 
This translates into the following condition on the toric diagram 
$\Delta$:
\begin{itemize}
 \item Each face of $\Delta$ is a triangle.
 \item There are no lattice points internal to any edge or face of $\Delta$.
\end{itemize}
These are necessary and sufficient\footnote{The proof is left as an exercise for the reader. 
However, we present here an argument in dimension three, to give an idea. In this case 
the toric diagram $\Delta=\Delta_2$ is a convex lattice polytope in $\IR^2\supset\IZ^2$. The external 
vertices are dual to the facets (codimension one faces) of the cone $\mathcal{C}^*_3$, 
and in this case the goodness condition is vacuous. On the other hand, two external 
vertices $w_1, w_2\in \IZ^2$ are joined by an external edge $E$ of $\Delta_2$ if and only if the dual 
facets meet at an edge of the cone $\mathcal{C}^*_3$. 
Using the shift symmetry of the problem, we may suppose that $w_1=(0,0)$ is at the origin, so $v_1=(1,0,0)$. 
Then $v_1, v_2$ can be extended to a $\IZ$-basis of $\IZ^3$ if and only if $w_2$ can be extended to a 
$\IZ$-basis of $\IZ^2$. But this is true if and only if the components $w_1^1,w_1^2$ of $w_2$ 
satisfy $\mathrm{gcd}(w_1^1,w_1^2)=1$, which is true if and only if $w_2$ cannot be written as $nw$ with 
$w\in\IZ^2$ and $n>1$ an integer, {\it i.e.} there is no lattice point in the interior of the edge $E$.
} 
for the ``link'' $Y_7$ to be a smooth manifold. It was proven recently in 
\cite{FOW} that all such toric Sasakian manifolds admit a unique 
Sasaki-Einstein metric compatible with the complex structure of the cone. 

\subsection{The QCS forward algorithm}\label{sec:forward}

The Abelian vacuum moduli spaces $\CM_4$ of interest will be 
toric Calabi-Yau varieties, $\CM_4=X_4$, and so will be specified by a toric diagram $\Delta$. 
The gauge theory construction of $\CM_4$ 
as a K\"ahler quotient of the 
toric master space $\f_{G+2}$ by $U(1)^{G-2}$ is, however, highly non-minimal, 
and this results in \emph{multiplicities} of the lattice points in $\Delta$. 
The construction of the VMS outlined in Subsection \ref{sec:CSreview} was turned into an algorithm in 
\cite{Hanany:2008gx}, whose end product is precisely the lattice 
points of $\Delta$, together with their multiplicities. 
We summarize this algorithm in (\ref{Gt}).

\begin{equation}\label{Gt}
\begin{array}{lllllll}
\fbox{
\mbox{
\begin{tabular}{l}
INPUT 1: \\
~~Quiver
\end{tabular}
}}
& \rightarrow & d_{G \times E}	& \rightarrow	&
	(Q_D)_{(G-2) \times c} = 
	\ker(C)_{(G-2)\times G} \cdot \tilde{Q}_{G \times c} \ ; 
&&\\
\vspace{-0.5cm}&&& \nearrow & \qquad \mbox{ with } d_{G \times E} := \tilde{Q}_{G\times c}\cdot (P^T)_{c \times E} &&\\
\fbox{
\mbox{
\begin{tabular}{l}
INPUT 2: \\
~~CS Levels
\end{tabular}
}}
& \rightarrow & C_{2 \times G}	&&&&\\[-0.3cm]
&&& \nearrow &&&\\
\fbox{\mbox{
\begin{tabular}{l}
INPUT 3: \\
~~Superpotential
\end{tabular}
}}
& \rightarrow & P_{E \times c}	& \rightarrow
		& (Q_F)_{(c-G-2)\times c} = [\ker P]^t \ ; \\
&&&&~~~~~~\downarrow\\
&&&& 
\hspace{-1in}
(Q_t)_{(c-4) \times c} =
\left( \begin{array}{c}
(Q_D)_{(G-2)\times c} \\
(Q_F)_{(c-G-2) \times c} 
\end{array} \right) \rightarrow
\fbox{\mbox{
\begin{tabular}{l}
OUTPUT:  \\
~~$(G_t)_{4 \times c} = [{\rm Ker}(Q_t)]^t$\\
\end{tabular}
}}
\end{array}
\end{equation}

In the diagram  $C$ denotes the $2\times G$ matrix \eqref{C}, $d$ denotes the $G\times E$ 
incidence matrix of the quiver, where $E$ is the number of edges, 
and $P=K\cdot T$ is the $E\times c$ matrix constructed from the superpotential $W$. 
Here $K$ is an $E\times (G+2)$ matrix that encodes the F-terms derived from $W$, where $T$ denotes the dual cone. 
The integer $c$ is in fact the number of perfect matchings in the brane tiling description. 
We refer to \cite{Hanany:2008gx} for further details, and references 
therein. The key point is that the algorithm takes the data of the 
matter content (specified by the incidence matrix $d$), 
the Chern-Simons levels (specified by the matrix $C$), and 
the superpotential (specified by the matrix $K$, from which one derives the matrix $P$), 
and produces the single charge matrix $Q_t$. The kernel of this, 
$G_t$, is a $4\times c$ matrix that encodes the toric diagram 
of the Calabi-Yau four-fold. Here the Calabi-Yau condition 
is equivalent to the four-vector columns being coplanar, on a hyperplane at unit distance from the origin. The number of repetitions of a given vector in the
$c$ columns is defined to be the \emph{multiplicity} of the corresponding lattice point in $\Delta$.

%=====================================================
\section{Higgsing the non-chiral phase of $\IC^{2}/\IZ_{n}\times \IC^{2}/\IZ_{n}$}\label{s:dm}
%=====================================================
In this section we begin by introducing the ABJM QCS theory for an M2-brane in flat spacetime, which we refer to as $(\IC^4)_I$, as well as another QCS theory $(\IC^4)_{II}$ which has been conjectured to be dual to this. After briefly reviewing orbifold projections and Higgsing in QCS theories, as a warm-up we study in detail a $\IZ_n$ projection of the ABJM theory $(\IC^4)_I$, which is conjecturally dual to an M2-brane at a $\IC^{2}/\IZ_{n}\times \IC^{2}/\IZ_{n}$ singularity. 
%=====================================================
\subsection{The simplest pair: $(\IC^4)_I$ and $(\IC^4)_{II}$}\label{s:C4}
\begin{figure}[H]
\includegraphics[trim=0mm 120mm 0mm 100mm, clip, width=6.0in]{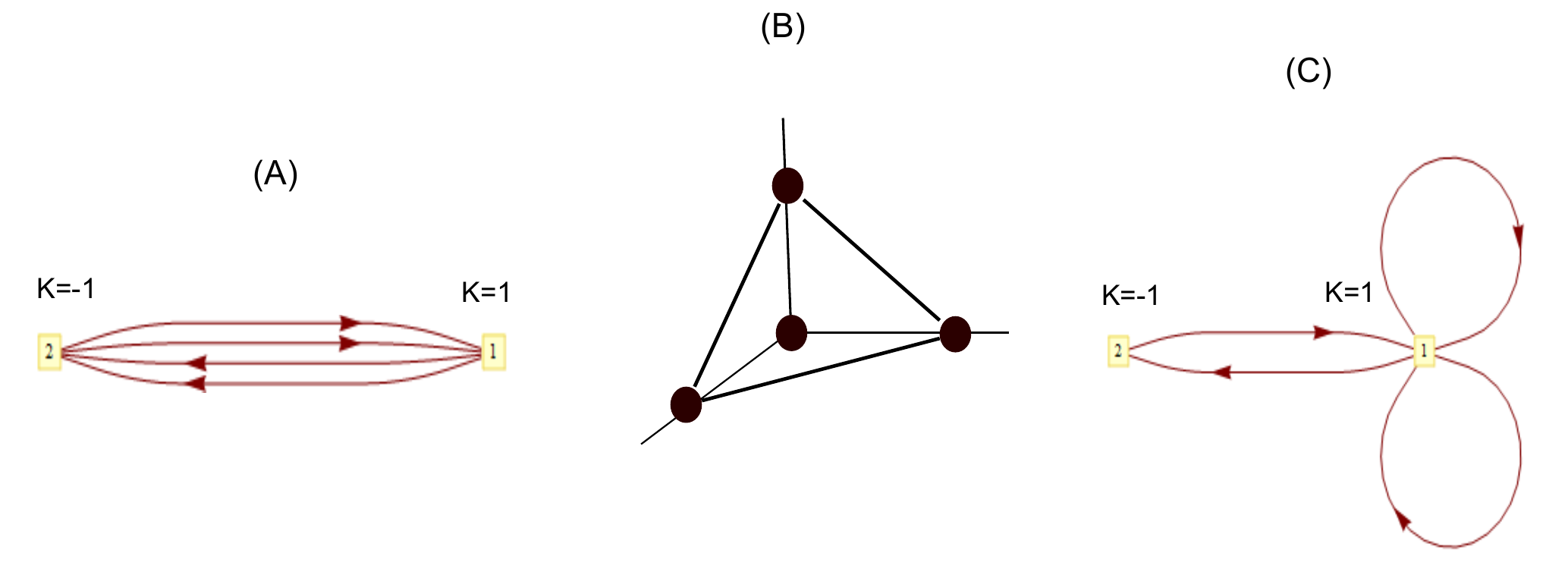}
\caption{{\sf (A) The quiver diagram for the $(\IC^4)_I$ theory (the Chern-Simons level $k$ associated with each node in the quiver is also shown). (C) The quiver diagram for the $(\IC^4)_{II}$ theory. (B) The toric diagram for the simplest Calabi-Yau four-fold $\IC^4$. }\label{f:c4}}
\end{figure}

%=====================================================
Let us begin with the simplest Calabi-Yau four-fold, namely $\IC^4$ equipped with a flat metric. 
In \cite{Aharony:2008ug,Hanany:2008fj,Franco:2008um,Hanany:2008gx,Davey:2009sr} there are two QCS theories presented which have $\IC^{4}$ as their VMS. The toric diagram is drawn in part (B) of Figure~\ref{f:c4}.

The first of the two phases is a special case of what has come to be known as the ABJM theory \cite{Aharony:2008ug} (and is also called, in the brane-tiling picture, the Chessboard model in \cite{Davey:2009sr}):  two gauge groups $U(N) \times U(N)$ with CS levels $(k,-k)$, four bifundamental fields, and the superpotential 
\begin{equation}
W=\tr(X_{1,2}^1 X_{2,1}^1 X_{1,2}^2 X_{2,1}^2 - X_{1,2}^1 X_{2,1}^2 X_{1,2}^2 X_{2,1}^1) \ . 
\end{equation}
The moduli space of this theory depends on the CS levels and is given by $\IC^{4}/\IZ_{k}$, where the generator of $\IZ_k$ acts with 
equal charge on each coordinate of $\IC^4$. We thus need to take $k=1$, and shall denote this theory $(\IC^4)_I$. The quiver diagram is given in part (A) of Figure~\ref{f:c4}.

The second phase is a special case of an example in \cite{Hanany:2008fj}, and was dubbed the 
One Double-Bonded One-Hexagon Model in \cite{Davey:2009sr}: two gauge groups $U(N) \times U(N)$ with CS levels $(k,-k)$, two bifundamental fields, two adjoint fields (one for each gauge group factor), and the superpotential
\begin{equation}
W=\tr(X_{2,1}[\phi_1^1 , \phi_1^2]X_{1,2}) \ .
\end{equation}
The moduli space of this theory is $(\IC^2/\IZ_{k}) \times \IC^{2}$. Thus, we again need to take $k=1$, and shall denote this theory $(\IC^4)_{II}$. The quiver diagram is given in part (C) of Figure~\ref{f:c4}.

%=====================================================
\subsection{Orbifold projections}\label{sec:orb}
%=====================================================
Next we review the well-known orbifold projection of Douglas-Moore (DM), which first brought the study of quiver theories to D-branes \cite{Douglas:1997de}. For branes in flat spacetime, the transverse direction is the trivial Calabi-Yau space $\IC^q$: in the case of D3-branes $q=3$, while in the present M2-brane case of interest $q=4$. We denote the complex coordinates of $\IC^q$ by $(x^1, \ldots, x^q)$. An orbifold is then a quotient space of the form $\IC^q / \Gamma$, where $\Gamma$ is an appropriate discrete group, its elements $\gamma\in\Gamma$ acting as matrices on the vector of coordinates. Indeed, $\Gamma$ needs to be a subgroup of $SU(q)$ to ensure that the orbifold is Calabi-Yau.
The induced orbifold projection on the spacetime fields on the brane worldvolume is by conjugation by the regular representation of $\Gamma$. Only the fields that are invariant under this action survive the orbifold projection.

It is straightforward to apply the above projection to the ABJM theory $(\IC^4)_I$ \cite{Hosomichi:2008jd,Benna:2008zy,Fuji:2008yj}. As an illustrative example we consider here a non-chiral theory where $\Gamma\cong \IZ_n$. Let $\mathcal{A}^1, \mathcal{A}^2$ be the gauge fields for the two $U(N)$ factors of the ABJM theory, and denote by ${\cal Z} = Z^A + \theta \zeta^A + \theta^2 F^A$ and ${\cal W} = W^A + \theta \omega^A + \theta^2 G^A$ the bifundamental and anti-bifundamental hypermultiplet superfields, respectively; these are $N \times N$ matrices, where the index $A=1,2$. In the Abelian $N=1$ case we may identify $Z^1$, $Z^2$, $W^{1\dagger}$ and $W^{2\dagger}$ with the four coordinates $x^a$ of $\IC^4$, and the $\IZ_n$ action on these is defined to be $x^a \to \me^{2 \pi \ii / n_a} x^a$ where $n_a = (n,n,-n,-n)$. In order to take the orbifold projection in the field theory, we begin with $nN$ M2-branes (so that the $Z$ and $W$ fields are now $nN \times nN$ matrices) and choose the regular $n$-dimensional representation of $\IZ_n$ to project back to a $U(N)$ theory. The orbifold action on the fields is then by conjugation:
\begin{eqnarray}
\nn
&& Z^A \to \me^{2\pi \ii / n} \Omega Z^A \Omega^{\dagger}~, \quad 
W^A \to \me^{2\pi \ii / n} \Omega W^A \Omega^{\dagger}~, \quad 
\zeta^A \to \me^{- 2\pi \ii / n} \Omega \zeta^A \Omega^{\dagger}~, \\
&& \omega^A \to \me^{- 2\pi \ii / n} \Omega \omega^A \Omega^{\dagger}~,\quad 
\mathcal{A}^1_\mu \to \Omega \mathcal{A}^1_\mu \Omega^{\dagger}~, \quad
\mathcal{A}^2_\mu \to \Omega \mathcal{A}^2_\mu \Omega^{\dagger}~,
\end{eqnarray}
where $\Omega := \mbox{diag}(I_{N \times N}, \me^{2\pi \ii /n} I_{N \times N},
\me^{4\pi \ii /n} I_{N \times N}, \ldots, \me^{2\pi (n-1) \ii /n} I_{N \times N}
)$.
Only fields invariant under this projection survive. Some explicit examples of such projections are presented in Appendix \ref{sec:orbifold}.

Naively, one might expect the Abelian $N=1$ moduli space of the resulting theory to be $\IC^4/\IZ_n$. However, this is not the case.
In order to apply the projection one needs to take the CS levels of the original theory to be a multiple of $n$, so that 
the levels are $(nk,-nk)$ for the two nodes. 
In the projected theory there are then $2n$ gauge nodes, all with CS levels $\pm k$, and the $N=1$ moduli space 
of the orbifold is instead $\IC^4/\IZ_n \times \IZ_{kn}$ \cite{Imamura:2008nn,Terashima:2008ba}. 
More generally, the CS levels should be quantized according to the order of the orbifold group $\Gamma$. This is where M2-brane orbifold projections differ from D3-brane orbifold projections, and is essentially the reason why
there currently does not exist a general method for constructing QCS theories for an arbitrary orbifold $\IC^4/\Gamma$.

%======================================================================
\subsection{Higgsing in QCS theories}
%======================================================================
By starting with a parent geometry and turning on FI parameters, we can (partially) resolve the singularity to derive new dualities between geometries and gauge theories. By turning on FI parameters, some of the chiral fields in the QCS theory acquire vacuum expectation values (VEVs), and this Higgses the theory at low energy. At the level of the VMS, the FI parameters (partially) resolve the singularity to $\pi:\hat{X}\rightarrow X$, and the choice 
of VEVs picks a point $p\in \hat{X}$ in this space; the low-energy limit is then a near-horizon limit of $p$ (giving the tangent cone at $p$). Partial resolution in $(2+1)$-dimensional QCS theories works very similarly to $(3+1)$-dimensional quiver Yang-Mills (QYM) theories ({\it q.v.}~\cite{Douglas:1997de,Morrison:1998cs,Beasley:1999uz,Feng:2000mi} for the latter). The key differences in the QCS case are that 
only $G-2$ FI parameters are relevant for resolving the singularity, and the CS levels of the gauge nodes being Higgsed should also be taken into account. 

To illustrate this last point, suppose we wish to give a VEV to the field $X_{1,2}$, a bifundamental $({N}_1,\bar{N}_2)$ under gauge nodes 1 and 2. 
The relevant part of the action is
\begin{equation}
S\supset\int \dd^3x\, \left(k_1 \epsilon^{\mu\nu\rho} {\cal A}^1_{\mu}\partial_{\nu}{\cal A}^1_{\rho}+k_2\epsilon^{\mu\nu\rho}{\cal A}^2_{\mu}\partial_{\nu}{\cal A}^2_{\rho} - |D_{\mu}X_{1,2}|^2 \right) \ ,
\label{revpart}
\end{equation}
where ${\cal A}^{1}$, ${\cal A}^2$ are the gauge fields for nodes 1 and 2, respectively, and the covariant derivative is
\begin{equation}
D_{\mu}X_{1,2}=\partial_{\mu}X_{1,2}-\ii({\cal A}^1_{\mu}-{\cal A}^2_{\mu})X_{1,2}\ ~.
\end{equation}
After giving $X_{1,2}$ a VEV, which we will denote as $M$, the combination ${\cal A}^1_{\mu}-{\cal A}^2_{\mu}$ becomes massive. 
If we define ${\cal A}^{\pm}=\frac{1}{2}({\cal A}^1\pm {\cal A}^2)$ and $k_{\pm}=k_1 \pm k_2$, we can rewrite \eqref{revpart} as follows  
\begin{equation}
S\supset\int \dd^3x\, \left(k_+\epsilon^{\mu\nu\rho} {\cal A}^+_{\mu}\partial_{\nu}{\cal A}^+_{\rho}+k_+\epsilon^{\mu\nu\rho}{\cal A}^-_{\mu}\partial_{\nu}{\cal A}^-_{\rho} +2k_-\epsilon^{\mu\nu\rho} {\cal A}^-_{\mu}\partial_{\nu}{\cal A}^+_{\rho}-4M^2({\cal A}^-_{\mu})^2 \right)\ .
\end{equation}
At energies well below the scale set by $M$, we can proceed to integrate out ${\cal A}^-$. Since in the IR this field is effectively constant, we have $\partial{\cal A}^-=0$. Solving the equations of motion we see that ${\cal A}^- \varpropto \frac{1}{M^2}$ and therefore terms that contain ${\cal A}^-$ can be deleted from the Lagrangian in the low-energy limit. As such, \eqref{revpart} reduces to
\begin{equation}
S\supset\int \dd^3x\, k_+\epsilon^{\mu\nu\rho} {\cal A}^+_{\mu}\partial_{\nu}{\cal A}^+_{\rho}.
\end{equation}
We therefore see that the CS level of the gauge node which survives in the IR is the \emph{sum} of the CS levels of the gauge nodes under which the field $X_{1,2}$ was charged.
%=====================================================
\subsection{Resolutions of $\IC^{2}/\IZ_{n}\times \IC^{2}/\IZ_{n}$}
%=====================================================
Let us put the above two techniques, orbifolding and Higgsing, into practice. For several reasons, the ``simplest'' orbifold of $\IC^4$ is perhaps when $\IC^4$ is thought of as two copies of $\IC^2$, with the orbifold group acting independently on these two copies. In this case, the singularity is simply the product of two $\IC^2$ orbifolds, and the latter have been studied to a great extent over the past decade. There is then also a standard Hanany-Witten type of brane configuration \cite{Hanany:1996ie} dual to the QCS theory. It is therefore natural to consider the space $\IC^{2}/\IZ_{n}\times \IC^{2}/\IZ_{n}$ as a demonstrative warm-up. 

\begin{figure}[H]
\centerline{\includegraphics[trim=0mm 220mm 0mm 10mm, clip, width=4.0in]{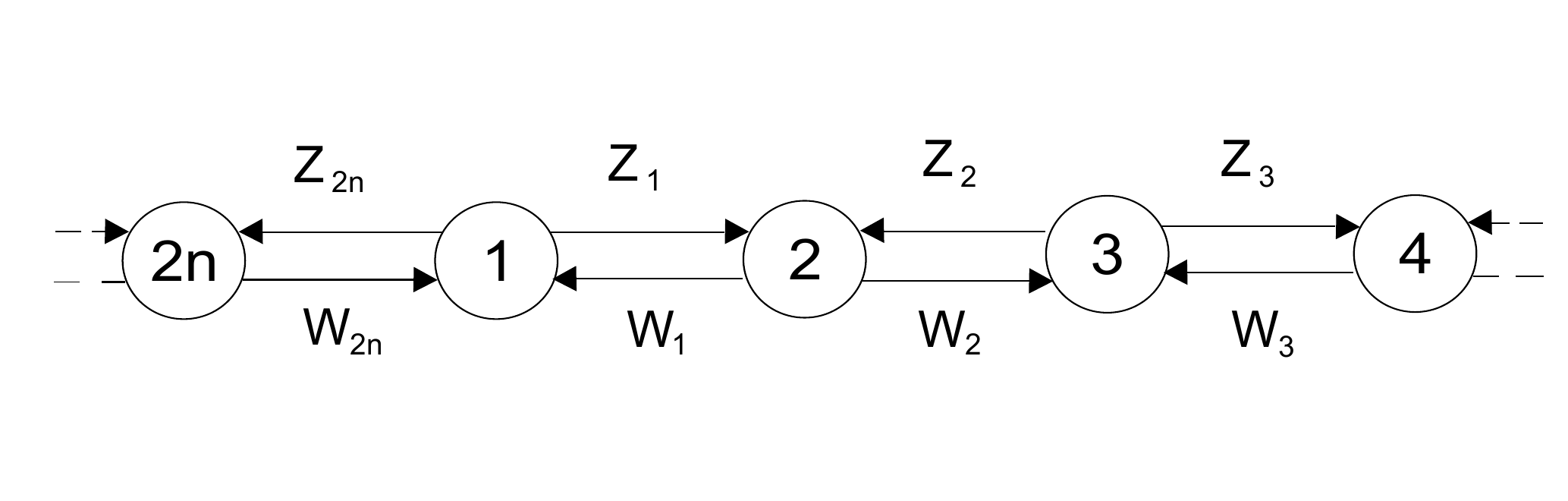}}
\caption{{\sf 
The quiver for the non-chiral $\IZ_n$ projection of the ABJM theory at CS level $k=n$. 
We close the chain by identifying node ${2n+1}$ with node ${1}$. The quiver is identical to that of a D-brane on the ALE space $\IC^2/\IZ_n$.
}
\label{f:znznquiver}}
\end{figure}

The theory for this orbifold has been studied already, and is the non-chiral theory first presented in \cite{Benna:2008zy}.  
It may be obtained by taking a $\IZ_{n}$ projection of the ABJM theory with CS levels $k=n$. The quiver is presented in Figure~\ref{f:znznquiver}, while 
the superpotential and CS matrix are as follows:
\begin{eqnarray}
W=\sum\limits_{l=1}^{n}
Z_{2l-1}W_{2l}Z_{2l}W_{2l-1}-Z_{2l}W_{2l}Z_{2l+1}W_{2l+1}~, \quad
C_{2\times 2n}=
\left(
\begin{array}{cccccc}
1 & 1 & 1 & 1 & \ldots & 1 \\
1 & $-1$ & $1$ & $-1$ & \ldots & $-1$ \\
\end{array}
\right) \ .
\label{CS}
\end{eqnarray}
\vspace{-1cm}

\begin{figure}[H]
\includegraphics[trim=0mm 180mm 0mm 40mm, clip, width=6.0in]{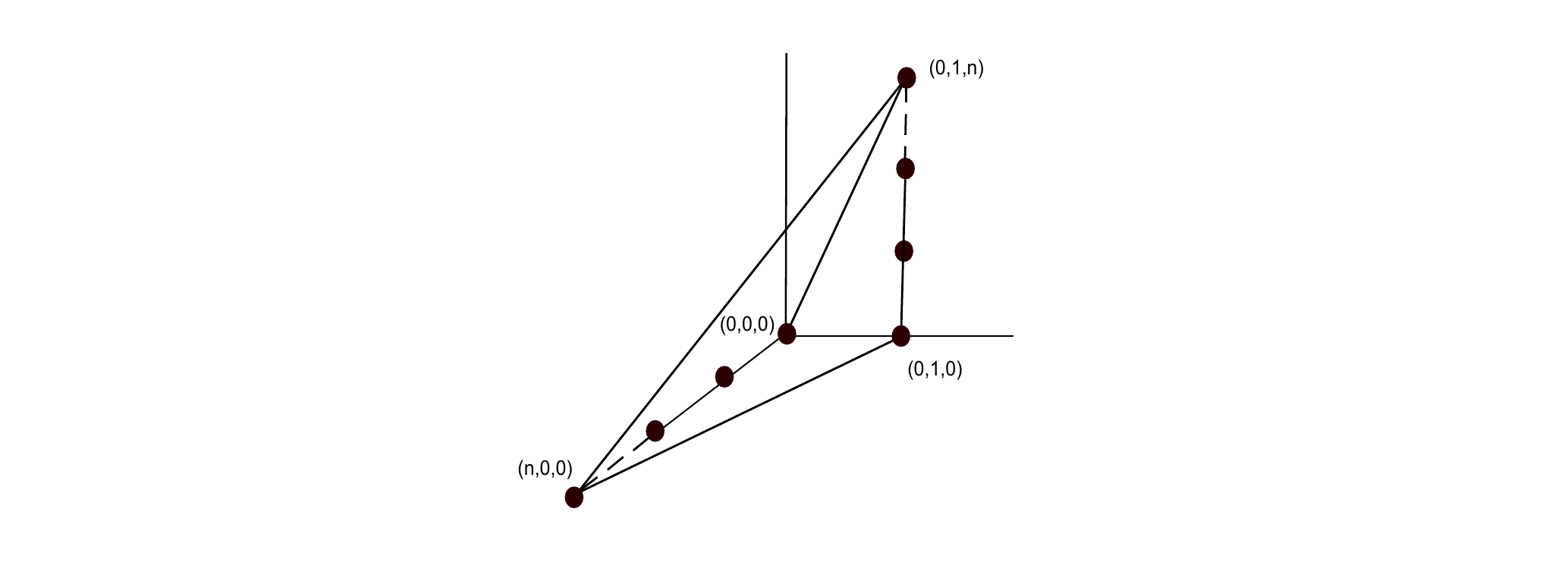}
\caption{{\sf The toric diagram of $\IC^{2}/\IZ_{n}\times \IC^{2}/\IZ_{n}$.}
\label{f:znzn}}
\end{figure}

We can compute the VMS using the forward algorithm,
with the above quiver, superpotential and Chern-Simons levels as input. 
The output is the toric diagram described by $G_t$, whose columns are the vertices of $\Delta$; we will present $G_t$ explicitly at the end of this subsection. 
Using Delzant's construction we can check that the moduli space of the theory is indeed $\IC^{2}/\IZ_{n}\times \IC^{2}/\IZ_{n}$. We present the toric diagram in Figure~\ref{f:znzn}. Notice that it has two external edges each containing $n-1$ lattice points, which do not intersect, and that there are no other lattice points inside the toric diagram. The former implies that the singularity is not isolated.

Let us study the moduli space in detail using the forward algorithm, reviewed briefly in Subsection \ref{sec:forward}. 
Recall from the flowchart \eqref{Gt} that each of the points in the toric diagram corresponds to a field $p$ in the gauged linear sigma model. 
The relation between the spacetime fields and the $p$-fields is encoded in the so-called perfect-matchings matrix $P$. 
To obtain $P$ we will first calculate the Kasteleyn matrix, the procedure for which we refer the reader to Appendix B of \cite{Hanany:2008gx}. 
The Kasteleyn matrix $\mathrm{Kas}$ and its determinant $D$ are computed to be
\begin{eqnarray}
\nn
\mathrm{Kas} &=&
\left(
\begin{array}{llllll}
Z_{2}+W_{2} & 0 & 0 & \ldots & 0 & Z_{1}+W_{1} \\
Z_{3}+W_{3} & Z_{4}+W_{4} & 0 & \ldots & 0 & 0 \\
0 & Z_{5}+W_{5} & Z_{6}+W_{6} & \ldots & 0 & 0 \\
\vdots & \vdots & \vdots & \ddots & \vdots & \vdots \\
0 & 0 & 0 & \ldots & Z_{2n-1}+W_{2n-1} & Z_{2n}+W_{2n}
\end{array}
\right)~,
\end{eqnarray}
\begin{eqnarray}
\nn D \equiv \det(\mathrm{Kas}) &= &(Z_{2}+W_{2})(Z_{4}+W_{4})\ldots(Z_{2n}+W_{2n}) \\
&& + (-1)^n(Z_{1}+W_{1})(Z_{3}+W_{3})\ldots(Z_{2n-1}+W_{2n-1})~.
\end{eqnarray}
We see that there are $2^{n+1}$ terms in $D$ and each field appears in $2^{n-1}$ terms. Moreover, the even and odd indexed fields do not mix in any product. 
Therefore $P$ is actually block-diagonal, one for the odd and one for the even indices; we shall denote the blocks by $P^{\mathrm{even}}$ and $P^{\mathrm{odd}}$, respectively, and similarly the charge matrix will also be a block matrix.
Henceforth we concentrate on the one pertaining to the even indexed fields, where $P^{\mathrm{even}}$ is given by
\bea
{\tiny
\left[
\begin{array}{l|cccccccccccccccc}
 & 
p_{1} & p_{2} & \ldots & p_{2^{n-2}} & p_{2^{n-2}+1} &\ldots & p_{2^{n-1}-1} & p_{2^{n-1}} & p_{2^{n-1}+1} & p_{2^{n-1}+2} & \ldots & p_{3*2^{n-2}} & p_{3*2^{n-2}+1} & \ldots & p_{2^{n}-1} & p_{2^{n}} \\ \hline
Z_{2} & 1 & 1 & \ldots & 1 & 1 & \ldots & 1 & 1 & 0 & 0 & \ldots & 0 & 0 & \ldots & 0 & 0 \\
W_{2} & 0 & 0 & \ldots & 0 & 0 & \ldots & 0 & 0 & 1 & 1 & \ldots & 1 & 1 & \ldots & 1 & 1 \\
Z_{4} & 1 & 1 & \ldots & 1 & 0 & \ldots & 0 & 0 & 1 & 1 & \ldots & 1 & 0 & \ldots & 0 & 0 \\
W_{4} & 0 & 0 & \ldots & 0 & 1 & \ldots & 1 & 1 & 0 & 0 & \ldots & 0 & 1 & \ldots & 1 & 1 \\
\vdots & \vdots & \vdots & \ddots & \vdots & \vdots & \ddots & \vdots & \vdots & \vdots & \vdots & \ddots & \vdots & \vdots & \ddots & \vdots & \vdots \\
Z_{2n} & 1 & 0 & \ldots & 0 & 1 & \ldots & 1 & 0 & 1 & 0 & \ldots & 0 & 1 & \ldots & 1 & 0 \\
W_{2n} & 0 & 1 & \ldots & 1 & 0 & \ldots & 0 & 1 & 0 & 1 & \ldots & 1 & 0 & \ldots & 0 & 1
\end{array}
\right]~.
}\nonumber
\eea
After writing explicitly the relevant part of the incidence matrix $d$ we easily compute the corresponding block in the charge matrix $\tilde{Q}$:
\begin{equation}
d_{\mathrm{even}}=
{\tiny
\left(
\begin{array}{l|ccccccc}
\mbox{node}/\mbox{field}
 & Z_{2} & W_{2} & Z_{4} & W_{4} & \ldots & Z_{2n} & W_{2n} \\ \hline
u_{1} & 0 & 0 & 0 & 0 & \ldots & $-1$ & 1 \\
u_{2} & 1 & $-1$ & 0 & 0 & \ldots & 0 & 0 \\
u_{3} & $-1$ & 1 & 0 & 0 & \ldots & 0 & 0 \\
u_{4} & 0 & 0 & 1 & $-1$ & \ldots & 0 & 0 \\
u_{5} & 0 & 0 & $-1$ & 1 & \ldots & 0 & 0 \\
\vdots & \vdots & \vdots & \vdots & \vdots & \ddots & \vdots & \vdots \\
u_{2n} & 0 & 0 & 0 & 0 & \ldots & 1 & $-1$ \\
\end{array}
\right)} \ , \qquad \
\tilde{Q}^{\mathrm{even}}_{i,j}= \frac{-(-1)^{P^{\mathrm{even}}_{i-1,j}}}{2^{n}}
\ ,
\end{equation}
where $1 \leqslant j \leqslant 2^n$, $1 \leqslant i \leqslant 2n$ and we identify $P^{\mathrm{even}}_{0,j}$ with $P^{\mathrm{even}}_{2n,j}$.
We can also compute, from \eqref{CS}, the kernel of the CS matrix $C$ explicitly. 
First we have that
\begin{equation}
\sum_j \ker(C)_{i,j}=0 \ , \qquad \sum_j (-1)^{j+1} \ker(C)_{i,j}=0 \ ,
\end{equation}
whence, 
\begin{equation}
\ker(C)_{i,1}=-\sum_{j=1}^{n-1} \ker(C)_{i,2j+1} \ , \qquad
\ker(C)_{i,2}=-\sum_{j=1}^{n-1} \ker(C)_{i,2(j+1)} \ .
\end{equation}
In a chosen basis we find that
\bea
\ker(C)_{i,j}= \left\{
\begin{array}{ll} 
\delta_{j,2(n+1)-i}-\delta_{j,2} &, \qquad i=2k \\
\delta_{j,2n-i}-\delta_{j,1} &, \qquad i=2k+1 
\end{array}\right.~.
\eea

Finally, by taking the null-space of the join of $Q_{d} = \ker(C) \cdot \tilde{Q}$ and $Q_{f}$ we obtain the desired matrix $G_{t}$:
\begin{align}
\includegraphics[trim=0mm 260mm 0mm 00mm, clip, width=5.9in]{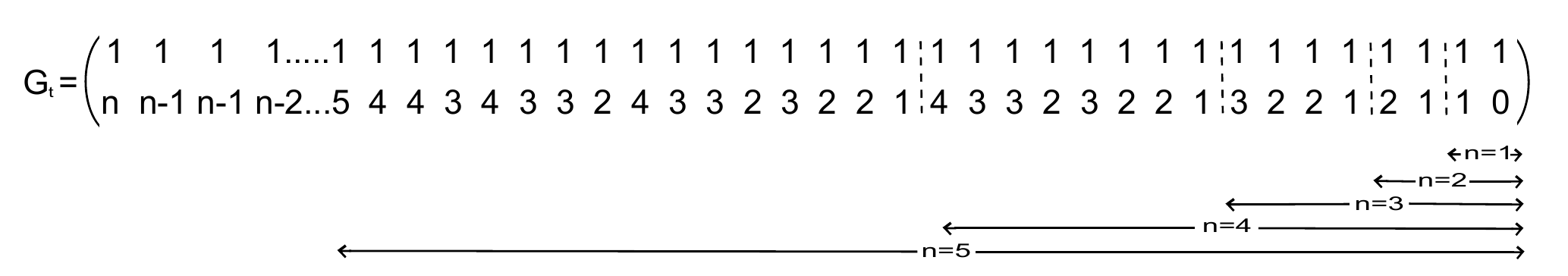}
\end{align}
The other two rows of this matrix are zero, as follows from the fact that $Q_{d}$ and $Q_{f}$ are block diagonal. 
The repetitions in the columns of $G_t$ indicate the multiplicities of the lattice points in the toric diagram. Notice here that these multiplicities are the numbers in Pascal's triangle; this was observed for the $\IC^2 / \IZ_n$ singularity in \cite{Feng:2002zw,Franco:2006gc}.

We next examine the partial resolutions of $\IC^{2}/\IZ_{n}\times \IC^{2}/\IZ_{n}$ obtained by Higgsing this theory. 
When Higgsing a spacetime field one needs to delete the corresponding $p$-fields  in the GLSM, as dictated by the matrix $P$ since each spacetime field is a specfic product of $p$-fields. The associated point in the toric diagram will either be deleted, or, if there are multiple $p$-fields for that point, {\it i.e.} repetitions in columns in $G_{t}$, the multiplicity will be reduced. Given the structure of the $P$ matrix, repeatedly Higgsing only one type of field ($Z$ or $W$) results in half of $G_{t}$ being deleted with each iteration. More specifically, the left half of $G_{t}$ will be deleted by Higgsing in the following order:
\begin{equation}
Z_{2} \to Z_{4} \to \ldots \to Z_{2n} \ . 
\end{equation}
This operation reduces the length of one of the lines in the toric diagram in Figure~\ref{f:znzn} by one at each Higgsing step, corresponding to a partial resolution of the $\IC^{2}/\IZ_{m}$ singularity to $\IC^{2}/\IZ_{m-1}$.

One might be concerned that during this proccess additional fields may acquire a VEV. 
However, each $Z$-field appears alone in one term in the superpotential, accompanied by $W$-fields only. 
Therefore, each $Z$-field corresponds to a unique $p$-field and as such cannot be Higgsed by Higgsing other $Z$-fields. 
Moreover, from the form of the superpotential it is guaranteed that Higgsing the $Z$-fields with even indices will not give mass to any of the other fields. 
This also holds for the odd indices. Thus a similar order of Higgsing for the odd-indexed fields will result in partial resolution of the second $\IC^{2}/\IZ_{n}$ in $\IC^{2}/\IZ_{n}\times \IC^{2}/\IZ_{n}$, which corresponds to the second line of lattice points in the toric diagram. 

In conclusion, therefore, QCS theories for \emph{all} toric sub-diagrams can be obtained by Higgsing the original theory, and these are all orbifolds of the form $\IC^2/\IZ_{l}\times \IC^2/\IZ_{m}$ for $1 \leq l,m\leq n$.

%=====================================================
\section{A complete family: resolutions of $\IC^4/\IZ_2^3$}
%=====================================================

In the previous section we have seen an example of a QCS theory which, via Higgsing, can generate QCS theories for \emph{all} toric sub-diagrams obtained by partial resolution of the parent. It is natural to wonder if, given an arbitrary toric Calabi-Yau four-fold $X$, there is a systematic way in which we can construct a QCS theory for $X$ via this method. Indeed, recall that in the case of four-dimensional gauge theories on D3-branes it has been shown  \cite{Douglas:1997de,Morrison:1998cs,Beasley:1999uz,Feng:2000mi} that a $d=(3+1)$, $\mathcal{N}=1$ quiver gauge theory on a D3-brane transverse to any toric Calabi-Yau three-fold can be obtained by partial resolution of an appropriate Abelian orbifold $\IC^3 / (\IZ_n)^2$, for sufficiently large $n \in \IZ$. 
The latter may then be constructed as a DM orbifold projection of $\mathcal{N}=4$ SYM \cite{Douglas:1996sw,Lawrence:1998ja,Hanany:1998sd}, as already mentioned. 

\begin{figure}[H]
\includegraphics[trim=0mm 5mm 0mm 5mm, clip, width=6in]{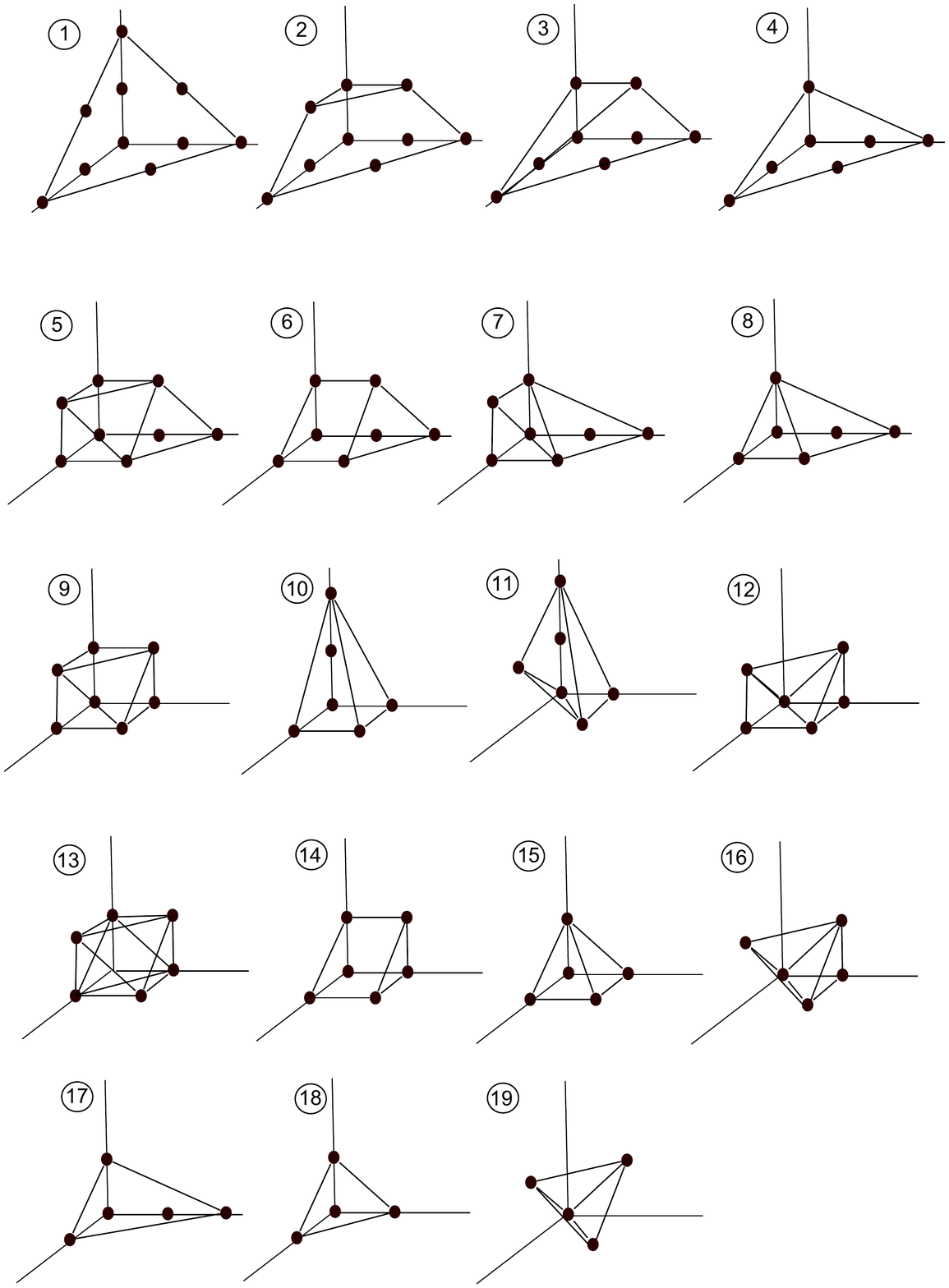}
\caption{{\sf All 18 toric sub-diagrams of that of the parent orbifold $\IC^4/\IZ_2^3$, including the latter itself which is presented in (1). Each diagram corresponds to an inequivalent affine toric Calabi-Yau four-fold, which is a partial resolution of 
$\IC^4/\IZ_2^3$.}}
\label{f:c4z2cube}
\end{figure}

Motivated by the works \cite{Beasley:1999uz, Feng:2000mi}, which obtained D3-brane gauge theories on all partial resolutions of $\IC^3/\IZ_2^2$ and $\IC^3/\IZ_3^2$ via Higgsing, we consider M2-branes probing the orbifold $\IC^4/\IZ_2^3$. Here the three $\IZ_2$ generators act 
on the four coordinates $(x^1,x^2,x^3,x^4)$ by multiplication by $(-1,-1,1,1)$, $(-1,1,-1,1)$, $(-1,1,1,-1)$, respectively. The toric diagram is 
presented in Figure~\ref{f:c4z2cube} (1). It has ten vertices and is simply a rescaling of the toric diagram of $\IC^4$ in each direction by a factor of two. 
The remaining diagrams in Figure~\ref{f:c4z2cube} represent all possible partial resolutions of $\IC^4/\IZ_2^3$, {\it i.e.} inequivalent toric sub-diagrams of that of $\IC^4/\IZ_2^3$. 
We have drawn the lattice points for the toric diagrams in a standard three-dimensional projection, so that, for example, the ten vertices for $\IC^4/\IZ_2^3$ are $(0, 0, 0)$, $(1, 0, 0)$, $(2, 0, 0)$, $(0, 1, 0)$, $(0, 2, 0)$, $(0, 0, 1)$, $(0,0, 2)$, $(1, 1, 0)$, $(1, 0, 1)$, $(0, 1, 1)$. 

There are 18 inequivalent children which have at least four vertices and which are non-coplanar; this is to guarantee that the geometry is really that of a Calabi-Yau four-fold, rather than say a three-fold. 
Moreover, recall from Subsection \ref{sec:toric} that two toric diagrams give an equivalent affine four-fold if and only if they are related by an $SL(4, \IZ)$-transformation. 
Each diagram in Figure~\ref{f:c4z2cube} is in a different equivalence class. Note that the list is exhaustive; that is, we have found \emph{all} possible $SL(4, \IZ)$-inequivalent sub-diagrams.

An immediate problem, already mentioned in Subsection \ref{sec:orb}, is that when taking a DM projection of a QCS theory, the order of the orbifold group must divide the CS levels of the parent. This means that a $(\IZ_2)^2$ DM quotient of the ABJM theory necessarily gives $\IC^4/(\IZ_2)^2\times\IZ_4$ as the minimal model \cite{Imamura:2008nn,Terashima:2008ba}. As far as the authors are aware, it is therefore not possible to obtain a QCS theory for $\IC^4/\IZ_2^3$ by a projection of the ABJM theory. We are thus naturally led to wonder if there are other methods by which we can find the QCS theory for an M2-brane probing $\IC^4/\IZ_2^3$, the natural analogue of $\IC^3/\IZ_2^2$  for D3-branes. 
We will use two different approaches. First, we will start by \emph{un}-Higgsing the two well-known theories which we called $(\IC^4)_I$, $(\IC^4)_{II}$ in Subsection \ref{s:C4}. This leads to a phase of the desired theory, which we will call $(\IC^4/\IZ_2^3)_I$. Second, we will examine another phase, $(\IC^4/\IZ_2^3)_{II}$, which will be constructed by lifting a parent theory from Type IIA to M-theory. 

%=====================================================
\subsection{The $(\IC^4/\IZ_2^3)_{I}$ theory}\label{sec:unhiggsing}
%=====================================================
In this section we wish to \emph{un}-Higgs $(\IC^4)_I$ to obtain a theory with $\IC^4/\IZ_2^3$ as VMS. We thus begin with a discussion of the known constraints on QCS theories, and then describe a general un-Higgsing algorithm. This is then applied to the ABJM theory to obtain a phase $(\IC^4/\IZ_2^3)_I$. We then study the Higgsing behaviour of the latter theory by giving VEVs to all possible combinations of bifundamental fields.
%=====================================================
\subsubsection{Calabi-Yau, toric and tiling conditions}
%=====================================================
We begin by reviewing the conditions which should be satisfied by a QCS theory on M2-branes probing a non-compact toric Calabi-Yau four-fold. 

In order that the VMS, and any (partial) resolution of it obtained by turning on FI parameters, is Calabi-Yau, we require that for each node in the quiver the number of arrows entering and leaving the node should be equal. This condition then guarantees that the $G_t$ matrix, which is the null-space of the charge matrix, can be put into a form with a row of $1$s by an appropriate $SL(4,\IZ)$ transformation -- see the discussion in Subsection~\ref{sec:toric}. Notice this is the same condition as gauge anomaly cancellation in the $(3+1)$-dimensional YM parent.

The superpotential satisfies the \emph{toric condition} if each chiral multiplet appears precisely twice in $W$: once with a positive sign and once with a negative sign. This ensures that the solution to the F-term equations is a toric variety.

The last condition that we want to impose is the so-called \emph{tiling condition}. All known quiver theories related to toric Calabi-Yaus, in both $(3+1)$ and $(2+1)$ dimensions, obey this condition due to their brane-tiling/dimer model description \cite{dimerrev,Hanany:2008cd,Hanany:2008fj}. This leads to the elegant condition
\begin{equation}\label{euler}
G - E + N_T = 0~,
\end{equation}
where $G$ is the number of nodes, $E$ is the number of fields and $N_T$ is the number of terms in the superpotential. It is intriguing that this relation, suggestive of a planar, rather than solid, tiling, still holds for all theories we have constructed in this paper. 
In the next subsection we will un-Higgs theories that obey this condition, and see that whenever the rule is broken in the resulting theory the dimension of its VMS is no longer four.
\begin{figure}[H]
\includegraphics[trim=0mm 250mm 0mm 0mm, clip, width=16in]{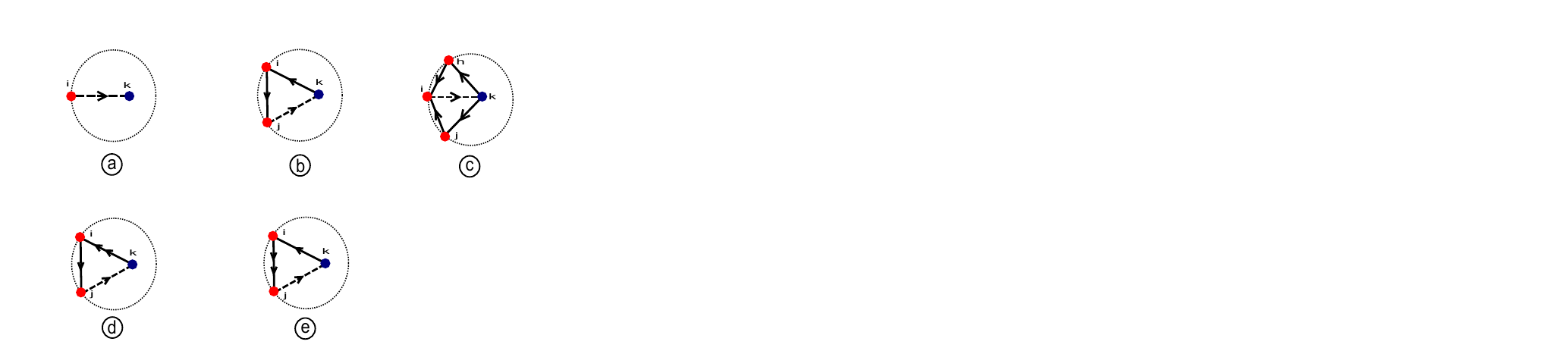}
\caption{{\sf The un-Higgsing process. Gauge nodes of the original theory (in red) appear on the boundary of the circle, while the gauge node that has been introduced in the un-Higgsing appears (in blue) inside the circle: (a) adding one field; (b) adding three fields; (c) adding five fields; (d) and (e) adding four fields.
}}
\label{f:unhiggsingdiagram}
\end{figure}
%
%=====================================================
\subsubsection{The un-Higgsing algorithm}
%=====================================================
The un-Higgsing procedure for quiver gauge theories was studied in \cite{Feng:2002fv} in the context of D3-branes probing complex cones over del Pezzo surfaces. Here we wish to systematize this method and use it as a guide for constructing QCS theories living on M2-brane worldvolumes. 
As we will explain shortly, the un-Higgsing process for theories with toric Calabi-Yau four-folds as VMS is quite restrictive.  The basic idea is that by adding one gauge node at a time we can obtain theories whose VMSs contain the original toric diagram as a sub-diagram; thus this will be a QCS form of ``blow-up''. 

Let us begin with the simplest case: un-Higgsing by adding one field to the quiver. This step is shown schematically in Figure~\ref{f:unhiggsingdiagram} (a). The gauge nodes which sit on the circumference of the dotted circle are those in the original theory which is being un-Higgsed. The gauge node sitting inside the circle is that being added to the theory. We have indicated the original node in red, indexed by $i$,  and the new node in blue, indexed by $k$. We shall say that $i$ \emph{participates} in the un-Higgsing process, because it will be attached to $k$, while all other nodes in the original theory are non-participatory.

Next, we add to the original quiver a bifundamental field $X_{i,k}$ charged under $({N_i},\bar{N_k})$; this is an arrow connecting node $i$ to node $k$. The key point in un-Higgsing is that we must be able to Higgs the new theory to the original one by letting $X_{i,k}$ acquire a non-zero VEV. To continue to satisfy the toric condition, the field $X_{i,k}$ must be added simultaneously to a positive and a negative term which already appears in the superpotential, and no extra terms should be introduced. In order to exhaust all possiblities for constructing new consistent theories the $i$ index should run over all values between $1$ and $G$, where $G$ is the number of gauge nodes in the original quiver. Moreover, the field $X_{i,k}$ must be inserted to all possible pairs of negative and positive terms in the superpotential. 

However, notice that after adding $X_{i,k}$ to the quiver, the Calabi-Yau condition mentioned in the previous subsection is broken: the number of arrows that enter node $i$ or node $k$ is not equal to the number of those that leave. To remedy this we need to relocate the heads and tails of arrows in the original quiver between node $i$ and node $k$. For example, for a three-noded quiver with nodes $i$, $i_1$ and $i_2$ we can do this by changing the tail of $X_{i,i_1}$ to $k$:
\begin{equation}
\includegraphics[trim=0mm 250mm 0mm 0mm, clip, width=6in]{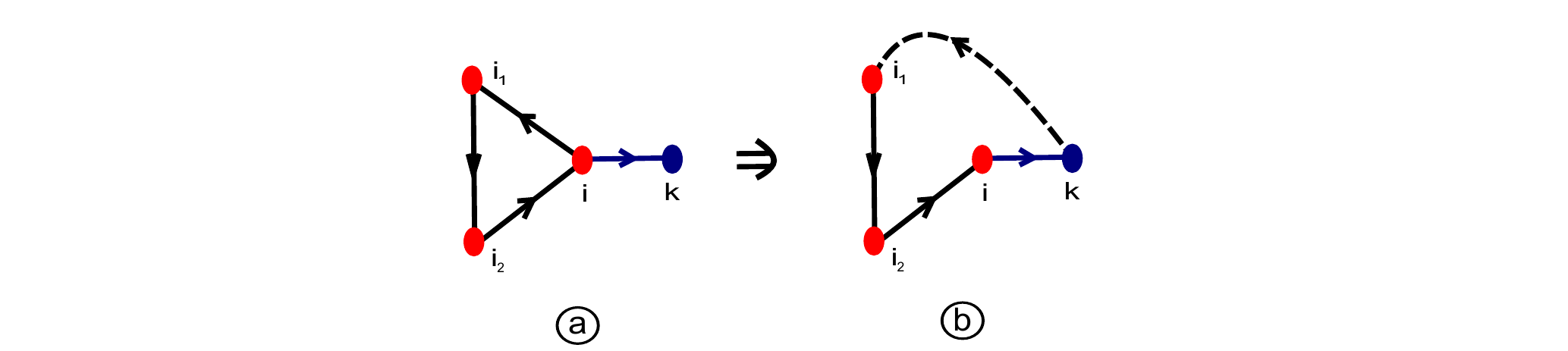}
\end{equation}
Finally, we assign CS levels to nodes $i$ and $k$ such that their sum is equal to the original CS level of node $i$.

Next we turn to more complicated un-Higgsing possiblities. Adding more than one field forces us to add terms to the superpotential, instead of simply adjoining the fields to existing terms, as was the case above; otherwise, it would be impossible to obtain the original quiver by Higgsing. The only possibility is that after introducing such new terms to the superpotential, some of the fields will become massive after the Higgsing and will be integrated out. Therefore, we see immediately that it is not possible to un-Higgs the theory by adding only two fields: insertion of a term that contains two fields is not a valid un-Higgsing step as these fields would be integrated out even before Higgsing because we would be adding a quadratic mass term.

Hence, let us move on to consider introducing three new fields. In accordance with the labelling in Figure~\ref{f:unhiggsingdiagram} (b), the three fields are denoted $X_{i,j}$, $X_{j,k}$ and $X_{k,i}$, where $X_{j,k}$ is the field which we wish to Higgs in order to reproduce the original theory in the IR. Since the three fields should disappear from the IR theory after Higgsing, there must be a cubic term in the superpotenital which contains all three. This new cubic term should be gauge invariant, and thus the fields which we add must form a closed loop. Notice that after Higgsing $X_{j,k}$ we are left with a term that contains two fields: $X_{i,j}$ and $X_{k,i}$. Those fields should be integrated out in the IR as they give rise to a quadratic mass term. 

To satisfy the toric condition, $X_{i,j}$, $X_{j,k}$ and $X_{j,k}$ should also appear in other terms in the superpotential and have opposite sign with respect to the cubic term. Furthermore, we must satisfy\footnote{{\it A priori}, violating this condition is not a problem. However, in all cases that we have studied the resulting theory will then have a five complex-dimensional VMS.} the tiling condition \eref{euler}. 
Now, since we have added one node and three fields, we must add two terms to the superpotential. The cubic term mentioned above is one of them. What about the other? There are two options: to add a new term or to split one of the existing terms into two. The first option would just be the cubic term with opposite sign, which would simply cancel in the Abelian theory and hence is ineffective. We must therefore take the second option and split an existing term, inserting $X_{i,j}$ and $X_{k,i}$ separately into the two split terms. This guarantees that after integrating out these fields the split terms are united. 
To see this in more detail, suppose the original superpotential contains a term $AB$, where $A$ and $B$ are monomials in bifundamental fields; that is: $W = AB + \ldots$. Then our procedure would change this superpotential to
$W = A\ X_{i,j} + B\ X_{k,i} - X_{i,j}X_{j,k}X_{k,i} + \ldots$. When $X_{j,k}$ acquires a VEV (say $\langle X_{j,k}\rangle=1$ for convenience), the equation of motion for $X_{i,j}$ becomes $A = X_{k,i}$, and the first and third terms cancel while the middle term becomes $AB$, as required.
Finally, $X_{j,k}$ can be added to an arbitrary term with the opposite sign to the cubic term. 

In order to exhaust all possibilities we split terms, insert fields, assign CS levels and vary $i$ and $j$ in all possible combinations (notice that $i$ could equal $j$ for the case of adjoint fields). Furthermore, we insert the cubic term both with positive and negative signs and allow relocation of heads or tails of arrows involving nodes $i$ and $k$ in ways that satisfy the Calabi-Yau condition, as in the case of adding one field.   

The next possibility is un-Higgsing by introducing four new fields. As shown in Figure~\ref{f:unhiggsingdiagram}  (d) and (e), this can be done in two different ways. Let us discuss (d) first. Notice that the only way this can be achieved is by insertion of two new cubic terms into the superpotential: $X_{j,k}X_{k,i}^1 X_{i,j}-X_{j,k}X^2_{k,i} X_{i,j}$. However, this violates the Calabi-Yau condition on both nodes $i$ and $k$. Since the field to be Higgsed is $X_{j,k}$, we can transform heads and tails of arrows between nodes $j$ and $k$ only and cannot fix the Calabi-Yau condition on node $i$. Therefore this un-Higgsing step is allowed only when $i$ is equal to $j$. The same analysis can be applied for (e), and the result is the same. With this constraint, since we have introduced four new fields, one gauge node, and two new terms to the superpotential, the tiling condition is violated. In the theories that we have checked this results in five-dimensional VMSs. We hence cannot introduce four fields.

The final un-Higgsing process involves insertion of five new fields. 
Careful examination implies that this can be done by introducing two cubic terms into the superpotential with opposite signs. If we use the notation of Figure~\ref{f:unhiggsingdiagram} (c), we can write the terms as follows: $X_{i,k}X_{k,h}X_{h,i}-X_{i,k}X_{k,j}X_{j,i}$ ($i$, $j$ and $h$ can be equal). 
Notice that by Higgsing $X_{i,k}$ we obtain two terms in the superpotential that contain two fields each, and therefore four fields should be integrated out. 
By a similiar analysis to the above, after satisfying the tiling condition by splitting terms in the superpotential it can be seen that $X_{k,h}$ and $X_{h,i}$ should appear in different split negative terms. 
Similarly, $X_{k,j}$ and $X_{j,i}$ should appear in different split positive terms. 

Finally, note that five fields is the maximum number of fields that can be introduced if one wants to obtain the original theory by Higgsing only one field. This concludes the discussion of our un-Higgsing algorithm.

%=====================================================
\subsubsection{Obtaining the $(\IC^4/\IZ_2^3)_I$ phase}
%=====================================================
With the aid of a computer we may apply the un-Higgsing algorithm described above to theory $(\IC^4)_I$ or theory $(\IC^4)_{II}$. In each step of the un-Higgsing we add one point to the toric diagram. 

\TABLE[h!t!b!]{
\caption{{\sf Stepwise un-Higgsing the $(\IC^4)_I$ theory to the $(\IC^4/\IZ_2^3)_I$ theory. At each step the fields added, quiver (numbered according to Figure~\ref{f:Unhiggsphase}), superpotential, toric diagram  of the vacuum moduli space $X$ (numbered according to Figure~\ref{f:c4z2cube}), and dual candidate (numbered according to Figure~\ref{f:duals}) are indicated.}}
\begin{tabular}{|c|c|c|c|c|c|} \hline
Step & Fields added & Quiver & Superpotential & $X$ & Duals\\ \hline\hline
0 & - & (a) & 
{\tiny
$\begin{array}{l}
X^1_{1,3} X^1_{3,1} X^2_{1,3} X^2_{3,1} - X^2_{1,3} X^1_{3,1} X^1_{1,3} X^2_{3,1}
\end{array}$
}
& (18) & \\
\hline
1 & ${X_{3,5}}$ & (e) & 
{\tiny
$\begin{array}{l}
X_{5,1} X^1_{1,3} X_{3,1} X^2_{1,3} X_{3,5} - X_{5,1} X^2_{1,3} X_{3,1} X^1_{1,3} X_{3,5}
\end{array}$
}
& (16) & (e$_2$)\\
\hline
2 & ${X_{2,1}}$ & (i) & 
{\tiny
$\begin{array}{l}
X_{5,1} X^1_{1,3} X_{3,2} X_{2,1} X^2_{1,3} X_{3,5} - \\
X_{5,1} X^2_{1,3} X_{3,2} X_{2,1} X^1_{1,3} X_{3,5}
\end{array}$
}
& (12) & (i$_{2-3}$)\\
\hline
3 & ${X_{1,7}}$ & (l) &  
{\tiny
$\begin{array}{l}
X_{5,1} X_{1,3} X_{3,2} X_{2,1} X_{1,7} X_{7,3} X_{3,5} - \\
X_{5,1} X_{1,7} X_{7,3} X_{3,2} X_{2,1} X_{1,3} X_{3,5}
\end{array}$
}
& (9) & (l$_{2-6}$)\\
\hline
4 & ${X_{2, 4}, X_{4, 3}, X_{3, 2}^1}$ & (p) &
{\tiny
$\begin{array}{l}
 X_{1, 7} X_{7, 3} X_{3, 2}^1 X_{2, 1} - X_{2, 4} X_{4, 3} X_{3, 2}^1 + \\
 X_{1, 4} X_{4, 3} X_{3, 2}^2 X_{2, 4} X_{4, 5} X_{5, 1} - \\
 X_{1, 4} X_{4, 5} X_{5, 1} X_{1, 7} X_{7, 3} X_{3, 2}^2 X_{2, 1}
\end{array}$
} 
& (5) & \\
\hline
5 & ${X_{2, 5}, X_{5, 6}, X_{6, 2}}$ & (s) & 
{\tiny
$\begin{array}{l}
 X_{2, 4} X_{4, 6} X_{6, 2} - X_{2, 4} X_{4, 3} X_{3, 2}^1 + \\
 X_{1, 7} X_{7, 3} X_{3, 2}^1 X_{2, 1} - X_{2, 5} X_{5, 6} X_{6, 2} + \\
 X_{1, 4} X_{4, 3} X_{3, 2}^2 X_{2, 5} X_{5, 6} X_{6, 1} - \\
 X_{1, 4} X_{4, 6} X_{6, 1} X_{1, 7} X_{7, 3} X_{3, 2}^2 X_{2, 1}
\end{array}$
}
& (2) &\\
\hline
6 & ${X_{2, 8}, X_{8, 4}, X_{4, 2}}$ & (t) & 
{\tiny
$\begin{array}{l}
 -X_{1, 4} X_{4, 2} X_{2, 1} - 
 X_{2, 4} X_{4, 3} X_{3, 2} + \\
 X_{2, 4} X_{4, 6} X_{6, 2} + 
 X_{2, 8} X_{8, 4} X_{4, 2} + \\
 X_{1, 7} X_{7, 3} X_{3, 2} X_{2, 1} - 
 X_{2, 8} X_{8, 5} X_{5, 6} X_{6, 2} - \\
 X_{1, 7} X_{7, 3} X_{3, 8} X_{8, 4} X_{4, 6} X_{6, 1} + \\
 X_{1, 4} X_{4, 3} X_{3, 8} X_{8, 5} X_{5, 6} X_{6,1}
\end{array}$
}
& (1) & \\ \hline
\end{tabular}
\label{t:a2t}
}

Let us describe the un-Higgsing of theory $(\IC^4)_I$, the standard ABJM theory. For the present purposes we will call this quiver (a). We find that by adding new fields stepwise we can indeed arrive at a theory whose VMS is $\IC^4/\IZ_2^3$, or diagram (1) in \fref{f:c4z2cube}. We present the intermediate results in Table~\ref{t:a2t}. Here we have listed the quiver, numbered according\footnote{Figure~\ref{f:Unhiggsphase} also summarizes results obtained later in this section.} to Figure~\ref{f:Unhiggsphase}, the superpotential of the non-Abelian theory, as well as the resulting toric moduli spaces, the latter numbered according to Figure~\ref{f:c4z2cube} above. 

Theory (e), and its dual (e$_2$), in Table~\ref{t:a2t} will be discussed in more detail later in the paper. Note that their VMS is $C(Y^{1,2}(\mathbb{CP}^2))$, where $Y^{1,2}(\mathbb{CP}^2)$ is one of the explicit Sasaki-Einstein seven-manifolds discussed in \cite{Martelli:2008rt} whose toric diagram is number (16) in our list. 
We shall also refer to these theories as $C(Y^{1,2}(\mathbb{CP}^2))_{Ia}$ and $C(Y^{1,2}(\mathbb{CP}^2))_{Ib}$, respectively (whenever the discussion is relevant for both phases we will omit the $a$($b$) subscript). 

Theories (s) and (p) have toric diagrams in which there are external vertices with multiplicities greater than one. This is an issue first raised in \cite{Feng:2000mi}:  it has been suggested that M2-brane theories, as well as D3-brane theories, should have external multiplicities equal to one \cite{Hanany:2008cd,Hanany:2008fj}. We have applied the algorithm together with the constraint that the external multiplicities in the toric diagrams are all equal to one. Although this produces QCS theories up to toric diagram (2), it is not possible to produce a theory for (1) this way, at least if we are limiting\footnote{Indeed, it is not possible to exhaustively un-Higgs without limiting the number of gauge nodes, as we can always un-Higgs a theory to a theory with the same VMS.} the number of gauge nodes to 8. We will briefly discuss this external multiplicity issue further in Subsection \ref{sec:multiplicity}. 
\subsubsection{Dualities}\label{sec:duality}
\begin{figure}[H]
\includegraphics[trim=0mm 75mm 0mm 00mm, clip, width=6in]{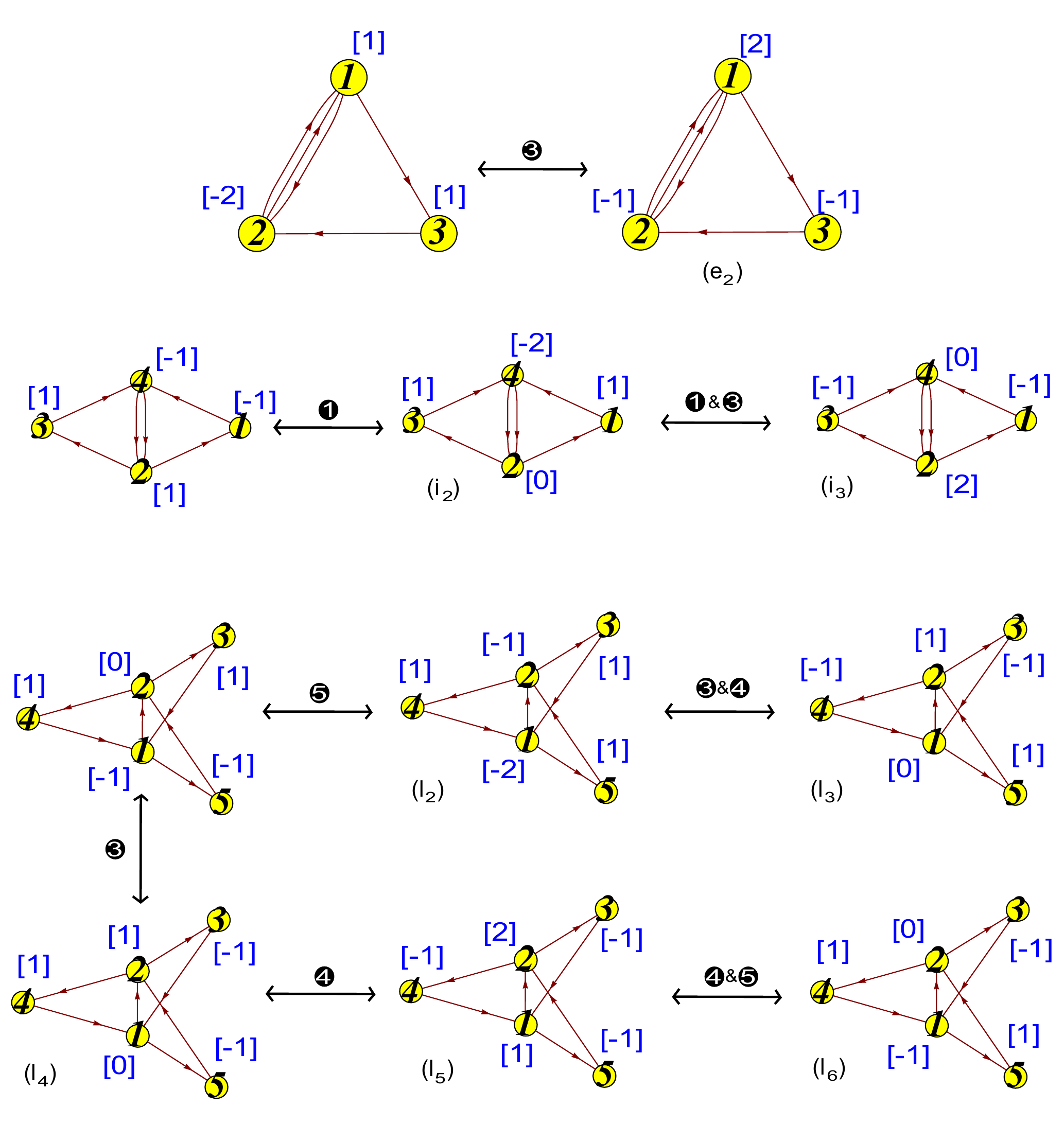}
\caption{{\sf Dual candidates. The quivers and superpotentials in each class are identical to those of the parent in Table 1. To transform between theories within the same class one performs the CS level transformation on gauge nodes indicated by black dots. CS levels are indicated in (blue) square brackets.
}}
\label{f:duals}
\end{figure}

In some steps in the un-Higgsing process more than one theory can be obtained with the same VMS. For toric diagrams with no internal vertices ({\it i.e.} (9), (12) and (16) in the case at hand) it is possible to exhaustively list these dual theories if we restrict the external multiplicites to one. Those theories are shown in Figure~\ref{f:duals}. 

In order to examine this in more detail, let us define an operation with respect to gauge node $i$ in the following way:
\bea
&& k_i \rightarrow -k_i \nn \\
&& k_j \rightarrow k_j+k_i~,
\label{rules}
\eea
where $j$ indexes nodes which are connected to node $i$. We will show that, with this operation, each class of dual theories form a closed system, {\it i.e.} by performing the operation that we have just defined on single-flavour nodes (nodes with one arrow entering and one arrow leaving) it is possible to obtain all other theories in the class, and no others. These transformation rules for the CS levels, as observed in \cite{Imamura:2008nn,Amariti:2009rb}, are related to changing the order of two $(1,p)$-branes on a circle in theories with Type IIB brane models. By examining the VMS equations we will show in general that the rule \eqref{rules} leaves the VMS invariant, provided one applies the transformation only to single-flavour nodes. Notice this is distinct from Seiberg duality in $(3+1)$ dimensions, and dualities in $(2+1)$ dimensions that were observed in \cite{Amariti:2009rb}, in that the quiver is unchanged, and only the CS levels are altered.
To see the above claim, let us concentrate on the following piece of quiver:
\begin{figure}[H]
\includegraphics[trim=0mm 255mm 0mm 10mm, clip, width=6in]{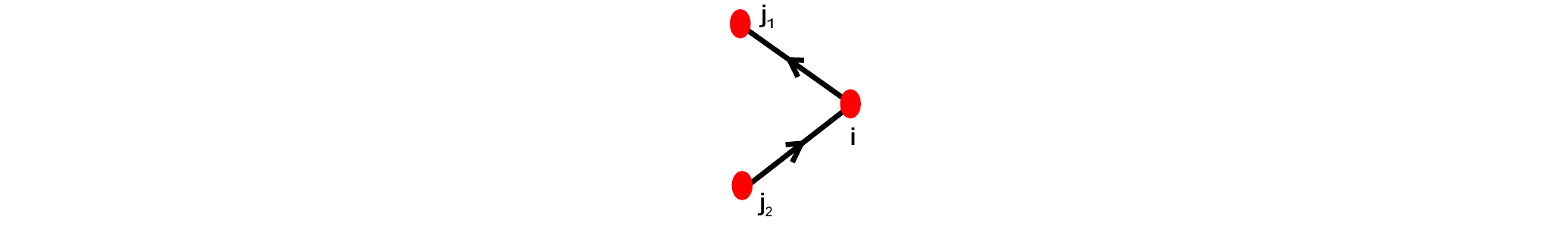}
\label{f:singleflavour}
\end{figure}
We write the Abelian VMS equations and concentrate on the branch in which the $\sigma$s are equal. The D-terms can be written as follows
\begin{align}
\nn
\mu _{i} &\equiv X_{i,j_1} X_{i,j_1}^\dagger - X_{j_2,i}^{\dagger} X_{j_2,i} =\frac{k_i}{2 \pi} \sigma~, \\ \nn
\mu _{j_1} &\equiv -X_{i,j_1}^\dagger X_{i,j_1} + \ldots =\frac{k_{j_1}}{2 \pi} \sigma~, \\
\mu _{j_2} &\equiv X_{j_2,i} X_{j_2,i}^{\dagger} + \dots =\frac{k_{j_2}}{2 \pi} \sigma~.
\end{align}
If we define
\bea
\nn
\tilde{\mu}_{i}\equiv-\mu_{i} \ , \quad \ \tilde{\mu}_{j_1}\equiv\mu_{j_1}+\mu_{i} \ , \quad \ \tilde{\mu}_{j_2}\equiv\mu_{j_2}+\mu_{i}~,
\eea
we can rewrite these D-terms as follows
\begin{align}
\nn
\tilde{\mu} _{i} &= -X_{i,j_1} X_{i,j_1}^\dagger + X_{j_2,i}^{\dagger} X_{j_2,i} =\frac{-k_i}{2 \pi} \sigma~, \\ \nn
\tilde{\mu} _{j_1} &= - X_{j_2,i}^{\dagger} X_{j_2,i} + \ldots =\frac{k_{j_1}+k_i}{2 \pi} \sigma~, \\
\tilde{\mu} _{j_2} &= X_{i,j_1} X_{i,j_1}^\dagger + \dots =\frac{k_{j_2}+k_i}{2 \pi} \sigma~.
\end{align}
If we now relabel $X_{i,j_1} \leftrightarrow X_{j_2,i}$ we obtain the same D-terms as before, only with the substitutions
\bea
k_i \rightarrow -k_i \ , \quad \ k_{j_1} \rightarrow k_{j_1}+k_i , \quad \ k_{j_2} \rightarrow k_{j_2}+k_i~.
\eea 
These are precisely the rules given in \eqref{rules}. Notice that the other vacuum equations \eqref{DF} are invariant under the relabelling. Indeed, the third equation in \eqref{DF} is invariant since all the $\sigma$s are equal. Moreover, the first equation, which is the F-term, is also invariant since the superpotential is invariant. To see this notice that $X_{i,j_1}$ and $X_{j_2,i}$ must appear in the same terms in the superpotential, otherwise the terms would not be gauge-invariant. 

%%%%
%=====================================================
\subsubsection{Higgsing $(\IC^4/\IZ_2^3)_I$}\label{s:higgsz2cubeI}
%=====================================================
%\begin{table}[h!b!p!]
\TABLE[h!b!p!]{
\caption{{\sf The toric diagrams and superpotentials for the possible Higgsings of theory (t), which is theory $(\IC^4/\IZ_2^3)_I$. Also listed are the corresponding toric moduli spaces $X$, numbered according to Figure~\ref{f:c4z2cube}. Note that all 19 toric sub-diagrams are obtained.}}
\begin{tabular}{|c|c|c|c|} \hline
& Toric diagram $G_t$ & Superpotential $W$ & $X$ \\ \hline \hline
(a) &
{\tiny
$\left (
\begin {array} {llll}
$-1$ & 0 & 0 & $-1$ \\[-1mm]
0 & 1 & 0 & 1 \\[-1mm]
0 & 0 & 1 & 1 \\[-1mm]
2 & 0 & 0 & 0
\end {array} \right)
$}
&
{\tiny
$\begin{array}{l}
 X_{1, 3}^1 X_{3, 1}^1 X_{1, 3}^2 X_{3, 1}^2 - 
 X_{1, 3}^1 X_{3, 1}^2 X_{1, 3}^2 X_{3, 1}^1
\end{array}$
}
& (19)
\\ \hline
(b) &
{\tiny
$\left (
\begin {array} {llll}
1 & 0 & 0 & 0 \\[-1mm]
0 & 1 & 0 & 0 \\[-1mm]
0 & 0 & 1 & 0 \\[-1mm]
0 & 0 & 0 & 1
\end {array} \right)
$}
&
{\tiny
$\begin{array}{l}
 X_{1, 5} X_{5, 1} X_{1, 1}^1 X_{1, 1}^2 - 
 X_{1, 5} X_{5, 1} X_{1, 1}^2 X_{1, 1}^1
\end{array}$
}
& (18)
\\ \hline
(c) &
{\tiny
$\left (
\begin {array} {llll}
1 & 0 & 0 & 0 \\[-1mm]
0 & 1 & 0 & 0 \\[-1mm]
0 & 0 & 1 & 0 \\[-1mm]
0 & 0 & 0 & 1
\end {array} \right)
$}
&
{\tiny
$\begin{array}{l}
 X_{1, 3}^1 X_{3, 1}^1 X_{1, 3}^2 X_{3, 1}^2 - 
 X_{1, 3}^1 X_{3, 1}^2 X_{1, 3}^2 X_{3, 1}^1
\end{array}$
}
& (18)
\\ \hline
(d) &
{\tiny
$\left(
\begin{array}{llllll}
 2 & 0 & 0 & 0 & 1 & 1 \\[-1mm]
 $-1$ & 0 & 0 & 1 & 0 & 0 \\[-1mm]
 0 & 0 & 1 & 0 & 0 & 0 \\[-1mm]
 0 & 1 & 0 & 0 & 0 & 0
\end{array}
\right)
$}
&
{\tiny
$\begin{array}{l}
 -X_{1, 3}^1 X_{3, 2}^1 X_{2, 1}^1 + 
 X_{1, 3}^2 X_{3, 2}^1 X_{2, 1}^2 + \\
 X_{1, 1} X_{1, 3}^1 X_{3, 2}^2 X_{2, 1}^1 - 
 X_{1, 1} X_{1, 3}^2 X_{3, 2}^2 X_{2, 1}^2 
\end{array}$
}
& (17)
\\ \hline
(e) &
{\tiny
$\left(
\begin{array}{lllll}
 $-1$ & 0 & 0 & 0 & 1 \\[-1mm]
 1 & 0 & 0 & 1 & 0 \\[-1mm]
 2 & 0 & 1 & 0 & 0 \\[-1mm]
 $-1$ & 1 & 0 & 0 & 0
\end{array}
\right)
$}
&
{\tiny
$\begin{array}{l}
 X_{1, 3}^1 X_{3, 1} X_{1, 3}^2 X_{3, 5} X_{5, 1} - \\
 X_{1, 3}^2 X_{3, 1} X_{1, 3}^1 X_{3, 5} X_{5, 1}
\end{array}$
}
& (16)
\\ \hline
(f) &
{\tiny
$\left(
\begin{array}{lllll}
 0 & 0 & 0 & 0 & 1 \\[-1mm]
 $-1$ & 0 & 0 & 1 & 0 \\[-1mm]
 1 & 0 & 1 & 0 & 0 \\[-1mm]
 1 & 1 & 0 & 0 & 0
\end{array}
\right)
$}
&
{\tiny
$\begin{array}{l}
 X_{1, 2} X_{2, 1} X_{1, 1} X_{1, 5} X_{5, 1} - \\
 X_{1, 5} X_{5, 1} X_{1, 1} X_{1, 2} X_{2, 1}
\end{array}$
}
& (15)
\\ \hline
(g) &
{\tiny
$\left(
\begin{array}{llllll}
 1 & 1 & 0 & 0 & 0 & 1 \\[-1mm]
 0 & 1 & 0 & 0 & 1 & 0 \\[-1mm]
 1 & 0 & 0 & 1 & 0 & 0 \\[-1mm]
 $-1$ & $-1$ & 1 & 0 & 0 & 0
\end{array}
\right)
$}
&
{\tiny
$\begin{array}{l}
 X_{1, 2} X_{2, 1} X_{1, 7} X_{7, 1} X_{1, 5} X_{5, 1} - \\
 X_{1, 7} X_{7, 1} X_{1, 2} X_{2, 1} X_{1, 5} X_{5, 1}
\end{array}$
}
& (14)
\\ \hline
(h) &
{\tiny
$\left(
\begin{array}{llllll}
 0 & 0 & $-1$ & 0 & 0 & 1 \\[-1mm]
 1 & 0 & 1 & 0 & 1 & 0 \\[-1mm]
 1 & 0 & 1 & 1 & 0 & 0 \\[-1mm]
 $-1$ & 1 & 0 & 0 & 0 & 0
\end{array}
\right)
$}
&
{\tiny
$\begin{array}{l}
 X_{3, 2} X_{2, 1} X_{1, 3}^1 X_{1, 5} X_{5, 1} X_{1, 3}^2 - \\
 X_{3, 2} X_{2, 1} X_{1, 3}^2 X_{1, 5} X_{5, 1} X_{1, 3}^1
\end{array}$
}
& (13)
\\ \hline
(i) &
{\tiny
$\left(
\begin{array}{llllll}
 $-1$ & 0 & 0 & 0 & 0 & 1 \\[-1mm]
 0 & 1 & 0 & 0 & 1 & 0 \\[-1mm]
 1 & $-1$ & 0 & 1 & 0 & 0 \\[-1mm]
 1 & 1 & 1 & 0 & 0 & 0
\end{array}
\right)
$}
&
{\tiny
$\begin{array}{l}
 X_{3, 2} X_{2, 1} X_{1, 3}^1 X_{1, 5} X_{5, 1} X_{1, 3}^2 - \\ 
 X_{3, 2} X_{2, 1} X_{1, 3}^2 X_{1, 5} X_{5, 1} X_{1, 3}^1
\end{array}$
}
& (12)
\\ \hline
(j) &
{\tiny
$\left(
\begin{array}{lllllll}
 1 & 0 & 0 & 1 & 2 & 0 & 1 \\[-1mm]
 0 & 0 & 0 & 0 & $-1$ & 1 & 0 \\[-1mm]
 1 & 0 & 1 & 0 & 0 & 0 & 0 \\[-1mm]
 $-1$ & 1 & 0 & 0 & 0 & 0 & 0
\end{array}
\right)
$}
&
{\tiny
$\begin{array}{l}
-X_{1, 3}^1 X_{3, 2}^1 X_{2, 1}^1 + 
 X_{1, 3}^1 X_{3, 2}^2 X_{2, 1}^1 X_{1, 5} X_{5, 1} + \\
 X_{1, 3}^2 X_{3, 2}^1 X_{2, 1}^2 -
 X_{1, 3}^2 X_{3, 2}^2 X_{2, 1}^2 X_{1, 5} X_{5, 1}
\end{array}$
}
& (11)
\\ \hline
(k) &
{\tiny
$\left(
\begin{array}{lllllll}
 2 & 0 & 0 & 1 & 2 & 0 & 1 \\[-1mm]
 $-1$ & 0 & 0 & 0 & $-1$ & 1 & 0 \\[-1mm]
 1 & 0 & 1 & 0 & 0 & 0 & 0 \\[-1mm]
 $-1$ & 1 & 0 & 0 & 0 & 0 & 0
\end{array}
\right)
$}
&
{\tiny
$\begin{array}{l}
 X_{1, 3} X_{3, 2}^1 X_{2, 1} - 
 X_{1, 3} X_{3, 2}^2 X_{2, 1} X_{1, 4} X_{4, 1} - \\
 X_{2, 4} X_{4, 3} X_{3, 2}^1 + 
 X_{1, 4} X_{4, 1} X_{4, 3} X_{3, 2}^2 X_{2, 4}
\end{array}$
}
& (10)
\\ \hline
(l) &
{\tiny
$\left(
\begin{array}{lllllll}
 1 & $-1$ & 0 & 0 & 0 & 0 & 1 \\[-1mm]
 1 & 0 & 1 & 0 & 0 & 1 & 0 \\[-1mm]
 $-1$ & 1 & $-1$ & 0 & 1 & 0 & 0 \\[-1mm]
 0 & 1 & 1 & 1 & 0 & 0 & 0
\end{array}
\right)
$}
&
{\tiny
$\begin{array}{l}
X_{3, 2} X_{2, 1} X_{1, 3} X_{3, 5} X_{5, 1} X_{1, 7} X_{7, 3} - \\
X_{3, 2} X_{2, 1} X_{1, 7} X_{7, 3} X_{3, 5} X_{5, 1} X_{1, 3}
\end{array}$
}
& (9)
\\ \hline
(m) &
{\tiny
$\left(
\begin{array}{lllllll}
 0 & 1 & 0 & 0 & 0 & 0 & 1 \\[-1mm]
 2 & 1 & 0 & 0 & 1 & 1 & 0 \\[-1mm]
 $-1$ & $-1$ & 0 & 1 & 0 & 0 & 0 \\[-1mm]
 0 & 0 & 1 & 0 & 0 & 0 & 0
\end{array}
\right)
$}
&
{\tiny
$\begin{array}{l}
 X_{1, 2} X_{2, 1} X_{2, 2} +
 X_{1, 4} X_{4, 2} X_{2, 4} X_{4, 5} X_{5, 1} -  \\
 X_{2, 2} X_{2, 4} X_{4, 2} - 
 X_{1, 2} X_{2, 1} X_{1, 4} X_{4, 5} X_{5, 1} 
\end{array}$
}
& (8)
\\ \hline
(n) &
{\tiny
$\left(
\begin{array}{llllllll}
 $-1$ & 1 & 0 & 0 & 0 & 0 & 0 & 1 \\[-1mm]
 1 & 1 & $-1$ & 1 & 0 & 0 & 1 & 0 \\[-1mm]
 0 & 0 & 1 & 0 & 0 & 1 & 0 & 0 \\[-1mm]
 1 & $-1$ & 1 & 0 & 1 & 0 & 0 & 0
\end{array}
\right)
$}
&
{\tiny
$\begin{array}{l}
 X_{1, 3} X_{3, 2}^1 X_{2, 1} + 
 X_{1, 4} X_{4, 3} X_{3, 2}^2 X_{2, 4} X_{4, 5} X_{5, 1} - \\
 X_{2, 4} X_{4, 3} X_{3, 2}^1 - 
 X_{1, 3} X_{3, 2}^2 X_{2, 1} X_{1, 4} X_{4, 5} X_{5, 1}
\end{array}$
}
& (7)
\\ \hline
(o) &
{\tiny
$\left(
\begin{array}{llllllll}
 $-1$ & $-1$ & 0 & 0 & 1 & 0 & 0 & 1 \\[-1mm]
 1 & 0 & 1 & 0 & 1 & 0 & 1 & 0 \\[-1mm]
 1 & 1 & 0 & 0 & $-1$ & 1 & 0 & 0 \\[-1mm]
 0 & 1 & 0 & 1 & 0 & 0 & 0 & 0
\end{array}
\right)
$}
&
{\tiny 
$\begin{array}{l}
-X_{2, 2} X_{2, 4} X_{4, 2} + 
 X_{1, 4} X_{4, 2} X_{2, 4} X_{4, 1} + \\
 X_{1, 7} X_{7, 2} X_{2, 2} X_{2, 1} - 
 X_{1, 4} X_{4, 1} X_{1, 7} X_{7, 2} X_{2, 1}
\end{array}$
}
& (6)
\\ \hline
(p) &
{\tiny
$\left(
\begin{array}{llllllllll}
 0 & 1 & 1 & 1 & 1 & 2 & 0 & 0 & 0 & 1 \\[-1mm]
 1 & 0 & $-1$ & 0 & 0 & $-1$ & 0 & 0 & 1 & 0 \\[-1mm]
 $-1$ & $-1$ & 0 & 0 & $-1$ & $-1$ & 0 & 1 & 0 & 0 \\[-1mm]
 1 & 1 & 1 & 0 & 1 & 1 & 1 & 0 & 0 & 0
\end{array}
\right)
$}
&
{\tiny
$\begin{array}{l}
 -X_{2, 4} X_{4, 3} X_{3, 2}^1 + 
 X_{1, 7} X_{7, 3} X_{3, 2}^1 X_{2, 1} + \\
 X_{1, 4} X_{4, 3} X_{3, 2}^2 X_{2, 4} X_{4, 5} X_{5, 1} - \\
 X_{1, 4} X_{4, 5} X_{5, 1} X_{1, 7} X_{7, 3} X_{3, 2}^2 X_{2, 1}
\end{array}$
}
& (5)
\\ \hline
(q) &
{\tiny
$\left(
\begin{array}{lllllllllll}
 $-1$ & $-1$ & $-1$ & $-1$ & 0 & 0 & 0 & 0 & 0 & 0 & 1 \\[-1mm]
 1 & 0 & 2 & 1 & 0 & 0 & 1 & 0 & 0 & 1 & 0 \\[-1mm]
 1 & 2 & 0 & 1 & 0 & 1 & 0 & 1 & 1 & 0 & 0 \\[-1mm]
 0 & 0 & 0 & 0 & 1 & 0 & 0 & 0 & 0 & 0 & 0
\end{array}
\right)
$}
&
{\tiny
$\begin{array}{l}
 X_{1, 3} X_{2, 5} X_{3, 2}^{1} X_{5, 1} + 
 X_{1, 1} X_{1, 3} X_{2, 5} X_{3, 2}^{2} X_{5, 1} + \\
 X_{1, 7} X_{2, 1} X_{3, 2}^{1} X_{7, 3} - 
 X_{1, 1} X_{1, 7} X_{2, 1} X_{3, 2}^{2} X_{7, 3}
\end{array}$
}
& (4)
\\ \hline
(r) &
{\tiny
$\left(
\begin{array}{lllllllllllll}
 0 & 1 & 0 & 1 & 0 & 1 & 1 & 0 & 1 & 2 & 0 & 0 & 1 \\[-1mm]
 1 & $-1$ & 2 & 1 & 0 & $-1$ & 0 & 0 & 1 & 0 & 0 & 1 & 0 \\[-1mm]
 0 & 0 & $-1$ & $-1$ & 0 & 0 & 0 & 0 & $-1$ & $-1$ & 1 & 0 & 0 \\[-1mm]
 0 & 1 & 0 & 0 & 1 & 1 & 0 & 1 & 0 & 0 & 0 & 0 & 0
\end{array}
\right)
$}
&
{\tiny
$\begin{array}{l}
 X_{1, 3} X_{3, 2} X_{2, 1} - 
 X_{1, 4} X_{4, 2} X_{2, 1} - \\
 X_{2, 4} X_{4, 3} X_{3, 2} + 
 X_{2, 4} X_{4, 5} X_{5, 2} + \\
 X_{2, 8} X_{8, 4} X_{4, 2} - 
 X_{1, 3} X_{3, 8} X_{8, 4} X_{4, 5} X_{5, 1} + \\
 X_{1, 4} X_{4, 3} X_{3, 8} X_{8, 5} X_{5, 1} - 
 X_{2, 8} X_{8, 5} X_{5, 2}
\end{array}$
}
& (3)
\\ \hline
(s) &
{\tiny
$\left(
\begin{array}{llllllllllllll}
 0 & 1 & 0 & 0 & 1 & 0 & 1 & 1 & 1 & 2 & 0 & 0 & 0 & 1 \\[-1mm]
 $-1$ & $-1$ & $-1$ & 0 & 0 & 0 & $-1$ & 0 & $-1$ & $-1$ & 0 & 0 & 1 & 0 \\[-1mm]
 2 & 1 & 1 & 1 & 0 & 0 & 0 & 0 & 1 & 0 & 0 & 1 & 0 & 0 \\[-1mm]
 0 & 0 & 1 & 0 & 0 & 1 & 1 & 0 & 0 & 0 & 1 & 0 & 0 & 0
\end{array}
\right)
$}
&
{\tiny
$\begin{array}{l}
-X_{2, 4} X_{4, 3} X_{3, 2}^1 + 
 X_{2, 4} X_{4, 6} X_{6, 2} + \\
 X_{1, 7} X_{7, 3} X_{3, 2}^1 X_{2, 1} - 
 X_{2, 5} X_{5, 6} X_{6, 2} - \\
 X_{1, 4} X_{4, 6} X_{6, 1} X_{1, 7} X_{7, 3} X_{3, 2}^2 X_{2, 1} + \\
 X_{1, 4} X_{4, 3} X_{3, 2}^2 X_{2, 5} X_{5, 6} X_{6, 1}
\end{array}$
}
& (2)
\\ \hline
(t) &
{\tiny
$\left(
\begin{array}{llllllllllllllllllll}
 0 & 1 & 0 & 1 & 0 & 0 & 0 & 1 & 1 & 0 & 0 & 0 & 1 & 1 & 1 & 2 & 0 & 0
   & 0 & 1 \\[-1mm]
 0 & 1 & 0 & 0 & 1 & 1 & 1 & 1 & 0 & 1 & 1 & 2 & 1 & 1 & 0 & 0 & 0 & 0
   & 1 & 0 \\[-1mm]
 1 & 0 & 2 & 1 & 0 & 1 & 1 & 0 & 0 & 1 & 1 & 0 & 0 & 0 & 1 & 0 & 0 & 1
   & 0 & 0 \\[-1mm]
 1 & 1 & 1 & 1 & 1 & 1 & 1 & 1 & 1 & 1 & 1 & 1 & 1 & 1 & 1
   & 1 & 1 & 1 & 1 & 1
\end{array}
\right)
$}
&
{\tiny
$\begin{array}{l}
 X_{1, 7} X_{7, 3} X_{3, 8} X_{8, 4} X_{4, 6} X_{6, 1} -
 X_{2, 4} X_{4, 3} X_{3, 2} + \\
 X_{2, 4} X_{4, 6} X_{6, 2} + 
 X_{2, 8} X_{8, 4} X_{4, 2} + \\
 X_{1, 7} X_{7, 3} X_{3, 2} X_{2, 1} - 
 X_{2, 8} X_{8, 5} X_{5, 6} X_{6, 2} - \\
 X_{1, 4} X_{4, 3} X_{3, 8} X_{8, 5} X_{5, 6} X_{6,1} -
 X_{1, 4} X_{4, 2} X_{2, 1}
\end{array}$
}
& (1)
\\ \hline
\end{tabular}
\label{t:unhiggsphase}
}

We have now arrived at our ``parent'' theory, namely the $(\IC^4/\IZ_2^3)_I$ theory. This is expected to be an M2-brane QCS theory because we have obtained it, via the un-Higgsing algorithm, from ABJM theory on an M2-brane in flat spacetime. 
We would now like to determine all possible Higgsings of this theory, and hence find QCS theories for the 18 sub-diagrams in Figure~\ref{f:c4z2cube}, corresponding to all toric partial resolutions of $\IC^4/\IZ_2^3$. 
Specifically, we will give VEVs to all possible subsets of the fields, and determine the resulting low-energy theories at scales well below the scale set by the VEVs. We can then compute the moduli spaces of these theories using the forward algorithm (\ref{Gt}), and compare their toric diagrams to sub-diagrams of that of the parent. As one might imagine, there are hundreds of thousands of possibilities; we have executed these exhaustively with the aid of a computer.

\TABLE[h!t!b!]{
\caption{{\sf The list of fields which acquire VEVs in Higgsing theory (t), {\it i.e.} $(\IC^4/\IZ_2^3)_I$, in order to obtain the various partial resolutions.}}
$\begin{array}{|l|l|}\hline
\mbox{Theory} & \mbox{Higgsed fields} \\ \hline
(a) &\{X_{2, 1}, X_{4, 3}, X_{7, 3}, X_{2, 8}, X_{5, 6}, X_{4, 6}\} \\
(b) &\{X_{2, 1}, X_{4, 3}, X_{7, 3}, X_{2, 8}, X_{5, 6}, X_{3, 8}\} \\
(c) &\{X_{2, 1}, X_{4, 3}, X_{7, 3}, X_{2, 8}, X_{5, 6}, X_{6, 1}\}\\
(d) &\{X_{1, 4}, X_{4, 6}, X_{7, 3}, X_{2, 8}, X_{5, 6}\}\\
(e) &\{X_{2, 1}, X_{4, 3}, X_{7, 3}, X_{2, 8}, X_{5, 6}\}\\
(f) &\{X_{1, 4}, X_{4, 3}, X_{7, 3}, X_{2, 8}, X_{5, 6}\}\\
(g) &\{X_{1, 4}, X_{4, 3}, X_{2, 8}, X_{5, 6}\}\\
(h) &\{X_{4, 3}, X_{1, 7}, X_{2, 8}, X_{5, 6}\}\\
(i) &\{X_{4, 3}, X_{7, 3}, X_{2, 8}, X_{5, 6}\}\\
(j) &\{X_{1, 4}, X_{7, 3}, X_{2, 8}, X_{5, 6}\}
\\ \hline
\end{array}
\qquad
\begin{array}{|l|l|}\hline
\mbox{Theory} & \mbox{Higgsed fields} \\ \hline
(k) &\{X_{4, 6}, X_{7, 3}, X_{2, 8}, X_{5, 6}\}\\
(l) &\{X_{4, 3}, X_{2, 8}, X_{5, 6}\}\\
(m) &\{X_{7, 3}, X_{2, 8}, X_{3, 8}, X_{5, 6}\}\\
(n) &\{X_{7, 3}, X_{2, 8}, X_{5, 6}\}\\
(o) &\{X_{2, 8}, X_{3, 8}, X_{6, 1}, X_{5, 6}\}\\
(p) &\{X_{2, 8}, X_{5, 6}\}\\
(q) &\{X_{2, 8}, X_{4, 6}, X_{6, 1}\}\\
(r) &\{X_{7, 3}, X_{5, 6}\}\\
(s) &\{X_{2, 8}\}
\\ \hline
\end{array}$
\label{t:vevfieldsI}
}

\begin{figure}[h!t!b!]
\includegraphics[trim=0mm 0mm 0mm 0mm, clip, width=6in]{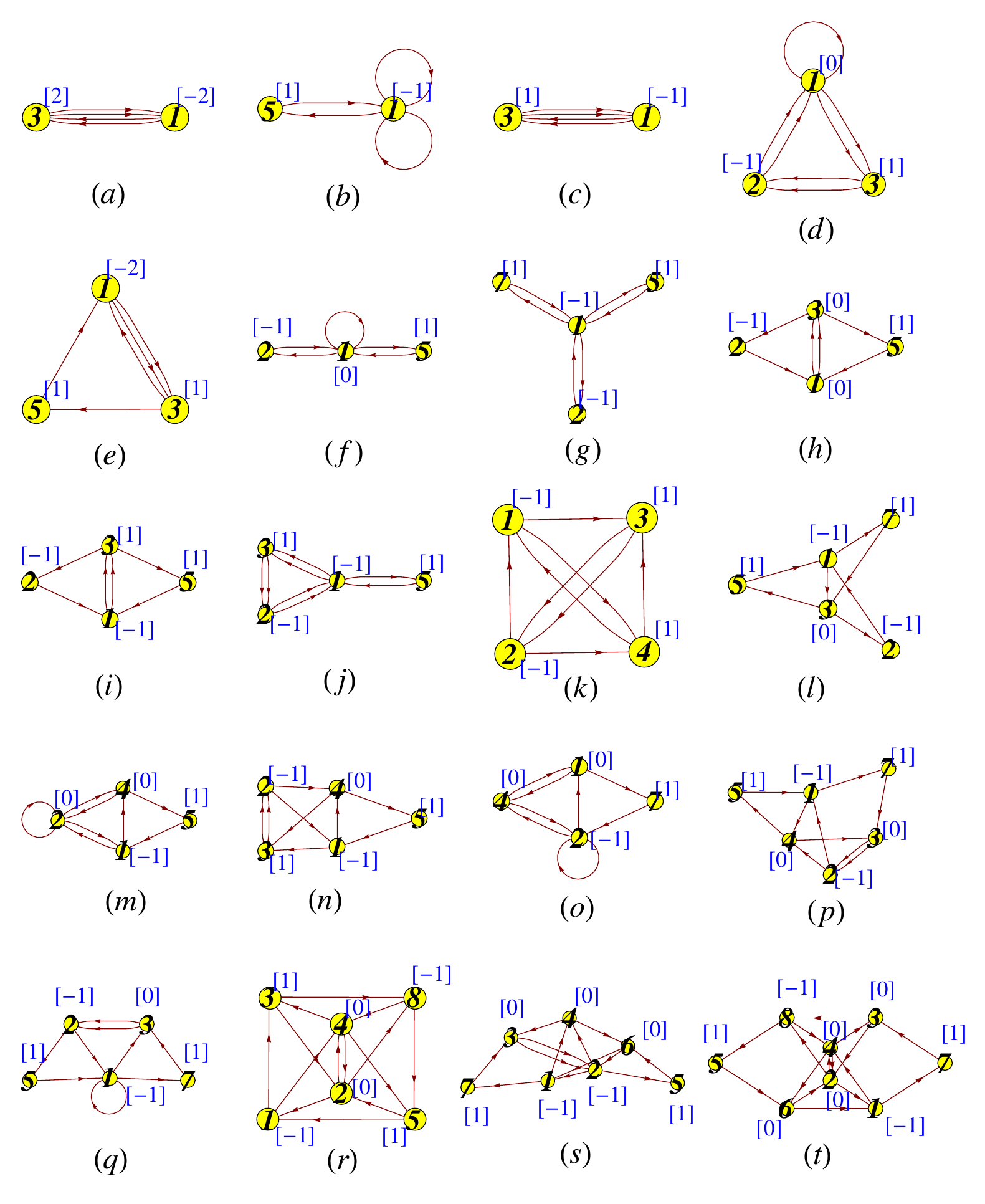}
\caption{{\sf Quiver diagrams for the 19 Higgsed theories obtained from (t), which is $(\IC^4/\IZ_2^3)_I$.
We have labelled the Chern-Simons levels with (blue) square brackets, and have kept an absolute numbering of the nodes with respect to the theory (t) from which all are derived.
}}
\label{f:Unhiggsphase}
\end{figure}

The results are summarized in Table~\ref{t:unhiggsphase}. Here we have applied the forward algorithm (\ref{Gt}) to each low-energy theory, with given quiver, superpotential and Chern-Simons levels inherited from the parent. The output is the matrix $G_t$, whose columns are the vertices of the toric diagram $\Delta$, with the number of repetitions of a column being the multiplicity. It turns out that there are typically many inequivalent QCS theories with a given Calabi-Yau four-fold moduli space $X$ -- in other words, different phases for $X$ -- and so for reasons of space we have in general presented only one such theory for each possible partial resolution in Table~\ref{t:unhiggsphase} (examples of this non-uniqueness of phases may be found in Appendix \ref{sec:plethora}). The quiver diagrams for the various theories are presented in Figure~\ref{f:Unhiggsphase}. Theory (t) in Figure~\ref{f:Unhiggsphase} is our parent $(\IC^4/\IZ_2^3)_I$. By Higgsing it, we find a total of 19 inequivalent affine toric Calabi-Yau four-folds, including the parent; we denote the corresponding theories as (a) to (t). Theories (b) and (c) correspond to toric diagram (18), and are shown in order to emphasize that both $(\IC^4)_{I}$ and $(\IC^4)_{II}$ can be obtained from Higgsing the same parent theory. 

We see that the entire list of toric sub-diagrams of $\IC^4/\IZ_2^3$ in Figure~\ref{f:c4z2cube} is obtained via Higgsing the parent theory $(\IC^4/\IZ_2^3)_I$. We have therefore constructed QCS theories for an entire family of partial resolutions, as promised.
For completeness, the list of fields acquiring VEVs for each theory (with respect to theory (t)) is presented in Table~\ref{t:vevfieldsI}. 
In fact, a stronger claim can be made. 
Each of the theories in Figure~\ref{f:Unhiggsphase} can be Higgsed to obtain theories that correspond to all their own toric sub-diagrams. 
Notice that theories (p), (q), (r) and (s) correspond to toric diagrams with external multiplicities greater than one, and of these theories (p), (q) and (s), as opposed to (r), can be further Higgsed in order to reduce all external multiplicities to one.

\begin{figure}[H]
\includegraphics[trim=0mm 0mm 0mm 0mm, clip, width=6in]{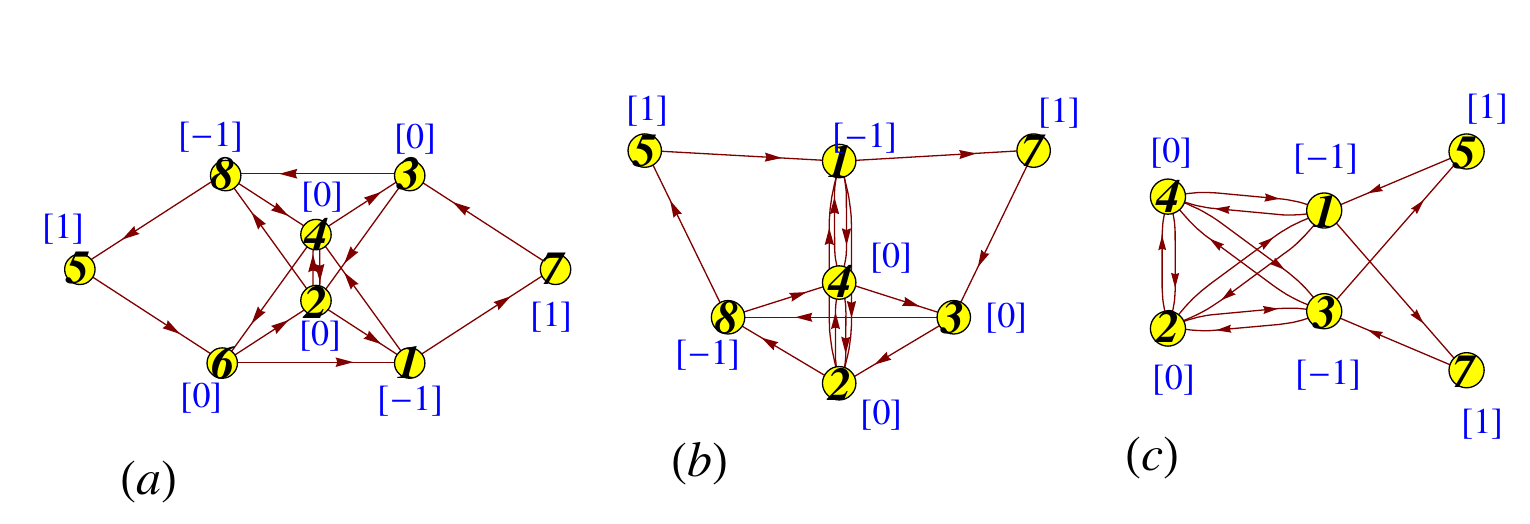}
\caption{{\sf Quiver diagrams for the two Higgsed theories obtained from (a), or theory $(\IC^4/\IZ_2^3)_I$. 
We have labelled the CS levels with (blue) square brackets, and have kept an absolute numbering of the nodes with respect to the theory (a) from which all are derived.
}}
\label{f:c4z2cube-3phases}
\end{figure}

Interestingly, we see that one can Higgs away fields $X_{6,1}$, or $\{X_{6,1}, X_{3,8}\}$, from theory (t) to obtain theories which are \emph{different phases}, {\it i.e.} share the same VMS, of the parent orbifold $\IC^4/\IZ_2^3$. 
Indeed, note these have different numbers of nodes in the quiver (respectively 8, 7 and 6) but still have the same moduli space. 
The quivers for these theories are shown in Figure~\ref{f:c4z2cube-3phases}, while the superpotential and $G_t$ matrix are presented in Table~\ref{t:c4z2cube-3phases}.

\TABLE[h!b!p!]{
\caption{{\sf The $G_t$ matrix and superpotentials for the two $\IC^4/\IZ_2^3$ daughter phases obtained by Higgsing the parent $(\IC^4/\IZ_2^3)_I$. Theories (b) and (c) are obtained by giving VEVs to $X_{6,1}$ and $\{X_{6,1},X_{3,8}\}$, respectively.}}
\begin{tabular}{|c|c|c|} \hline
& Toric diagram $G_t$ & Superpotential $W$ \\ \hline \hline
(b) &
{\tiny
$\left(
\begin{array}{llllllllllllllllll}
 0 & 1 & 0 & 1 & 1 & 1 & 0 & 0 & 0 & 1 & 0 & 1 & 2 & 1 & 0 & 0 & 0 & 1
   \\
 0 & 1 & 0 & 1 & 0 & 0 & 1 & 1 & 1 & 1 & 2 & 0 & 0 & 1 & 0 & 0 & 1 & 0
   \\
 1 & 0 & 2 & 0 & 1 & 0 & 0 & 1 & 1 & 0 & 0 & 1 & 0 & 0 & 0 & 1 & 0 & 0
   \\
 1 & 1 & 1 & 1 & 1 & 1 & 1 & 1 & 1 & 1 & 1 & 1 & 1 & 1 & 1
   & 1 & 1 & 1
\end{array}
\right)
$}
&
{\tiny
$\begin{array}{l}
X_{1,2} X_{2,4} X_{4,1} - X_{1,4} X_{4,2} X_{2,1} - \\
X_{2,4} X_{4,3} X_{3,2}  + X_{1,7} X_{7,3} X_{3,2} X_{2,1} + \\
X_{2,8} X_{8,4} X_{4,2}  - X_{1,7} X_{7,3} X_{3,8} X_{8,4} X_{4,1} - \\
X_{1,2} X_{2,8} X_{8,5} X_{5,1}  + X_{1,4} X_{4,3} X_{3,8} X_{8,5} X_{5,1}
\end{array}$
}
\\ \hline
(c) &
{\tiny
$\left(
\begin{array}{llllllllllllllll}
 $-1$ & $-1$ & $-1$ & $-1$ & $-1$ & $-1$ & $-1$ & $-1$ & $-1$ & 0 & 0 & 0 & 0 & 0 & 0 &
   1 \\
 0 & 1 & 1 & 0 & 0 & 0 & 2 & 1 & 1 & 1 & 0 & 0 & 0 & 0 & 1 & 0 \\
 1 & 0 & 1 & 0 & 2 & 1 & 0 & 0 & 1 & 0 & 1 & 0 & 0 & 1 & 0 & 0 \\
 1 & 1 & 0 & 2 & 0 & 1 & 0 & 1 & 0 & 0 & 0 & 1 & 1 & 0 & 0 & 0
\end{array}
\right)
$}
&
{\tiny
$\begin{array}{l}
X_{1,2} X_{2,4} X_{4,1} - X_{1,4} X_{4,2} X_{2,1}  + \\
X_{2,3} X_{3,4} X_{4,2} - X_{2,4}  X_{4,3} X_{3,2} - \\
X_{1,2} X_{2,3} X_{3,5} X_{5,1} + X_{1,4} X_{4,3} X_{3,5} X_{5,1} + \\
X_{1,7} X_{7,3} X_{3,2} X_{2,1} - X_{1,7} X_{7,3} X_{3,4} X_{4,1} 
\end{array}$
}
\\ \hline
\end{tabular}
\label{t:c4z2cube-3phases}
}

Another theory obtained from Higgsing $(\IC^4/\IZ_2^3)_I$, which we have not presented in Figure~\ref{f:Unhiggsphase}, is a theory which is dual to theory (h). We will refer to this theory as $C(Q^{1,1,1})_I$. Geometrically, this is a cone over $Q^{1,1,1}$ and the toric diagram is number (13) in our list.
The quiver and superpotential for these theories are the same, while the CS levels are different. To obtain $C(Q^{1,1,1})_I$ from theory (h) we need to apply the duality rules, discussed in Subsection \ref{sec:duality}, on one of the single-flavour nodes. The CS levels that are obtained are then $(-1,1,-1,1)$. These dual theories were first presented in \cite{Franco:2008um}, and $C(Q^{1,1,1})_I$ was studied in detail in \cite{Franco:2009sp}. In the latter reference it was shown that the manifest global symmetry of the gauge theory is $U(1)_R\times SU(2) \times U(1)$, which is strictly smaller than the isometry group of $Q^{1,1,1}$.
It was conjectured that the gauge theory at CS levels $(-k,k,-k,k)$ is dual to AdS$_4 \times Q^{1,1,1}/\IZ_k$, where the action of $\IZ_k$ precisely breaks the isometry group to $U(1)_R \times SU(2)\times U(1)$. Moreover, the simplest chiral operators in this gauge theory were analyzed \cite{Franco:2009sp}, and shown to match the Kaluza-Klein harmonics on AdS$_4 \times Q^{1,1,1}/\IZ_k$, thus proving further tests of this gauge theory as a theory on M2-branes at the $C(Q^{1,1,1})$ singularity.

%=====================================================
\subsection{The $(\IC^4/\IZ_2^3)_{II}$ theory}
%=====================================================
%=====================================================
We have just seen two more phases of the parent $\IC^4/\IZ_2^3$ theory.
It is therefore natural to ask whether we could obtain other phases.
In this subsection we shall see that this is indeed so.
We shall succeed in constructing yet another phase, $(\IC^4/\IZ_2^3)_{II}$, using a rather different method.
\subsubsection{Obtaining the $(\IC^4/\IZ_2^3)_{II}$ phase from a $PdP_5$ parent theory}
%=====================================================
It is by now well-known that it is possible to generate a $(2+1)$-dimensional QCS theory with toric Calabi-Yau four-fold moduli space by starting from the quiver and superpotential of a $(3+1)$-dimensional theory \cite{Martelli:2008si,Hanany:2008cd,Ueda:2008hx,crystal,Hanany:2008gx,Davey:2009sr,Hanany:2009vx}. 
The $(2+1)$-dimensional theory can be obtained by appropriate assignment of CS levels to the nodes of the $(3+1)$-dimensional parent theory.
The resulting four-dimensional moduli space, which has a three-dimensional toric diagram $\Delta=\Delta_3$, can be seen to be an ``inflated'' version of the original two-dimensional parent toric diagram $\Delta_2$; more precisely, the latter is a projection of the former. 
\begin{figure}[H]
\includegraphics[trim=0mm 220mm 0mm 10mm, clip, width=6.0in]{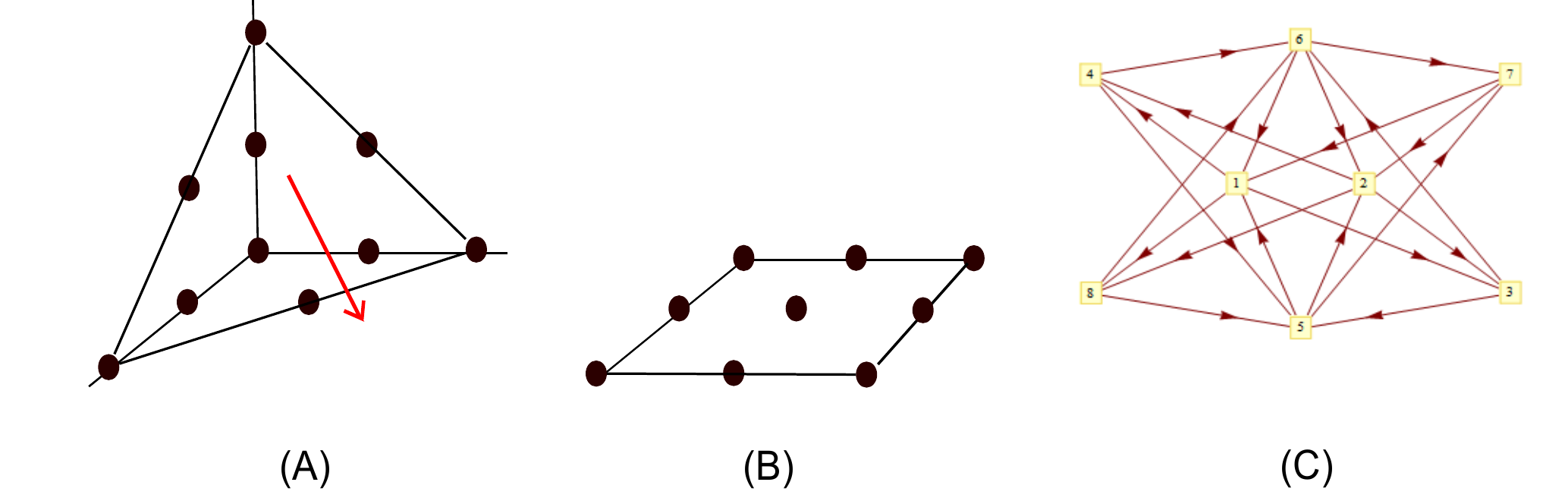}
\caption{{\sf The toric diagram (A) for the $\IC^4/\IZ_2^3$ theory can be projected to that of the Pseudo del Pezzo 5 singularity in (B). The quiver for the associated $(3+1)$-dimensional theory for $PdP_5$ is presented in part (C).}
\label{f:projection}}
\end{figure}
In order to find a potential $(3+1)$-dimensional parent theory for $\IC^4/\IZ_2^3$ we must first find an appropriate projection of its three-dimensional toric diagram, which we recall is diagram (1) in Figure \ref{f:c4z2cube}, or diagram (A) in Figure~\ref{f:projection}. This may be achieved as in Figure~\ref{f:projection}, where in part (B) we have shown the resulting two-dimensional toric diagram. This is the Pseudo Del Pezzo 5 ($PdP_5$) geometry of \cite{Feng:2002fv}, which is a complex cone over a non-generic blow-up of $\mathbb{CP}^2$ and is, in fact, an orbifold of the conifold. The quiver for the corresponding $(3+1)$-dimensional theory is presented in part (C) of Figure~\ref{f:projection}, while the superpotential is
\begin{eqnarray}
\nn
W&=&-X_{1,3} X_{3,5} X_{5,1}+ 
   X_{1,4} X_{4,6} X_{6,1}+ 
   X_{5,2} X_{2,4} X_{4,5}- 
   X_{6,2} X_{2,3} X_{3,6}+ 
   X_{5,1} X_{1,8} X_{8,5}- \\
\nn
&&
   X_{6,1} X_{1,8} X_{8,6}- 
   X_{5, 2} X_{2, 8} X_{8, 5}+
   X_{6, 2} X_{2, 8} X_{8, 6}+ 
   X_{3, 5} X_{5, 7} X_{7, 2} X_{2, 3}- \\
&&
\label{WPdP5}
   X_{4, 6} X_{6, 7} X_{7, 2} X_{2, 4}+
   X_{6, 7} X_{7, 1} X_{1, 3} X_{3, 6}- 
   X_{5, 7} X_{7, 1} X_{1, 4} X_{4, 5}~.
\end{eqnarray}

\TABLE[h!b!p!]{
\caption{{\sf The toric diagrams and superpotentials for the possible Higgsings of theory (a), {\it i.e.} $(\IC^4/\IZ_2^3)_{II}$. Also listed are the corresponding toric moduli spaces $X$, numbered according to Figure~\ref{f:c4z2cube}. Note that all except toric diagram (19), corresponding to
$\IC^4/\IZ_2$, are obtained.}}
\begin{tabular}{|c|c|c|c|} \hline
& Toric diagram $G_t$ & Superpotential $W$ & $X$ \\ \hline \hline
(a) &
{$\tiny
\left (
\begin{array}{ccccccllll}
\times 7 & \times 2 & \times 2 & \times 2 & \times 2 & \times 7 & & & & \\ [-1ex]
0 & 1 & 1 & 0 & 0 & 1 & 0 & 0 & 2 & 0 \\ [-1ex]
1 & 1 & 0 & 1 & 0 & 0 & 0 & 2 & 0 & 0 \\ [-1ex]
1 & 0 & 1 & 0 & 1 & 0 & 2 & 0 & 0 & 0 \\ [-1ex]
1 & 1 & 1 & 1 & 1 & 1 & 1 & 1 & 1 & 1
\end {array}
\right)$}
&
{\tiny
$\begin{array}{l}
-X_ {1, 3} X_ {3, 5} X_ {5, 1} + 
X_ {1, 4} X_ {4, 6} X_ {6, 1} + \\
X_ {5, 2} X_ {2, 4} X_ {4, 5} - 
X_ {6, 2} X_ {2, 3} X_ {3, 6} + \\
X_ {5, 1} X_ {1, 8} X_ {8, 5} - 
X_ {6, 1} X_ {1, 8} X_ {8, 6} - \\
X_ {5, 2} X_ {2, 8} X_ {8, 5} + 
X_ {6, 2} X_ {2, 8} X_ {8, 6} + \\
X_ {3, 5} X_ {5, 7} X_ {7, 2} X_ {2, 3} - 
X_ {4, 6} X_ {6, 7} X_ {7, 2} X_ {2, 4} + \\
X_ {6, 7} X_ {7, 1} X_ {1, 3} X_ {3, 6} - 
X_ {5, 7} X_ {7, 1} X_ {1, 4} X_ {4, 5}
\end{array}$
}
& (1)
\\ \hline
(b) &
{\tiny
$\left (
\begin {array} {llllllllllllllllll}
1 & 1 & 1 & 1 & 1 & 1 & 1 & 1 & 1 & 1 & 1 & 1 & 1 & 1 & 1 & 1 & 1 & 1 \\ [-1ex]
0 & 0 & 0 & 0 & 0 & 0 & 1 &  $-1$  & $-1$  & 0 & 0 & $-1$  &  $-1$  & 0 & 0 & 0 & 1 & 0 \\ [-1ex]
1 & 1 & 1 & 1 & 0 & 1 & 0 & 1 & 2 & 1 & 1 & 0 & 1 & 0 & 0 & 1 & 0 & 0 \\ [-1ex]
1 & 1 & 1 & 1 & 1 & 1 & 1 & 1 & 1 & 1 & 1 & 1 & 1 & 1 & 0 & 0 & 0 & 0
\end {array} \right)
$}
&
{\tiny
$\begin{array}{l}
-X_ {1, 3} X_ {3, 5} X_ {5, 1} + 
X_ {1, 2} X_ {2, 4} X_ {4, 5} X_ {5, 1} - \\
X_ {1, 2} X_ {2, 3} X_ {3, 6} X_ {6, 1} + 
X_ {1, 4} X_ {4, 6} X_ {6, 1} - \\
X_ {1, 4} X_ {4, 5} X_ {5, 7} X_ {7, 1} + 
X_ {1, 3} X_ {3, 6} X_ {6, 7} X_ {7, 1} + \\
X_ {2, 3} X_ {3, 5} X_ {5, 7} X_ {7, 2} - 
X_ {2, 4} X_ {4, 6} X_ {6, 7} X_ {7, 2}
\end{array}$
}
& (2)
\\ \hline
(c) &
{\tiny
$\left (
\begin {array} {llllllllllllll}
$-1$  & 0 & $-1$ & $-1$ & 0 & $-2$ & $-1 $ & $-1 $ & $-1$ & 0 & 0 & 1 & 0 & 1 \\ [-1ex]
1 & 0 & 1 & 1 & 0 & 2 & 1 & 1 & 1 & 0 & 0 & 0 & 1 & 0 \\ [-1ex]
1 & 1 & 1 & 1 & 2 & 0 & 0 & 1 & 0 & 0 & 1 & 0 & 0 & 0 \\ [-1ex]
0 & 0 & 0 & 0 & $-1$ & 1 & 1 & 0 & 1 & 1 & 0 & 0 & 0 & 0
\end {array} \right)
$}
&
{\tiny
$\begin{array}{l}
-X_ {1, 2} X_ {2, 3} X_ {3, 1} + 
X_ {1, 4}  X_ {4, 3} X_ {3, 1} - \\
X_ {1, 3} X_ {3, 5} X_ {5, 1} + 
X_ {1, 2} X_ {2, 4} X_ {4, 5} X_ {5, 1} + \\
X_ {1, 3} X_ {3, 7} X_ {7, 1} - 
X_ {1, 4} X_ {4, 5} X_ {5, 7} X_ {7, 1} - \\
X_ {2, 4}  X_ {4, 3} X_ {3, 7} X_ {7, 2} + 
X_ {2, 3} X_ {3, 5} X_ {5, 7} X_ {7, 2} 
\end{array}$
}
& (3)
\\ \hline
(d) &
{\tiny
$\left (
\begin {array} {lllllllllll}
1 & 0 & 0 & 0 & 1 & 0 & 1 & 2 & 0 & 0 & 1 \\ [-1ex]
0 & 0 & $-1 $ & 0 & $-1 $ & 0 & $-1 $ & $-1 $ & 0 & 1 & 0 \\ [-1ex]
0 & 0 & 2 & 1 & 1 & 0 & 1 & 0 & 1 & 0 & 0 \\ [-1ex]
0 & 1 & 0 & 0 & 0 & 1 & 0 & 0 & 0 & 0 & 0
\end {array} \right)
$}
&
{\tiny
$\begin{array}{l}
X_ {1, 2} X_ {2, 6} X_ {6, 1} - 
X_ {1, 5} X_ {5, 6} X_ {6, 1} - \\
X_ {1, 6} X_ {6, 2} X_ {2, 1} + 
X_ {2, 5} X_ {5, 6} X_ {6, 2} - \\
X_ {1, 2} X_ {2, 5} X_ {5, 7} X_ {7, 1} + 
X_ {1, 6} X_ {6, 7} X_ {7, 1} + \\
X_ {1, 5} X_ {5, 7} X_ {7, 2} X_ {2, 1} - 
X_ {2, 6} X_ {6, 7} X_ {7, 2}
\end{array}$
}
& (4)
\\ \hline
(e) &
{\tiny
$\left (
\begin {array} {llllllllllllll}
1 & $-1 $ & 1 & 1 & 1 & $-1 $ & 1 & 1 & 0 & 2 & 0 & 0 & 1 & 1 \\ [-1ex]
0 & 0 & 0 & 0 & 0 & 1 & $-1 $ & 0 & 0 & $-1$ & 0 & 1 & 0 & 0 \\ [-1ex]
0 & 1 & 0 & 0 & 0 & 0 & 1 & 0 & 0 & 0 & 1 & 0 & 0 & 0 \\ [-1ex]
0 & 1 & 0 & 0 & 0 & 1 & 0 & 0 & 1 & 0 & 0 & 0 & 0 & 0
\end {array} \right)
$}
&
{\tiny
$\begin{array}{l}
-X_ {1, 3} X_ {3, 5} X_ {5, 2}^{1} X_ {2, 1}^{1} + 
X_ {2, 4} X_ {4, 5} X_ {5, 2}^{1} + \\
X_ {2, 3} X_ {3, 5} X_ {5, 2}^{2} - 
X_ {1, 4} X_ {4, 5} X_ {5, 2}^{2} X_ {2, 1}^{2} - \\
X_ {2, 3} X_ {3, 6} X_ {6, 2}^{2} + 
X_ {1, 4} X_ {4, 6} X_ {6, 2}^{2} X_ {2, 1}^{1} + \\
X_ {1, 3}  X_ {3, 6} X_ {6, 2}^{1} X_ {2, 1}^{2} - 
X_ {2, 4} X_ {4, 6} X_ {6, 2}^{1}
\end{array}$
}
& (5)
\\ \hline
(f) &
{\tiny
$\left (
\begin {array} {llllllll}
1 & 1 & $-1$ & 0 & 0 & 0 & 0 & 1 \\ [-1ex]
$-1$ & $-1$ & 2 & 1 & 0 & 0 & 1 & 0 \\ [-1ex]
0 & 1 & 0 & 0 & 0 & 1 & 0 & 0 \\ [-1ex]
1 & 0 & 0 & 0 & 1 & 0 & 0 & 0
\end {array} \right)
$}
&
{\tiny
$\begin{array}{l}
X_ {1, 2}  X_ {2, 4} X_ {4, 2} X_ {2, 1} - 
X_ {1, 2}  X_ {2, 1} X_ {1, 6} X_ {6, 1} - \\
X_ {2, 4} X_ {4, 6} X_ {6, 4} X_ {4, 2} + 
X_ {1, 6}  X_ {6, 4} X_ {4, 6} X_ {6, 1} 
\end{array}$
}
& (6)
\\ \hline
(g) &
{\tiny
$\left (
\begin {array} {llllllllll}
1 & 1 & 1 & 1 & 0 & 0 & 2 & 1 & 0 & 1 \\ [-1ex] 
0 & 0 & 0 & $-1$ & 0 & 0 & $-1$ & 0 & 1 & 0 \\ [-1ex]
0 & 0 & 1 & 0 & 0 & 1 & 0 & 0 & 0 & 0 \\ [-1ex]
0 & 0 & $-1$ & 1 & 1 & 0 & 0 & 0 & 0 & 0
\end {array} \right)
$}
&
{\tiny
$\begin{array}{l}
 X_{1, 2} X_{2, 4} X_{4, 5}^{2} X_{5, 7}^{2} X_{7, 1} - 
 X_{1, 2} X_{2, 5} X_{5, 1} + \\
 X_{1, 4} X_{4, 5}^{1} X_{5, 1} - 
 X_{1, 4} X_{4, 5}^{2} X_{5, 7}^{1} X_{7, 1} - \\
 X_{2, 4} X_{4, 5}^{1} X_{5, 7}^{2} X_{7, 2} + 
 X_{2, 5} X_{5, 7}^{1} X_{7, 2}
\end{array}$
}
& (7)
\\ \hline
(h) &
{\tiny
$\left (
\begin {array} {lllllll}
1 & 0 & 0 & 0 & 0 & 0 & 1 \\ [-1ex]
0 & 0 & 0 & 0 & 0 & 1 & 0 \\ [-1ex]
1 & $-1$ & 0 & 0 & 1 & 0 & 0 \\ [-1ex]
$-1$ & 2 & 1 & 1 & 0 & 0 & 0
\end {array} \right)
$}
&
{\tiny
$\begin{array}{l}
X_ {1, 2} X_ {2, 1} X_ {1, 5}  X_ {5, 7} X_ {7, 1} -
X_ {1, 2}  X_ {2, 2} X_ {2, 1} + \\
X_ {2, 2} X_ {2, 5} X_ {5, 2} -
X_ {1, 5} X_ {5, 2} X_ {2, 5} X_ {5, 7} X_ {7, 1}
\end{array}$
}
& (8)
\\ \hline
(i) &
{\tiny
$\left (
\begin {array} {llllllllll}
$-1$ & $-1$ & 1 & 1 & 0 & $-1$ & 0 & 1 & 0 & 1 \\ [-1ex]
1 & 1 & 0 & 0 & 0 & 1 & 0 & 0 & 1 & 0 \\ [-1ex]
1 & 2 & $-1$ & 0 & 0 & 1 & 1 & 0 & 0 & 0 \\ [-1ex]
0 & $-1$ & 1 & 0 & 1 & 0 & 0 & 0 & 0 & 0
\end {array} \right)
$}
&
{\tiny
$\begin{array}{l}
X_ {1, 4} X_ {4, 6} X_ {6, 1} -  
X_ {1, 8}^{1} X_ {8, 4} X_ {4, 6} X_ {6, 7} X_ {7, 1}^{2} - \\
X_ {1, 4} X_ {4, 7} X_ {7, 1}^{1} + 
X_ {1, 8}^{2} X_ {8, 4} X_ {4, 7} X_ {7, 1}^{2}  - \\
X_ {1, 8}^{2} X_ {8, 6} X_ {6, 1} + 
X_ {1, 8}^{1}  X_ {8, 6} X_ {6, 7} X_ {7, 1}^{1}
\end{array}$
}
& (9)
\\ \hline
(j) &
{\tiny
$\left (
\begin {array} {llllllll}
0 & 0 & $-1$ & 1 & 0 & 0 & 0 & 1 \\ [-1ex]
0 & 0 & 1 & $-1$ & 0 & 0 & 1 & 0 \\ [-1ex]
0 & 0 & 1 & $-1 $ & 0 & 1 & 0 & 0 \\ [-1ex]
1 & 1 & 0 & 2 & 1 & 0 & 0 & 0
\end {array} \right)
$}
&
{\tiny
$\begin{array}{l}
-X_ {1, 5} X_ {5, 2}^{3} X_ {2, 1}^{2} + 
X_ {1, 5} X_ {5, 2}^{2} X_ {2, 1}^{1} + \\
X_ {2, 8} X_ {8, 5}^{2} X_ {5, 2}^{3} - 
X_ {1, 8} X_ {8, 5}^{2} X_ {5, 2}^{1} X_ {2, 1}^{1} - \\
X_ {2, 8} X_ {8, 5}^{1} X_ {5, 2}^{2} + 
X_ {1, 8} X_ {8, 5}^{1} X_ {5, 2}^{1} X_ {2, 1}^{2}
\end{array}$
}
& (10)
\\ \hline
(k) &
{\tiny
$\left (
\begin {array} {llllllll}
1 & 1 & 1 & 0 & 0 & 2 & 0 & 1 \\ [-1ex]
0 & 0 & 0 & 0 & 0 & $-1$ & 1 & 0 \\ [-1ex]
0 & 0 & 1 & 0 & 1 & 0 & 0 & 0 \\ [-1ex]
0 & 0 & $-1$ & 1 & 0 & 0 & 0 & 0
\end {array} \right)
$}
&
{\tiny
$\begin{array}{l}
X_ {1, 5}^{2} X_ {5, 6}^{2} X_ {6, 2}^{1} X_ {2, 1}^{1} - 
X_ {1, 5}^{2} X_ {5, 6}^{1} X_ {6, 2}^{1} X_ {2, 1}^{2} - \\
X_ {1, 5}^{1} X_ {5, 6}^{2} X_ {6, 2}^{2} X_ {2, 1}^{1} + 
X_ {1, 5}^{1} X_ {5, 6}^{1} X_ {6, 2}^{2} X_ {2, 1}^{2}
\end{array}$
}
& (11)
\\ \hline
(l) &
{\tiny
$\left (
\begin {array} {llllllll}
$-2$ & $-2$ & $-1$ & 0 & 1 & 0 & 0 & 1 \\ [-1ex]
1 & 1 & 1 & 0 & 0 & 0 & 1 & 0 \\ [-1ex]
1 & 1 & 1 & 0 & 0 & 1 & 0 & 0 \\ [-1ex]
1 & 1 & 0 & 1 & 0 & 0 & 0 & 0
\end {array} \right)
$}
&
{\tiny
$\begin{array}{l}
-X_ {1, 4} X_ {4, 2}^{3} X_ {2, 1}^{2} + 
X_ {1, 4} X_ {4, 2}^{1} X_ {2, 1}^{1} + \\
X_ {2, 8} X_ {8, 4}^{2} X_ {4, 2}^{3} - 
X_ {1, 8} X_ {8, 4}^{2} X_ {4, 2}^{2} X_ {2, 1}^{1} - \\
X_ {2, 8} X_ {8, 4}^{1} X_ {4, 2}^{1} + 
X_ {1, 8} X_ {8, 4}^{1} X_ {4, 2}^{2} X_ {2, 1}^{2}
\end{array}$
}
& (12)
\\ \hline
(m) &
{\tiny
$\left (
\begin {array} {llllllll}
$-1$ & 0 & 0 & 1 & $-1$ & 0 & 0 & 1 \\ [-1ex]
1 & 1 & 0 & 0 & 1 & 0 & 1 & 0 \\ [-1ex]
1 & 1 & 0 & 0 & 1 & 1 & 0 & 0 \\ [-1ex]
0 & $-1$ & 1 & 0 & 0 & 0 & 0 & 0
\end {array} \right)
$}
&
{\tiny
$\begin{array}{l}
X_ {1, 8}^{1} X_ {8, 4}^{1} X_ {4, 7}^{1} X_ {7, 1}^{2} - 
X_ {1, 8}^{2} X_ {8, 4}^{1} X_ {4, 7}^{2} X_ {7, 1}^{2} - \\
X_ {1, 8}^{1} X_ {8, 4}^{2} X_ {4, 7}^{1} X_ {7, 1}^{1} + 
X_ {1, 8}^{2} X_ {8, 4}^{2} X_ {4, 7}^{2} X_ {7, 1}^{1}
\end{array}$
}
& (13)
\\ \hline
(n) &
{\tiny
$\left (
\begin {array} {llllll}
1 & $-1$ & 0 & 0 & 0 & 1 \\ [-1ex]
$-1$ & 1 & 0 & 0 & 1 & 0 \\ [-1ex]
1 & 0 & 0 & 1 & 0 & 0 \\ [-1ex]
0 & 1 & 1 & 0 & 0 & 0
\end {array} \right)
$}
&
{\tiny
$\begin{array}{l}
-X_ {1, 2} X_ {2, 2} X_ {2, 1} + 
X_ {1, 2} X_ {2, 1} X_ {1, 8} X_ {8, 1} + \\
X_ {2, 2} X_ {2, 8} X_ {8, 2} - 
X_ {1, 8} X_ {8, 2} X_ {2, 8} X_ {8, 1}
\end{array}$
}
& (14)
\\ \hline
(o) &
{\tiny
$\left (
\begin {array} {lllll}
1 & 0 & 0 & 0 & 1 \\ [-1ex]
0 & 0 & 0 & 1 & 0 \\ [-1ex]
1 & 0 & 1 & 0 & 0 \\ [-1ex]
$-1$ & 1 & 0 & 0 & 0
\end {array} \right)
$}
&
{\tiny
$\begin{array}{l}
X_{1, 5}^1 X_{5, 1} X_{1, 5}^2 X_{5, 7} X_{7,1} - 
X_{1, 5}^2 X_{5, 1} X_{1, 5}^1 X_{5, 7} X_{7,1}
\end{array}$
}
& (15)
\\ \hline
(p) &
{\tiny
$\left (
\begin {array} {llllll}
1 & $-2$ & 0 & 0 & 0 & 1 \\ [-1ex]
0 & 1 & 0 & 0 & 1 & 0 \\ [-1ex]
0 & 1 & 0 & 1 & 0 & 0 \\ [-1ex]
0 & 1 & 1 & 0 & 0 & 0
\end {array} \right)
$}
&
{\tiny
$\begin{array}{l}
-X_ {1, 4}^{1} X_ {4, 2}^{1} X_ {2, 1}^{3} + 
X_ {1, 4}^{3} X_ {4, 2}^{1} X_ {2, 1}^{1} - \\
X_ {1, 4}^{2} X_ {4, 2}^{2} X_ {2, 1}^{1} + 
X_ {1, 4}^{1} X_ {4, 2}^{2} X_ {2, 1}^{2} + \\
X_ {1, 4}^{2} X_ {4, 2}^{3} X_ {2, 1}^{3} - 
X_ {1, 4}^{3} X_ {4, 2}^{3} X_ {2, 1}^{2}
\end{array}$
}
& (16)
\\ \hline
(q) &
{\tiny
$\left (
\begin {array} {llllll}
0 & 0 & 0 & 0 & 0 & 1 \\ [-1ex]
0 & 0 & 0 & 0 & 1 & 0 \\ [-1ex]
1 & 2 & 0 & 1 & 0 & 0 \\ [-1ex]
0 & $-1$ & 1 & 0 & 0 & 0
\end {array} \right)
$}
&
{\tiny
$\begin{array}{l}
X_ {1, 8}^{1} X_ {8, 5}^{2} X_ {5, 1}^{2} - 
X_ {1, 8}^{2} X_ {8, 5}^{2} X_ {5, 1}^{1} - \\
X_ {1, 1} X_ {1, 8}^{1} X_ {8, 5}^{1} X_ {5, 1}^{2} + 
X_ {1, 1} X_ {1, 8}^{2} X_ {8, 5}^{1} X_ {5, 1}^{1}
\end{array}$
}
& (17)
\\ \hline
(r) &
{\tiny
$\left (
\begin {array} {llll}
0 & 0 & 0 & 1 \\ [-1ex]
0 & 0 & 1 & 0 \\ [-1ex]
1 & 0 & 0 & 0 \\ [-1ex]
0 & 1 & 0 & 0
\end {array} \right)
$}
&
{\tiny
$\begin{array}{l}
X_{1, 5}^1 X_{5, 1}^1 X_{1, 5}^2 X_{5, 1}^2 - \\
X_{1, 5}^2 X_{5, 1}^1 X_{1, 5}^1 X_{5, 1}^2
\end{array}$
}
& (18)
\\ \hline
\end{tabular}
\label{t:phaseIII}
}

To obtain the second phase of $\IC^4/\IZ_2^3$ we assign CS levels $k=(0 \ldots 0,-1,1)$ to the $PdP_5$ quiver in Figure~\ref{f:projection}. 
The forward algorithm may be used to verify that the moduli space is indeed $\IC^4/\IZ_2^3$, as desired. We shall call this theory $(\IC^4/\IZ_2^3)_{II}$. Notice that, as in $(\IC^4/\IZ_2^3)_{I}$, the external multiplicities of the lattice points in the toric diagram are all equal to one.
%%%
%=====================================================
\subsubsection{Higgsing $(\IC^4/\IZ_2^3)_{II}$}
%=====================================================

We now take $(\IC^4/\IZ_2^3)_{II}$ as our parent theory and determine all possible Higgsings thereof, precisely as in Subsection 
\ref{s:higgsz2cubeI}. We will see that the situation here is more subtle than that for theory $(\IC^4/\IZ_2^3)_I$, in that certain toric sub-diagrams cannot be obtained by Higgsing the parent theory. We thus see that it is possible for different phases to lead to different sets of toric sub-diagrams. Again, we have executed this exhaustively with the aid of a computer.

\begin{figure}[H]
\includegraphics[trim=0mm 0mm 0mm 0mm, clip, width=6in]{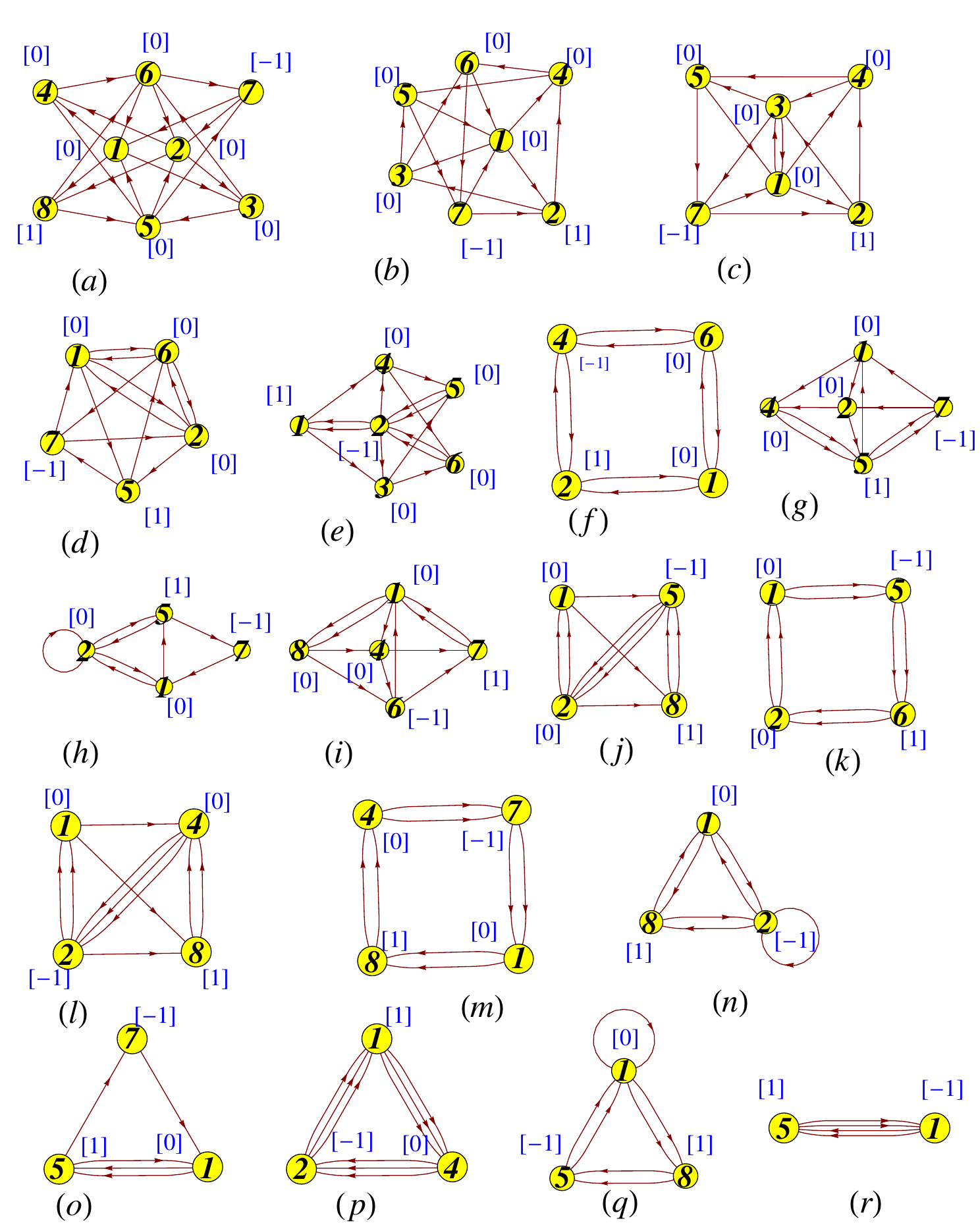}
\caption{{\sf Quiver diagrams for the 18 Higgsed theories from $(\IC^4/\IZ_2^3)_{II}$. 
}}
\label{f:phaseIIIhiggs}
\end{figure}

The resulting toric diagrams, specified by $G_t$, and superpotentials $W$ for the Higgsed theories are summarized in Table~\ref{t:phaseIII}.
We find a total of 18 inequivalent affine toric Calabi-Yau four-folds, including the parent; we denote the corresponding theories as (a) to (r). 
The list of fields acquiring VEVs for each theory is shown in Table~\ref{t:vevfieldsII}. The quiver diagrams are shown in 
Figure~\ref{f:phaseIIIhiggs}. Notice that we get a different set of theories from those obtained from the Higgsing of $(\IC^4/\IZ_2^3)_{I}$. 
Observe, in particular, theory (p), which we will refer to as $C(Y^{1,2}(\mathbb{CP}^2))_{II}$, and compare to the theory we called $C(Y^{1,2}(\mathbb{CP}^2))_I$ in Subsection \ref{s:higgsz2cubeI}.  In fact $C(Y^{1,2}(\mathbb{CP}^2))_{II}$ appeared first in the literature in 
reference \cite{Martelli:2008si}, while $C(Y^{1,2}(\mathbb{CP}^2))_I$ is new.
In theories (b), (c), (d), (i), (l), (m) and (p) we encounter toric diagrams with external points which have multiplicity greater than one; moreover, these theories cannot be further Higgsed to reduce these multiplicities. We will return to a more systematic discussion of this point later. We see that we only obtain a \emph{partial} list of the possible toric sub-diagrams; in particular we are missing geometry (19) in Figure~\ref{f:c4z2cube}, which is the orbifold $\IC^4/\IZ_2$. 

\begin{figure}[H]
\includegraphics[trim=0mm 210mm 0mm 0mm, clip, width=6in]{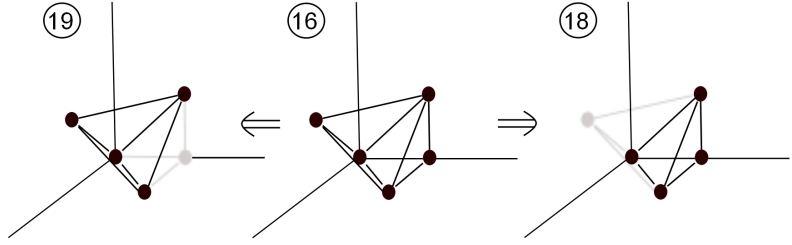}
\caption{{\sf Resolving toric diagram (16), which is $C(Y^{1,2}(\mathbb{CP}^2))$, to (18), which is simply $\IC^4$, and to (19), which is $\IC^4/\IZ_2$.}}
\label{f:resolving16}
\end{figure}

By examining the toric diagrams in Figure \ref{f:c4z2cube}, we see that the ``missing'' theory for $\IC^4/\IZ_2$ should be obtained by Higgsing a theory corresponding to toric diagram (16). 
We present this toric diagram, together with its two partial resolutions to (18) and (19), in Figure~\ref{f:resolving16}. In the next section we will examine in detail the Higgsing behaviour of the dual candidates $C(Y^{1,2}(\mathbb{CP}^2))_{I}$,  $C(Y^{1,2}(\mathbb{CP}^2))_{II}$ to this geometry.

\TABLE[h!]{
\caption{{\sf The list of fields which acquire VEVs in Higgsing theory $(\IC^4/\IZ_2^3)_{II}$ to obtain the various partial resolutions.}}
$\begin{array}{|l|l|}\hline
\mbox{Theory} & \mbox{Higgsed fields} \\ \hline
(b) &\{X_ {2, 8}\} \\
(c) &\{X_ {2, 8}, X_ {3, 6}\}\\
(d) &\{X_ {1, 3}, X_ {2, 4}, X_ {8, 5}\}\\
(e) &\{X_ {1, 8}, X_ {7, 2}\}\\
(f) &\{X_ {1, 3}, X_ {2, 8}, X_ {4, 5}, X_ {5, 7}\}\\
(g) &\{X_ {2, 3}, X_ {8, 5}, X_ {8, 6}\}\\
(h) &\{X_ {1, 3}, X_ {2, 4}, X_ {4, 6}, X_ {8, 5}\}\\
(i) &\{X_ {1, 3}, X_ {2, 3}, X_ {4, 5}\}\\
(j) &\{X_ {1, 3}, X_ {1, 4}, X_ {5, 7}, X_ {6, 7}\}\\
\hline
\end{array}
\qquad
\begin{array}{|l|l|}\hline
\mbox{Theory} & \mbox{Higgsed fields} \\ \hline
(k) &\{X_ {1, 3}, X_ {1, 4}, X_ {6, 7}, X_ {8, 5}\}\\
(l) &\{X_ {1, 3}, X_ {4, 5}, X_ {4, 6}, X_ {7, 2}\}\\
(m) &\{X_ {1, 3}, X_ {2, 3}, X_ {4, 5}, X_ {4, 6}\}\\
(n) &\{X_ {1, 3}, X_ {1, 4}, X_ {4, 5}, X_ {6, 7}, X_ {7, 2}\}\\
(o) &\{X_ {1, 3}, X_ {1, 4}, X_ {2, 4}, X_ {3, 6}, X_ {8, 5}\}\\
(p) &\{X_ {1, 3}, X_ {1, 8}, X_ {4, 5}, X_ {4, 6}, X_ {7, 2}\}\\
(q) &\{X_ {1, 3}, X_ {1, 4}, X_ {2, 4}, X_ {5, 7}, X_ {6, 7}\}\\
(r) &\{X_ {1, 3}, X_ {1, 4}, X_ {2, 4}, X_ {3, 6}, X_ {7, 2}, X_ {8, 5}\}
\\ \hline
\end{array}$
\label{t:vevfieldsII}
}

%%%%%%%%%%%%%%%%%%%%%%%%%%%%%%%%%%%%%%%%%%%%%%%%%%%%%%%%%%%
%=====================================================
\section{Torsion $G$-flux and Higgsing}\label{sec:Gflux}
%=====================================================

In this section we discuss the effect of adding \emph{torsion $G$-flux} to the AdS background AdS$_4\times Y_7$, where $Y_7$ is a Sasaki-Einstein seven-manifold. 
For the ABJM theory, where $Y_7=S^7/\IZ_k$, this has been conjectured to be dual to changing the \emph{ranks} from $U(N)_k\times U(N)_{-k}$ to $U(N+l)_k\times U(N)_{-k}$, where $0\leq l<k$ is identified with the torsion flux in $H^4(Y_7/\IZ_k,\IZ)\cong\IZ_k$ \cite{Aharony:2008gk}. One thus expects to find a similar behaviour in other QCS theories. Here we point out that adding such torsion flux non-trivially affects the supergravity dual of the Higgsing. More precisely, Higgsing a superconformal QCS theory leads to an RG flow, and in the supergravity dual of this flow one needs to appropriately extend the non-zero $G$-flux on the UV boundary at infinity. This is an interesting problem, and leads to non-trivial predictions about the Higgsing patterns expected in the field theory. 

We begin by explaining this in a general context in the next subsection, and then proceed to study the example $Y_7=Y^{1,2}(\mathbb{CP}^2)$ in detail, which is toric diagram number (16) in Figure~\ref{f:c4z2cube}. This has two inequivalent choices of $G$-flux, corresponding to the two elements in the group $\IZ_2\cong H^4(Y^{1,2}(\mathbb{CP}^2),\IZ)$.
We show that the behaviour of Phase $C(Y^{1,2}(\mathbb{CP}^2))_{Ib}$ (theory (e$_2$) in Figure~\ref{f:duals}) under Higgsing, with different choices of ranks, is precisely as expected from the dual supergravity solutions with the two choices of torsion $G$-flux. On the other hand, we find that the behaviour of Phase $C(Y^{1,2}(\mathbb{CP}^2))_{II}$, which recall we obtained here by Higgsing $(\IC^4/\IZ_2^3)_{II}$, does \emph{not} seem to match the supergravity analysis. Indeed, we show that there are various related puzzles in interpreting this theory as an M2-brane QCS theory, despite the fact that it has a Type IIA construction \cite{Aganagic:2009zk}.

%It is then 
%also reasonable to conjecture that this effect is responsible for the difference in Higgsings between $(\IC^4/\IZ_2^3)_I$ and $(\IC^4/\IZ_2^3)_{II}$ 
%observed in the previous section, although since the singularity is not isolated it seems difficult to test this hypothesis. This leads to the interesting 
%question of whether there are Seiberg-like dualities in which a toric QCS theory with unequal ranks, dual to a non-trivial torsion $G$-flux background, is dual to a toric QCS theory with a \emph{different} quiver but equal ranks. Since Seiberg dualities in $(2+1)$ dimensions are currently not well-understood, we leave this as a tantalizing speculation.

%=====================================================
\subsection{$G$-flux and the supergravity dual of Higgsing}\label{sec:generalGflux}
%=====================================================
We begin by discussing more carefully the supergravity backgrounds of interest. Thus, consider the M-theory Freund-Rubin background AdS$_4\times Y_7$, where $Y_7$ is a Sasaki-Einstein seven-manifold. The M-theory flux $G$ is quantized, satisfying
\begin{equation}
\IZ\ni N = \frac{1}{(2\pi l_p)^6}\int_{Y_7} *_{11} G~.
\end{equation}
As is well-known, this background may be interpreted as the near-horizon limit of $N$ M2-branes placed at the singularity of the Calabi-Yau four-fold cone\footnote{The reason for the bar over $X$ will become apparent later.} $\bar{X} = C(Y_7)$. Such backgrounds are very similar to their $d=(3+1)$ cousins in Type IIB supergravity, where one has $N$ D3-branes placed at the singularity of a Calabi-Yau three-fold cone $C(Y_5)$. However, at least for \emph{toric} geometries, for which the field theories are currently best understood, there is a key difference:
for a simply-connected toric Sasaki-Einstein five-manifold $Y_5$ there is no torsion in the cohomology of $Y_5$ \cite{Lerman2}, while for the corresponding geometries in seven dimensions typically $H^4(Y_7,\IZ)$ has non-trivial torsion.
Because of this latter fact, we may turn on a flat torsion $G$-flux without affecting the supergravity equations of motion, or the supersymmetry of the background. 
Since these are physically inequivalent M-theory backgrounds, the SCFTs will also be physically distinct, and should therefore display different properties. 
This was first discussed in the context of QCS theories for the ABJM theory in \cite{Aharony:2008gk}, although here the torsion in $H^4(S^7/\IZ_k,\IZ)\cong\IZ_k$ is due to the $\IZ_k$ quotient, giving $\pi_1(S^7/\IZ_k)\cong\IZ_k$.
More generally there are examples in which the torsion $G$-flux is \emph{not} associated to the CS level quotient by $\IZ_k$ -- for example, the $Y^{p,k}$ geometries discussed in detail in \cite{Martelli:2008rt}.

For our discussion, it is useful to think of the AdS background instead as the warped product $\IR^{1,2}\times X_0$, where
$X_0=\{r>0\}\subset \bar{X}$ is the cone $\bar{X}$ minus the singular apex.
Here one may think of $r$ as either the cone coordinate on
$X_0\cong\IR_+\times Y_7$, where the cone metric on $X_0$ is $g_{X_0}=\mathrm{d} r^2 + r^2 g_{Y_7}$, or as
the radial coordinate in AdS$_4$ in a Poincar\'e slicing. In this picture the warping is due
to the near-horizon limit of the harmonic function, $1/r^6$, sourced by the presence of the $N$ M2-branes at $\{r=0\}$.
Consider adding $G$-flux to this background, in a way that preserves the AdS$_4$ symmetry $SO(3,2)$ and supersymmetry. The former implies that $G$ is the pull-back
of a flux on $Y_7$. On the other hand, supersymmetry requires $G$ to be self-dual on $X_0$ \cite{Becker:1996gj}. These two facts together hence imply that $G$ is flat, and thus
the different choices of $G$-flux in the AdS background are classified by the torsion cohomology class
$[G]\in H^{4}_{\mathrm{tor}}(Y_7,\IZ)$.

Suppose now that one has a field theory dual to the above gravity solution; for example, we may take this to be the superconformal  fixed point of a
QCS theory for concreteness. Consider Higgsing this theory by giving non-zero VEVs to some of the matter fields. As usual,
this typically requires one to turn on FI parameters in the field theory in order to satisfy the D-term equations, and this
in turn gives a (partial) resolution of the VMS. For the theory on a single M2-brane, for which the
VMS is the Calabi-Yau singularity $\bar{X}$, the VMS with the given FI parameters is thus some (partial)
Calabi-Yau resolution $\pi:\hat X\rightarrow\bar{X}$. If we give corresponding diagonal VEVs in the $U(N)^G$ theory,
we pick a point in the VMS which is the image of
the diagonal $(p,p,\ldots,p)\in \hat{X}\times \hat{X}\times \cdots \times \hat{X}$, where $p\in\hat{X}$.
At the same time this introduces a scale into the theory, and thus an RG flow.

The supergravity dual of this RG flow was first discussed in the Type IIB context by Klebanov-Witten \cite{KW}, and has been further elucidated in \cite{KM, baryonic, condensate}, the latter in particular discussing this for general D3-brane quivers and Calabi-Yau three-folds. 
The M-theory discussion is precisely analogous: the dual supergravity solution to the RG flow induced by the Higgsing involves replacing the Calabi-Yau cone $\bar{X}$ by the (partial) resolution $\hat{X}$, which is no longer a cone and thus breaks the scaling symmetry of the supergravity solution. 

We should equip $\hat{X}$ with a Ricci-flat K\"ahler metric which is asymptotic at large $r$ to the cone metric on $X_0$, so that in the UV $r=\infty$ we obtain the AdS$_4\times Y_{\mathrm{UV}}$ geometry, where we have now denoted the
original Sasaki-Einstein seven-manifold as $Y_7=Y_{\mathrm{UV}}$. There has been recent mathematical work proving existence of complete asymptotically conical Ricci-flat K\"ahler metrics on such manifolds -- see \cite{craig1, craig2, craig3} and references therein.
In particular, there is a general existence theorem for \emph{toric} singularities.
For partial resolutions with residual singularities, we may take an appropriate limit of the smooth metrics by varying the K\"ahler class.
The diagonal Higgsing described above is then dual to placing all $N$ M2-branes
at the point $p$ in $\hat{X}$ in the supergravity solution. (Non-diagonal Higgsings of course correspond to separating the stack of M2-branes.) 

\begin{figure}[H]
\includegraphics[trim=0mm 210mm 0mm 0mm, clip, width=6.0in]{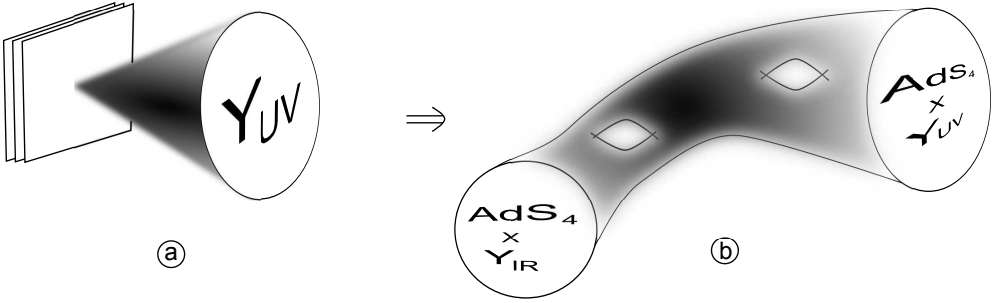}
\caption{{\sf (a) A stack of $N$ M2-branes transverse to the Calabi-Yau cone singularity $C(Y_{\mathrm{UV}})$; (b) the supergravity geometry
describing an RG flow dual to a diagonal Higgsing. The eight-manifold is $X=\hat{X}\setminus\{p\}$, where the $N$ M2-branes
are placed at the point $p$ on the partial resolution $\hat{X}$.}
\label{fig:UVIR}}
\end{figure}

The full supergravity solution, including the back-reaction of the M2-branes, requires us to find a solution to the Green's function on $\hat X$, with source at $p$, decaying as $1/r^6$ at infinity. Again, there are general existence and uniqueness
theorems implying we can always do this, discussed in \cite{baryonic}. Once we include the back-reaction of the M2-branes at the point $p\in \hat X$, the latter point is sent to infinity (by the Green's function), and the spacetime has two boundaries: AdS$_4\times Y_{\mathrm{UV}}$ in the UV, and AdS$_4\times Y_{\mathrm{IR}}$ near the point $p$. This is shown in Figure \ref{fig:UVIR}.
Here the tangent space at $p$ is the cone $C(Y_{\mathrm{IR}})$. Thus if $p$ is a smooth point, $Y_{\mathrm{IR}}=S^7$. We will be more interested in partial resolutions, and placing the M2-branes at a residual singular point $p$.

It is expedient to briefly summarize the spaces which we study and the relationships amongst them:

\vspace{2mm}

\begin{tabular}{lll}
$Y_7$: &$ = Y_{UV}$ & The Sasaki-Einstein seven-fold \\
$\bar{X}$: & $= C(Y_7)$ & Singular Calabi-Yau cone over $Y_7$ \\
$X_0$: & $=\{r>0\} \subset \bar{X}$ & Cone $\bar{X}$ minus the apex \\
$\hat{X}$: & & (partial) Calabi-Yau resolution of singularity $\pi:\hat{X}\rightarrow \bar{X}$ \\
$Y_{IR}$: && Near-horizon limit of $p\in \hat{X}$, close to the M2-branes \\
$X$: & $=\hat{X}\setminus\{p\}$ & M2-branes are placed at $p \in \hat{X}$ \\
\end{tabular}

The above discussion implies that, for \emph{zero} $G$-flux on $Y_7=Y_{\mathrm{UV}}$, we expect a supergravity solution to exist for \emph{any} choice
of Higgsing in the field theory. Conversely, since for any partial resolution of $\bar{X}$ we have a supergravity solution,
there should exist a field theory dual to this given by an appropriate Higgsing pattern. This suggests, for example, that
Phase $(\IC^4/(\IZ_2)^3)_I$ is dual to having no torsion $G$-flux on the boundary $Y_{\mathrm{UV}}=S^7/(\IZ_2)^3$, although
this example is complicated by the fact that the latter is not a smooth manifold.

More interesting is when we turn on torsion $G$-flux on $Y_{\mathrm{UV}}$. In this case, to obtain a supergravity solution we
must extend $G$ over the (partial) resolution $\hat X$, satisfying the appropriate supersymmetric equations of motion.
The key point here is that when $G=0$ on $Y_{UV}$, we may obviously extend this as $G=0$ on the partial resolution,
while for non-trivial torsion $G$ the process of completing the supergravity solution is much more involved.
There are two steps: first, it must be possible to extend the cohomology class
$[G]\in H^4(Y_{\mathrm{UV}},\IZ)$ to a cohomology class\footnote{We assume here that the membrane anomaly
on $X$ is zero. This will be true in the example that we shall study. The membrane anomaly on
$Y$ is automatically zero, as it is zero on any oriented spin seven-manifold.} in $H^4(X,\IZ)$, where we have defined
$X=\hat{X}\setminus\{p\}$ -- {\it a priori} there might be topological obstructions to this; second, if this is possible, we must choose a flux in this cohomology class to satisfy the supersymmetry
conditions
\cite{Becker:1996gj}, which require
that $G$ must be primitive, so $G\wedge \omega=0$ where $\omega$ is the K\"ahler form, and have Hodge type $(2,2)$ with respect to the complex structure (which implies it is self-dual). 

This leads to two issues: (i) if the choice of $G$-flux on
$Y_{\mathrm{UV}}$ cannot be so extended then the supergravity solution \emph{does not exist}, and therefore
the SCFT dual to $Y_{\mathrm{UV}}$ with this $G$-flux cannot be Higgsed to the partial resolution
corresponding to $\hat{X}$, (ii) the choice of $G$-flux may not be unique, meaning that
the SCFT should be Higgsable to the partial resolution but with
potentially more than one choice of torsion $G$-flux in $H^4(Y_{\mathrm{IR}},\IZ)$.
Indeed, notice that choosing an extension of $[G]$ over $X$ immediately leads by restriction
to a choice of $G$-flux in $H^{4}(Y_{IR},\IZ)$, and thus a torsion $G$-flux in the IR theory dual to
AdS$_4\times Y_{IR}$.

To conclude, one expects M2-brane QCS theories dual to torsion $G$-flux backgrounds to display different behaviour to those without $G$-flux --
namely, one should see obstructions to Higgsings to certain partial resolutions in theories with $G$-flux. We shall investigate
this in detail in the remainder of this section for a particular example, and show that this behaviour is indeed realized.

%It is also reasonable to suppose that this is responsible for the different Higgsing patterns observed in Phases
%I and II of $\IC^4/(\IZ_2)^3$ studied in section 4, although
%testing this would require a better understanding of Seiberg-like dualities in $d=(2+1)$ QCS theories.
%=====================================================
\subsection{$Y^{1,2}(\mathbb{CP}^2)$: Gravity results}\label{sec:gravity}
%=====================================================
We now investigate the above discussion in detail in a particular example:
the toric Calabi-Yau cone
$\bar{X}=C(Y^{1,2}(\mathbb{CP}^2))$ \cite{Martelli:2008rt}. This is precisely toric diagram number (16) in Figure~\ref{f:c4z2cube}.
Here the Sasaki-Einstein metric on
$Y_{\mathrm{UV}}=Y^{1,2}(\mathbb{CP}^2)$ is known explicitly,
and was constructed in \cite{Gauntlett:2004hh}.
The complex structure of the cone singularity $\bar X$ may be described
as the affine holomorphic quotient of $\IC^5$
by $\IC^*$ with charges $(1,2,-1,-1,-1)$.
This is of course the same complex structure induced by the
K\"ahler quotient, at moment map level zero, of $\IC^5$ by $U(1)$ with the same charges.
There are precisely two (partial) Calabi-Yau resolutions of this singularity,
given by taking the moment map level $\zeta<0$ or $\zeta>0$.

To describe these partial resolutions, let $z_1,\ldots,z_5$ denote coordinates on $\IC^5$.
The moment map/GLSM D-term equation is then
\bea\label{momentum}
|z_1|^2+2|z_2|^2 - |z_3|^2 -|z_4|^2-|z_5|^2 = \zeta~.
\eea
For $\zeta<0$ this describes the smooth Calabi-Yau four-fold
$\hat{X}_-\equiv$ total space of $\mathcal{O}(-1)\oplus\mathcal{O}(-2)\rightarrow
\mathbb{CP}^2$. The zero-section, which is a copy
of $\mathbb{CP}^2$, is at $\{z_1=z_2=0\}$, while the boundary $
\partial \hat{X}_-=Y_{\mathrm{UV}}=Y^{1,2}(\mathbb{CP}^2)$. In fact, note that an \emph{explicit}
Ricci-flat K\"ahler metric on this manifold was constructed in
\cite{metrics}.
Since $\hat{X}_-$ is contractible to $\mathbb{CP}^2$, it follows
that $H^0(\hat{X}_-,\IZ)\cong H^2(\hat{X}_-,\IZ)\cong H^4(\hat{X}_-,\IZ)\cong\IZ$, with
all other cohomology vanishing. Moreover, since
$\hat{X}_-$ is the total space of a rank four real vector bundle over
$\mathbb{CP}^2$, the Thom isomorphism implies that
$H^4(\hat{X}_-,Y_{\mathrm{UV}},\IZ)\cong H^6(\hat{X}_-,Y_{\mathrm{UV}},\IZ)\cong H^8(\hat{X}_-,Y_{\mathrm{UV}},\IZ)\cong\IZ$, where
the generator of $H^4(\hat{X}_-,Y_{\mathrm{UV}},\IZ)$ is the Thom class.
It follows that the image of the generator
in the map
$H^4(\hat{X}_-,Y_{\mathrm{UV}},\IZ)\rightarrow H^4(\hat{X}_-,\IZ)\cong H^4(\mathbb{CP}^2,\IZ)$ is the Euler class of the bundle $\mathcal{O}(-1)\oplus\mathcal{O}(-2)$.
Denoting $H$ the hyperplane class that generates $H^2(\mathbb{CP}^2,\IZ)\cong\IZ$, we have
\bea
e(\mathcal{O}(-1)\oplus\mathcal{O}(-2))&=&c_2(\mathcal{O}(-1)\oplus\mathcal{O}(-2))=c_1(\mathcal{O}(-1))\cup c_1(\mathcal{O}(-2))\nonumber\\
&=& (-H) \cup (-2H) = 2\in H^4(\mathbb{CP}^2,\IZ)\cong\IZ~.
\eea
Recall here that $H\cup H$ generates $H^4(\mathbb{CP}^2,\IZ)\cong \IZ$.
Thus the long exact sequence
\bea
H^4(\hat{X}_-,Y_{\mathrm{UV}},\IZ)\stackrel{f}{\rightarrow} H^4(\hat{X}_-,\IZ)\rightarrow H^4(Y_{\mathrm{UV}},\IZ)\rightarrow
H^5(\hat{X}_-,Y_{\mathrm{UV}},\IZ)\cong 0
\eea
implies, since the first ``forgetful'' map $f$ is multiplication
by the Euler number $e=2$, that $H^4(Y_{\mathrm{UV}},\IZ)\cong \IZ_2$.
Thus we may turn on precisely one non-trivial
torsion $G$-flux on $Y_{\mathrm{UV}}=Y^{1,2}(\mathbb{CP}^2)$.
It is similarly straightforward to show that the only
other non-trivial cohomology groups of $Y_{\mathrm{UV}}$ are $H^2(Y_{\mathrm{UV}},\IZ)\cong H^5(Y_{\mathrm{UV}},\IZ)\cong\IZ$. Notice this agrees with
\cite{Martelli:2008rt}, where the
cohomology groups of $Y_{\mathrm{UV}}$ were computed
via a completely different method.

Now consider the other partial resolution, with $\zeta>0$ in (\ref{momentum}).
This may be described as $\hat{X}_+\equiv$ total space of $\mathcal{O}(-1)^3\rightarrow
\mathbb{WCP}^1_{[1,2]}$, where the zero-section weighted
projective space $\mathbb{WCP}^1_{[1,2]}$ is now at
$\{z_3=z_4=z_5=0\}$. $\hat{X}_+$ has a single, isolated singular
point at $p=\{z_1=z_3=z_4=z_5=0\}$, which has tangent cone
$\IC^4/\IZ_2$ where the $\IZ_2$ generator acts with equal charge on
each coordinate of $\IC^4$; thus this is the ABJM $k=2$
quotient. This tangent cone is precisely the partial resolution
19 in Figure~\ref{f:c4z2cube}. If we remove $p$ from $\hat{X}_+$, we obtain
a smooth eight-manifold $X_+$ with boundaries $Y_{\mathrm{UV}}=Y^{1,2}(\mathbb{CP}^2)$
and $Y_{\mathrm{IR}}=S^7/\IZ_2$, as in Figure~\ref{fig:UVIR}.

From our general discussion in Subsection \ref{sec:generalGflux}, one
thus expects to be able to
Higgs the field theory dual to
AdS$_4\times Y_{\mathrm{UV}}$ with zero $G$-flux to
the ABJM theory with $k=1$ or $k=2$ (or a dual theory) and zero $G$-flux.
Here the latter corresponds to putting the $N$ M2-branes
at the singular point $p$ of $\hat{X}_+$, while putting the
M2-branes anywhere else on $\hat{X}_+$, or at any point on
$\hat{X}_-$, should have near horizon limit given by the ABJM theory at CS level $k=1$ (or a dual theory).
To investigate what happens \emph{with}
$G$-flux, we must extend the non-zero $G$ over either $X_\pm$, satisfying the
appropriate supersymmetry equations for the flux. Analysing this is
in fact quite technical, although for this relatively simple example
we will be able to provide a complete answer to the problem.
\subsubsection{The geometry $X_+$}
In order to know whether or not we can extend the non-zero torsion flux in $H^4(Y_{\mathrm{UV}},\IZ)\cong\IZ_2$, we need to know something about the cohomology of the smooth eight-manifold
$X_+\equiv \hat{X}_+\setminus \{p\}$.
This has boundary $\partial X_+=Y_{UV}\amalg Y_{IR}$
with two connected components
$Y_{\mathrm{UV}}=Y^{1,2}(\mathbb{CP}^2)$, and $Y_{IR}=S^7/\IZ_2$.
To extend the non-trivial element of $H^4(Y_{\mathrm{UV}},\IZ)\cong\IZ_2$
over $X_+$ we need to examine the exact sequence
\bea
H^4(X_+,\IZ)\rightarrow H^4(\partial X_+,\IZ)\rightarrow
H^5(X_+,\partial X_+,\IZ)~.
\eea
This says we may extend an element of
$H^4(\partial X_+,\IZ)$ over $X_+$ if and only if
it maps to zero in $H^5(X_+,\partial X_+,\IZ)$.
Thus we need to compute the latter group, and
also the map. By Poincar\'e-Lefschetz duality, notice that
$H^5(X_+,\partial X_+,\IZ)\cong H_3(X_+,\IZ)$.

We compute by covering $X_+\subset \hat{X}_+$ with two
open sets, and then using the resulting Mayer-Vietoris
sequence. We first define
$V_1=\{z_1\neq 0\}\cong\IC^*\times\IC^4\subset\IC^5$, with $\IC^5$ having coordinates $(z_1,\ldots,z_5)$.
The invariants under the $\IC^*$ action on $\IC^5$ with charges
$(1,2,-1,-1,-1)$ are
spanned by $x_1=z_2/z_1^2$,
$w_1=z_3z_1$, $w_2=z_4z_1$ and $w_3=z_5z_1$.
Thus $V_1/\IC^*\equiv U_1\cong \IC^4$, with coordinate functions
$x_1,w_1,w_2,w_3$. We similarly
define $V_2=\{z_2\neq 0\}\subset \IC^5$.
The invariants are now
spanned by the 10 functions $x_2=z_1^2/z_2$,
$y_1=z_3z_1$, $y_2=z_4z_1$, $y_3=z_5z_1$,
$y_4=z_3^2z_2$, $y_5=z_4^2z_2$, $y_6=z_5^2z_2$,
$y_7=z_3z_4z_2$, $y_8=z_4z_5z_2$, $y_9=z_3z_5z_2$.
These satisfy the 6 relations
\bea
&& y_1y_2=y_7x_2~, \qquad y_2y_3=y_8x_2~, \qquad y_1y_3=y_9x_2~,\nonumber\\
&& y_1y_7=y_2y_4~, \qquad y_2y_8=y_3y_5~, \qquad y_3y_9=y_1y_6~.
\eea
This precisely defines the affine variety $\IC^4/\IZ_2$,
and thus $V_2/\IC^*\equiv U_2\cong \IC^4/\IZ_2$.
Indeed, if $u_1,u_2,u_3,u_4$ denote standard
coordinates on $\IC^4$, with the $\IZ_2$ action multiplication by
$-1$ on all coordinates, then the invariants
are $u_1^2$, $u_2^2$, $u_3^2$, $u_4^2$, $u_1u_2$, $u_1u_3$, $u_1u_4$,
$u_2u_3$, $u_2u_4$, $u_3u_4$. We may identify
$x_2=u_1^2$, $y_1=u_1u_2$, $y_2=u_1u_3$, $y_3=u_1u_4$,
$y_4=u_2^2$, $y_5=u_3^2$, $y_6=u_4^2$,
$y_7=u_2u_3$, $y_8=u_3u_4$, $y_9=u_2u_4$.

The two coordinate patches $U_1\cong\IC^4$, $U_2\cong\IC^4/\IZ_2$ in fact now cover
$\hat{X}_+$, since one cannot have both $z_1=0$ \emph{and}
$z_2=0$ -- such points violate the moment map equation
(\ref{momentum}) for $\zeta>0$
(in holomorphic language, these points are \emph{unstable} in the GIT quotient).
Hence $X_+=\hat{X}_+\setminus\{p\}$ is covered
by $A_1\equiv U_1\cong\IC^4$ and $A_2\equiv U_2\setminus\{p\}\cong \IR\times S^7/\IZ_2$.
The coordinate patch $U_1$ overlaps
$U_2$ where $z_2\neq 0$. In $U_1$, this is the subset
$\{x_1\neq 0\}$. Thus $U_1\cap U_2\cong A_1\cap A_2\cong
\IC^*\times\IC^3\cong S^1\times \IR^7$,
where the first $\IC^*$ coordinate is $x_1$.

Consider now the Mayer-Vietoris sequence:
\bea\label{MVA}
\nn
0\cong H_3(A_1\cap A_2,\IZ)\rightarrow
H_3(A_1,\IZ)\oplus H_3(A_2,\IZ)\rightarrow
H_3(X_+,\IZ)\rightarrow H_2(A_1\cap A_2,\IZ)\cong0~.
\\
\eea
Since $H_3(A_2,\IZ)\cong H_3(S^7/\IZ_2,\IZ)\cong\IZ_2$,
it thus follows that $H_3(X_+,\IZ)\cong\IZ_2$, which
is the homology group of interest.
Moreover, $U_2\cong\IC^4/\IZ_2$ is the tangent cone to
the singular point $p$, whose link is thus
$Y_{\mathrm{IR}}=S^7/\IZ_2$. The generator of
$H_3(S^7/\IZ_2,\IZ)\cong\IZ_2$ thus trivially
maps to the generator of $H_3(A_2,\IZ)$, whose image via inclusion we have
shown generates $H_3(X_+,\IZ)\cong\IZ_2$.
The Poincar\'e-Lefschetz dual of this is thus
that we have shown that the map
\bea\label{seq}
\IZ_2\oplus\IZ_2\cong H^4(S^7/\IZ_2,\IZ)\oplus H^4(Y_{\mathrm{UV}},\IZ)
\rightarrow H_3(X_+,\IZ)\cong \IZ_2
\eea
takes $(1,0)\in \IZ_2\oplus\IZ_2$ to $1\in\IZ_2$.
To determine the map completely, we need to also
know the image of $(0,1)$. This is the image
of $H_3(Y_{\mathrm{UV}},\IZ)$ in $H_3(X_+,\IZ)$ under inclusion. We may compute this
with a slight modification of the above argument.

Let $X_0=\IR_+\times Y_{\mathrm{UV}}$. This is $\hat{X}_+$ minus the
$\mathbb{WCP}^1_{[1,2]}$ zero-section, which recall
is $\{z_3=z_4=z_5=0\}$. We would like to remove
these points from $\hat{X}_+$ to obtain $X_0$.
In terms of the coordinate
patches, this gives
$B_1\equiv U_1\setminus\{w_1=w_2=w_3=0\}\cong
\IC\times \IR\times S^5\cong \IR^3\times S^5$
and $B_2\equiv U_2\setminus \{y_i=0, i=1,\ldots,9\}\cong
(\IC\times \IR\times S^5)/\IZ_2$, where the $\IZ_2$
acts as $-1$ on $\IC$, and is the $\IZ_2\subset U(1)$ in the Hopf $U(1)$ action on $S^5$
(and thus acts freely of course).
Now $B_1\cap B_2$ is still $\{x_1\neq 0\}\subset B_1$,
which gives $B_1\cap B_2\cong \IC^*\times \IR\times
S^5\cong S^1\times S^5\times \IR^2$.
The Mayer-Vietoris sequence for
$X_0=B_1\cup B_2$ is hence
\bea
0\cong H_3(B_1\cap B_2,\IZ)\rightarrow
H_3(B_1,\IZ)\oplus H_3(B_2,\IZ)\rightarrow
H_3(X_0,\IZ)\rightarrow H_2(B_1\cap B_2,\IZ)\cong0~.\nonumber
\eea
Again, $H_3(B_2,\IZ)\cong\IZ_2$. If we coordinatize
$B_2$ with the coordinates $u_1,u_2,u_3,u_4$,
the generator may be taken to be
$u_1=u_2=0$, $|u_3|^2+|u_4|^2=1$,
which is a copy of $S^3/\IZ_2\subset B_2$.
The above sequence thus proves
that $H_3(X_0,\IZ)\cong H_3(Y_{\mathrm{UV}},\IZ)\cong \IZ_2$, which
of course we already knew. However,
the key point is that this shows that the
generator of $H_3(Y_{\mathrm{UV}},\IZ)$ is represented by the above copy of $S^3/\IZ_2$. But this is also contained in $A_2\supset B_2$,
and similarly generates $H_3(A_2,\IZ)\cong\IZ_2$, and the
Mayer-Vietoris sequence (\ref{MVA}) thus proves that the generator
of $H_3(Y_{\mathrm{UV}},\IZ)\cong\IZ_2$ maps to the generator of $H_3(X_+,\IZ)\cong\IZ_2$ under inclusion. Hence $(0,1)\in\IZ_2\oplus\IZ_2$ in (\ref{seq}) also maps to $1\in\IZ_2$, and thus
the map (\ref{seq}) is simply addition of the two
factors.

All this rather abstract algebraic topology thus shows that
zero $G$-flux on the UV boundary $Y_{\mathrm{UV}}\cong Y^{1,2}(\mathbb{CP}^2)$ lifts to a (necessarily torsion) $G$-flux on the RG flow manifold $X_+$ only if there is \emph{zero} $G$-flux on the IR boundary $Y_{\mathrm{IR}}$.
On the other hand, non-trivial torsion $G$-flux on the UV
boundary, where $H^4(Y_{\mathrm{UV}},\IZ)\cong\IZ_2$,
lifts to a $G$-flux on the RG flow manifold only if there is
\emph{non-trivial} $G$-flux on the IR boundary $Y_{\mathrm{IR}}$,
where $H^4(Y_{\mathrm{IR}},\IZ)\cong\IZ_2$. In the field
theory, it follows that the dual to
$Y_{\mathrm{UV}}$ with/without torsion $G$-flux can be Higgsed to
the CS level $k=2$ ABJM theory (or a dual theory)
with/without torsion $G$-flux, \emph{respectively}. 
This is summarized in Table \ref{HiggsTable}.
\subsubsection{The geometry $X_-$}
Finally, we consider the resolution $\hat{X}_-\cong \mathcal{O}(-1)\oplus\mathcal{O}(-2)\rightarrow\mathbb{CP}^2$. In this case zero $G$-flux
on $Y_{\mathrm{UV}}\cong \partial \hat{X}_-$ clearly extends
as zero $G$-flux over $\hat{X}_-$, but for non-zero flux we must necessarily
turn on a \emph{non-torsion} $G$-flux in $H^4(\hat{X}_-,\IZ)\cong\IZ$. More precisely, we should pick 
a point $p\in \hat{X}_-$ and extend $G$ in $H^4(X_-,\IZ)$ where $X_-=\hat{X}_-\setminus\{p\}$, although the difference
between $\hat{X}_-$ and $X_-$ will not affect our discussion of flux, since removing $p$ does not affect the cohomology of interest.
The flux in turn must be primitive and type $(2,2)$ in order
to satisfy the supersymmetry equations (and hence equations of motion).

To see the existence of such a flux, we may appeal to the
results of \cite{Hausel}. The latter reference proves that
for a complete asymptotically
conical manifold $(X,g_{X})$ of real dimension $m$, we have
\bea
\label{tamas}
\mathcal{H}^k_{L^2}(X,g_{X}) \cong \left\{
\begin{array}{ll} H^k(X,\partial X,\IR), & k<m/2 \\
f(H^{m/2}( X,\partial X,\IR))\subset H^{m/2}( X,\IR), & k=m/2 \\
H^k( X,\IR), & k>m/2\end{array}\right.~,
\eea
where $f$ denotes the ``forgetful'' map, forgetting that a class is relative. Thus the space of $L^2$ harmonic forms $\mathcal{H}^k_{L^2}(X,g_{X})$ is topological.
In particular, we showed earlier that $H^4(\hat{X}_-,\partial \hat{X}_-,\IR)\cong H^4(\hat{X}_-,\IR)\cong \IR$
under the forgetful map (which is multiplication by 2),
and we thus learn that there is a unique
$L^2$ harmonic four-form $G$ on $\hat{X}_-$, up to scale.
If we normalize $(2\pi l_p)^{-3}\int_{\mathbb{CP}^2} G = 2M+1$ to be
odd, then this maps under reduction modulo 2 to the generator of $H^2(Y_{\mathrm{UV}},\IZ)\cong\IZ_2$, for any $M\in\IZ$.
Next note that $\omega\wedge G=0$, where $\omega$ is the K\"ahler form on $\hat{X}_-$,  follows since if $\omega\wedge G$ were not zero
it would be an $L^2$ normalizable harmonic six-form on $\hat{X}_-$, and
there are not any of these by (\ref{tamas}).
Thus the $L^2$ harmonic four-form on $\hat{X}_-$ is necessarily primitive. Next, all of the cohomology
on a toric Calabi-Yau four-fold is of Hodge type $(2,2)$.
Each Hodge type is separately harmonic, and thus
again we see that $G$ has to be purely type $(2,2)$
(any other type would be topologically trivial and harmonic,
and thus zero by (\ref{tamas})). Recall that the explicit asymptotically conical Ricci-flat
K\"ahler metric on $\hat{X}_-$ is known \cite{metrics}, and so in principle
one should be able to construct this harmonic four-form $G$ explicitly.

The Higgsing behaviour expected from this supergravity analysis is summarized in Table \ref{HiggsTable}.
We should see this behaviour in the candidate dual Chern-Simons
quiver theories, to which we turn next.

\TABLE[h!t!]{
\caption{{\sf From the supergravity point of view: a summary of the geometries $Y_{\mathrm{IR}}$ and $Y_{\mathrm{UV}}$, with and without $G$-flux, and whether (yes/no) the corresponding supergravity solution exists.
}}
\begin{tabular}{|l||l|l|}
\hline
Partial resolution & $Y_{\mathrm{UV}}$ without $G$-flux & $Y_{\mathrm{UV}}$ with $G$-flux \\ \hline\hline
$X_-$, near horizon $Y_{\mathrm{IR}}=S^7$ & yes & yes \\
\hline
$X_+$, near horizon $Y_{\mathrm{IR}}=S^7/\IZ_2$, without $G$-flux & yes & no\\
\hline
$X_+$, near horizon $Y_{\mathrm{IR}}=S^7/\IZ_2$, with $G$-flux & no & yes\\
\hline
\end{tabular}
\label{HiggsTable}
}
%=====================================================
\subsection{$Y^{1,2}(\mathbb{CP}^2)$: Field theory results}
%=====================================================
In this subsection we present QCS theories which we conjecture to be dual to AdS$_4 \times Y^{1,2}(\mathbb{CP}^2)$, with and without torsion $G$-flux 
in $H^4(Y^{1,2}(\mathbb{CP}^2),\IZ)\cong\IZ_2$. 
Recall from Table \ref{t:a2t} and Figure \ref{f:duals} that within the first phase $C(Y^{1,2}(\mathbb{CP}^2))_{I}$ for toric diagram number (16) there are two dual phases, denoted as $Ia$ and $Ib$. The analysis for both theories is similiar and gives the same results.
For convenience we fix on the $C(Y^{1,2}(\mathbb{CP}^2))_{Ib}$ phase. 
We conjecture that equal ranks for the three gauge group factors corresponds to backgrounds without torsion $G$-flux, and that by a certain rank deformation (which will be described in Figure~\ref{f:phase0}) the theory with torsion $G$-flux is obtained. This can be motivated by similar reasoning to \cite{Aharony:2008gk} 
for the ABJM theory. As we demonstrate, Higging can be used as a further check for matching $G$-flux backgrounds to particular choices of ranks 
in the dual theory. 

%=====================================================
\subsubsection{Equal ranks: $U(N)\times U(N)\times U(N)$ for $C(Y^{1,2}(\mathbb{CP}^2))_{Ib}$}
\begin{figure}[H]
\includegraphics[trim=0mm 170mm 0mm 0mm, clip, width=6in]{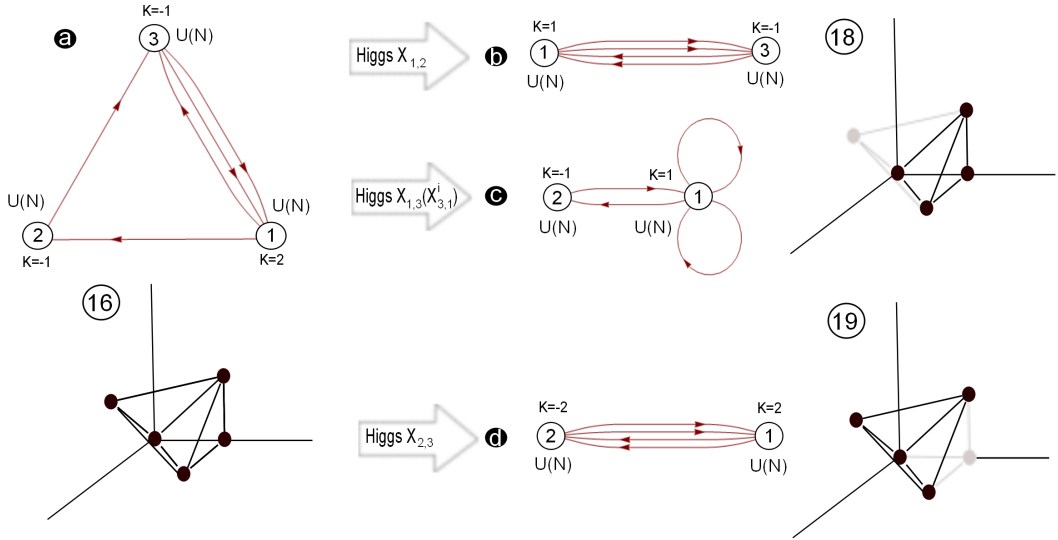}
\caption{{\sf Quiver diagrams for the Higgsed theories obtained from $C(Y^{1,2}(\mathbb{CP}^2))_{Ib}$ with equal ranks $U(N)$ for all three nodes.
}}
\label{f:Tphase0}
\end{figure}

%=====================================================
In this subsection we discuss our candidate for the $Y^{1,2}(\mathbb{CP}^2)$ Freund-Rubin background without torsion $G$-flux. 
This is simply the $C(Y^{1,2}(\mathbb{CP}^2))_{Ib}$ phase with equal ranks, whose quiver is shown in Figure~\ref{f:Tphase0}. 
The superpotential of the theory we recall, from Table \ref{t:a2t}, to be:
\begin{equation}\label{WIb}
W=X_ {3, 1}^{1} X_ {1, 2} X_ {2, 3} X_ {3, 1}^{2} X_ {1, 3} -
X_ {3, 1}^{2} X_ {1, 2} X_ {2, 3} X_ {3, 1}^{1} X_ {1, 3} \ .
\end{equation}
Here, and in the rest of the section, we leave traces implicit. Note that $W$ vanishes in the Abelian case of a single brane ($N=1$). 

%=====================================================
\subsubsection*{Moduli space}
%=====================================================
We first show explicitly, by computing the VMS for $N=1$, that the moduli space for the theory  $C(Y^{1,2}(\mathbb{CP}^2))_{Ib}$ has no extra branches. 
The VMS is determined by equations (\ref{VMSeqns}). 
The last equation in (\ref{VMSeqns}) can be written explicitly as
\begin{align}
\nn
\sigma_1 X_{1,2}-X_{1,2}\sigma_2 &=\sigma_1 X_{1,3}-X_{1,3}\sigma_3 =\sigma_2 X_{2,3}-X_{2,3}\sigma_3 =\\
\sigma_3 X_{3,1}^1-X_{3,1}^1\sigma_1 &=\sigma_3 X_{3,1}^2-X_{3,1}^2\sigma_1 =0 \ .
\label{SigmaEqu-toric0}
\end{align}
The D-terms are:
\begin{align}
\nn
\mu _{1} &=X_{1,2} X_{1,2}^\dagger +X_{1,3} X_{1,3}^\dagger -X_{3,1}^{1 \dagger} X_{3,1}^{1} -X_{3,1}^{2 \dagger} X_{3,1}^{2}=\frac{2}{2 \pi} \sigma_1~, \\ \nn
\mu _{2}&= X_{2,3}X_{2,3}^\dagger -X_{1,2}^\dagger X_{1,2}=-\frac{1}{2 \pi} \sigma_{2}~, \\
\mu _{3}&= X_{3,1}^1 X_{3,1}^{1 \dagger} + X_{3,1}^2 X_{3,1}^{2 \dagger} - X_{2,3}^\dagger X_{2,3} - X_{1,3}^\dagger X_{1,3}=-\frac{1}{2 \pi} \sigma_{3}~,
\label{dterm-toric0}
\end{align}
which we can sum to obtain $\sigma_1 = \frac{1}{2}(\sigma_2 + \sigma_3)$. 
We wish first to solve this equation together with \eqref{SigmaEqu-toric0}. 
By taking $\sigma_1=\sigma_2=\sigma_3$ we indeed obtain the four complex dimensional VMS which is $C(Y^{1,2}(\mathbb{CP}^2))$. 
The $G_t$ matrix which describes the toric diagram can easily be calculated with the aid of the forward algorithm, which gives
\bea
G_t=
\left(
\begin{array}{lllll}
p_1&p_2&p_3&p_4&p_5 \\
\hline
1 & 1 & 1 & 1 & 1 \\
0 & 1 & 0 & 1 & 1 \\
0 & 1 & 1 & 0 & 0 \\
0 & 0 & 1 & 1 & 0 
\end{array}
\right)~.
\eea
This is indeed toric diagram number (16). What about other solutions (branches)? 
A second solution could {\it a priori} come from setting $\sigma_2 \neq \sigma_3$, and since $\sigma_1$ is the average of $\sigma_2$ and $\sigma_3$ we then have $\sigma_1 \neq \sigma_2 \neq \sigma_3$. However, subtituting this into \eqref{SigmaEqu-toric0} we see that all the fields $X_{i,j}$ must vanish and therefore from \eqref{dterm-toric0} we obtain $\sigma_1=\sigma_2=\sigma_3=0$, which is a contradiction. 
Therefore, we see that the only possible solution to the VMS equations is the one in which all the $\sigma$s are equal, and there is hence one irreducible branch in the VMS which is $C(Y^{1,2}(\mathbb{CP}^2))$.
%=====================================================
\subsubsection*{Higgsing}
%=====================================================
The Higgsing of the theory is presented schematically in Figure~\ref{f:Tphase0}. 
Note that for $N=1$ this Higgsing is exhaustive. Since the Abelian superpotential vanishes there is a $1-1$ correspondence between points in the toric diagram and bifundamental fields. That is, the $P$ matrix which relates spacetime fields and the GLSM fields is equal to the identity matrix, and there are hence no multiplicities in the toric diagram.
Recall that when a bifundamental chiral field acquires a VEV, one should delete the corresponding point in the toric diagram. Thus, all toric sub-diagrams can be obtained in this case. Now, we claim that the theories which correspond to toric diagram (18), namely, theories (b) and (c) in Figure~\ref{f:Tphase0}, are dual to backgrounds with no $G$-flux. To see this recall that diagram (18) corresponds to $\IC^4 = C(S^7)$ and since $H^4_{\mathrm{tor}}(S^7,\IZ)\cong 0$, no torsion $G$-flux can be turned on. Theory (d), which corresponds to toric diagram (19) or $\IC^4/\IZ_2$, has equal ranks and was conjectured by \cite{Aharony:2008gk} to have zero torsion $G$-flux. These results are hence precisely in accordance with the calculation in the gravity dual in Subsection \ref{sec:gravity} (see Table \ref{HiggsTable}).
%=====================================================
\subsubsection{Unequal ranks: $U(N) \times U(N) \times U(N+1)$ for $C(Y^{1,2}(\mathbb{CP}^2))_{Ib}$}
%=====================================================
\begin{figure}[h!b!t!]
\includegraphics[trim=0mm 170mm 0mm 0mm, clip, width=6in]{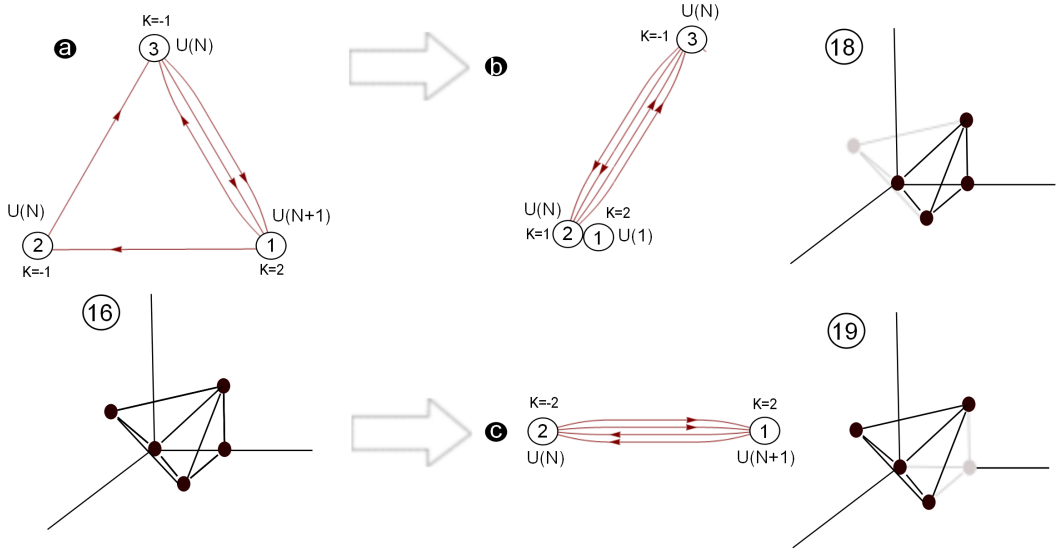}
\caption{{\sf Quiver diagrams for the Higgsed theories obtained from theory $C(Y^{1,2}(\mathbb{CP}^2))_{Ib}$ with unequal ranks.
}}
\label{f:phase0}
\end{figure}
In the previous subsection we analysed a possible candidate theory for the Freund-Rubin solution based on $Y^{1,2}(\mathbb{CP}^2)$ without torsion $G$-flux. 
To complete the discussion we now suggest a candidate dual theory to $Y^{1,2}(\mathbb{CP}^2)$ {\it with} torsion $G$-flux in $H^4(Y^{1,2}(\mathbb{CP}^2),\IZ)\cong\IZ_2$. 
Motivated by the work of \cite{Aharony:2008gk}, we suggest that this theory can be obtained by changing the gauge group ranks of the theory studied above. 
{\it A priori}, there are many ways in which this can be done. However, as we will explain, the analysis of \cite{Aharony:2008gk}  suggests that the only possibility here is
changing the rank of the gauge group at the node with CS level $k=2$ to $U(N+1)$, as shown in Figure~\ref{f:phase0}(a). 
We show that the Higgsing constraints obtained earlier in Subsection \ref{sec:gravity} can serve as a further guide; that is, the Higgsing behaviour of the desired theory should fit the allowed partial resolutions with $G$-flux, as shown in Table \ref{HiggsTable}. 
Note that the superpotential of the theory remains as in \eqref{WIb}.
%=====================================================
\subsubsection*{Moduli space}
%=====================================================
Let us begin by computing explicitly the VMS for our candidate theory with $N=1$ (note this is non-Abelian since the gauge group at node 1 is now $U(2)$); this will serve as a first test of our proposal by confirming that the change in ranks does not alter the \emph{classical} VMS with respect to the theory with equal ranks. 

The classical VMS is determined, as usual, by the equations (\ref{VMSeqns}). 
It will be convenient to use a gauge transformation to set, for the non-Abelian node 1 in the quiver, $\sigma_{1}= \left(
\begin{array}{ll}
\sigma_1^1 & 0 \\
0 & \sigma_1^2
\end{array}
\right)$, where $\sigma _2$ and $\sigma _3$ are real scalars. 
The bifundamental fields involving node 1 are:
\begin{equation}
X_{1,2}=
\left(
\begin{array}{l}
a_1 \\
a_2
\end{array}
\right) \ , \quad \
X_{1,3}=
\left(
\begin{array}{l}
b_1 \\
b_2
\end{array}
\right) \ , \quad \
X_{3,1}^{1}=
\left(
\begin{array}{ll}
c_1 & c_2
\end{array}
\right) \ , \quad \
X_{3,1}^{2}=
\left(
\begin{array}{ll}
d_1 & d_2
\end{array}
\right) \ .
\label{fields}
\end{equation}
Hence, the last equation in \eqref{VMSeqns} can be written explicitly in the following way:
\begin{align}
\nn
0&=
\left(
\begin{array}{l}
(\sigma_1^1 - \sigma_2)a_1 \\
(\sigma_1^2 - \sigma_2)a_2 
\end{array}
\right) =
\left(
\begin{array}{l}
(\sigma_1^1 - \sigma_3)b_1 \\
(\sigma_1^2 - \sigma_3)b_2 
\end{array}
\right)~, \\ \nn
0&=\left(
\begin{array}{ll}
(\sigma_3 - \sigma_1^1)c_1, &
(\sigma_3 - \sigma_1^2)c_2 
\end{array}
\right) =
\left(
\begin{array}{ll}
(\sigma_3 - \sigma_1^1)d_1, &
(\sigma_3 - \sigma_1^2)d_2 
\end{array}
\right)~,\\
0&=X_{2,3} (\sigma_2 - \sigma_3) \ .
\label{SigmaEqu}
\end{align}
The D-terms become
\bea
\nn
\left(
\begin{array}{ll}
\vert a_1 \vert ^2 & a_1a_2^* \\
a_2a_1^* & \vert a_2 \vert ^2 
\end{array}
\right) +
\left(
\begin{array}{ll}
\vert b_1 \vert ^2 & b_1b_2^* \\
b_2b_1^* & \vert b_2 \vert ^2 
\end{array}
\right) -
\left(
\begin{array}{ll}
\vert c_1 \vert ^2 & c_1^*c_2 \\
c_2^*c_1 & \vert c_2 \vert ^2 
\end{array}
\right) -
\left(
\begin{array}{ll}
\vert d_1 \vert ^2 & d_1^*d_2 \\
d_2^*d_1 & \vert d_2 \vert ^2 
\end{array}
\right) &=& 
\frac{2}{2 \pi} \left(
\begin{array}{ll}
\sigma_1^1 & 0 \\
0 & \sigma_1^2
\end{array}
\right)~,\\ \nn
\vert X_{2,3} \vert ^2 - \vert a_1 \vert ^2 - \vert a_2 \vert ^2 &=& -\frac{1}{2 \pi} \sigma _{2}~, \\
\vert c_1 \vert ^2 + \vert c_2 \vert ^2 + \vert d_1 \vert ^2 + \vert d_2 \vert ^2 - \vert X_{2,3} \vert ^2 
- \vert b_1 \vert ^2 - \vert b_2 \vert ^2 &=& -\frac{1}{2 \pi} \sigma _{3}~.
\label{dterm2}
\eea
There are three different solutions to these equations:
\begin{align}
\nn (1) \qquad & 
{\sigma_1^2 = \sigma_3=\sigma_2 = \sigma~, \quad \sigma_1^1 = 0~, \quad a_1 = 0~, \quad b_1 = 0~, \quad c_1 = 0~, \quad d_1 = 0}~, \\
\nn (2) \qquad & 
{\sigma_1^1 = \sigma_3=\sigma_2 = \sigma~, \quad \sigma_1^2 = 0~, \quad a_2 = 0~, \quad b_2 = 0~, \quad c_2 = 0~, \quad d_2 = 0}~,\\
(3) \qquad & {\sigma_1^1 = \sigma_1^2 = \sigma_2 = \sigma_3 = 0}~.
\label{solutions}
\end{align}
Solutions (1) and (2) are related by the $\IZ_2$ Weyl symmetry in the gauge group $U(2)$. 
It is easy to see that by setting $\sigma_1^1=\sigma_2=\sigma_3=\sigma$ and taking $a_2=b_2=c_2=d_2=0$ we satisfy all the equations, obtaining the same set of equations (including vanishing F-term) as for the $U(1)^3$ theory. 
Therefore, after identifying by the $\IZ_2$ Weyl symmetry, this branch of the VMS is precisely $C(Y^{1,2}(\mathbb{CP}^2))$. 

To prove there are no extra branches we now show that solution (3) is a sub-locus of solution (2) (or equivalently solution (1) using the $\IZ_2$ Weyl symmetry).
Since the constraints (3) in \eqref{solutions} are a subset of (2), we just need to show that the constraints on the bifundamental fields in branch (3) are those in (2). Notice that for branch (3) we are free to use the complete $U(2)$ gauge symmetry because $\sigma_1=0$. The first case to examine is $X_{2,3}=0$, the substitution of which into the D-term gives us $a_1=a_2=0$. 
Using a gauge transformation we set $b_2=0$, and, from \eqref{dterm2}, $c_2=d_2=0$. 
Therefore, we see that for $\sigma_1=\sigma_2=\sigma_3=0$ and $X_{2,3}=0$ the constraints in (2) and (3) are the same. 
Next, we repeat this calculation for $X_{2,3}\neq 0$. We can first use a gauge transformation to set $d_2=0$. 
Let us consider $d_1=0$ and $d_1\neq 0$ separately. If $d_1=0$, we can use a gauge transformation to set $c_2=0$, and we see from the first equation in \eqref{dterm2} that $a_2=b_2=0$. 
If $d_1\neq 0$ the F-term reduces to
\begin{align}
\nn b_1 a_2 = a_1 b_2~, \quad b_2 c_2 = 0~, \quad b_1 c_2 = 0~, \quad a_2 c_2 = 0~, \quad a_1 c_2 = 0~,
\end{align}
which we can solve in two different ways. First, we could set $c_2=0$, and from the first equation in \eqref{dterm2}, we obtain $a_2=0$ and $b_2=0$. 
Second, we could set $a_1=a_2=b_1=b_2=0$, and from \eqref{dterm2}, we then have $c_1=c_2=X_{2,3}=0$. 
Hence, in both cases the constraints are at least $a_2=b_2=c_2=d_2=0$. We thus conclude that the only branch in the moduli space is $C(Y^{2,1}(\mathbb{CP}^2))$, as claimed. 

Let us discuss briefly the implications of the moduli space computation on the allowed configurations of ranks. 
At low energy on the VMS we get $N$ copies of a $U(1)_2 \times U(1)_{-1} \times U(1)_{-1}$ theory, together with a pure $U(1)_2$ ${\cal N} = 3$ supersymmetric CS theory. 
The fields that are charged under the latter are massive, as can be seen from the following term in the action
\begin{align}\label{S-massive}
S&\supset \int \dd^3x\,\sum_{X_{i,j}} (\sigma _i 
X_{i,j} - X_{i,j} \sigma_j)(\sigma _i X_{i,j} - X_{i,j} \sigma_j)^\dagger \ .
\end{align}
As discussed in \cite{Aharony:2008gk}, an ${\cal N} = 3$ supersymmetric $U(\ell)_k$ pure CS theory does not exist as a unitary theory for $\ell > k$. 
This implies that in our case the allowed ranks are $\ell \leq 2$ for the node with CS level $k=2$ and $\ell \leq 1$ for the other nodes. 
If we assume that the $\ell=k$ and $\ell=0$ cases correspond to equivalent theories, which is true for the ABJM theory and the non-toric theories 
discussed in \cite{Martelli:2009ga}, we see that the rank deformation that we have chosen is the only one possible. 
We shall discuss this further at the end of the Higgsing analysis in the next subsection.
%=====================================================
\subsubsection*{Higgsing}
%=====================================================
Finally, we examine the Higgsing behaviour of the proposed theory, and compare it to that expected from the dual supergravity analysis.  As we shall 
see, the two precisely match with the choice of ranks we have made. 
The theories that can be obtained by Higgsing are presented in Figure~\ref{f:phase0}. 

To explain these results, let us start by recalling the action of the theory:
\begin{eqnarray}
\nonumber
S&=\int \dd^3x\, \Big[ \sum_i k_i \epsilon^{\mu\nu\rho} ({\cal A}^i_{\mu}\partial_{\nu}{\cal A}^i_{\rho} + \frac{2}{3} {\cal A}^i_{\mu} {\cal A}^i_{\nu} {\cal A}^i_{\rho}) - \sum\limits_{X_{i,j}} (D_{\mu} X_{i,j})^\dagger.(D_{\mu} X_{i,j}) \\
&+ \frac{1}{2 \pi} \sum\limits_i k_i \sigma_i D_i - \sum\limits_i D_i \mu_i(X) \\
\nn
& -\sum\limits_{X_{i,j}} (\sigma_i X_{i,j} - X_{i,j} \sigma_j)(\sigma_i X_{i,j} - X_{i,j} \sigma_j)^\dagger -\sum\limits_{X_{i,j}} |\partial_{X_{i,j}} W|^2 \Big]\ ,
\end{eqnarray}
where $\mu_i(X)$ is the D-term  for the $i$th gauge group:
\begin{equation} 
\mu_i(X)\equiv \sum_j X_{i,j} X_{i,j}^\dagger - \sum_k X_{k,i}^\dagger X_{k,i}~. %+ [X_{i,i},X_{i,i}^\dagger] \ .
\end{equation}

To derive theory (b) in Figure~\ref{f:phase0} we let $X_{1,2}$ acquire the following VEV
$X_{1,2}=m
\left(
\begin{array}{l}
1_{N \times N} \\
0
\end{array}
\right)$.
As a check, notice that with this VEV it is still possible to satisfy the VMS equations. 
The F-term can be satisfied by setting all other chiral field VEVs to zero, and the D-terms 
\begin{align}
\nn
\mu _{1} &\equiv X_{1,2} X_{1,2}^\dagger +X_{1,3} X_{1,3}^\dagger -X_{3,1}^{1 \dagger} X_{3,1}^{1} -X_{3,1}^{2 \dagger} X_{3,1}^{2}=\frac{2}{2 \pi} \sigma_1 + \epsilon_1~,\\ \nn
\mu _{2}&\equiv  X_{2,3}X_{2,3}^\dagger -X_{1,2}^\dagger X_{1,2}=-\frac{1}{2 \pi} \sigma_{2} + \epsilon_2~,\\
\mu _{3}&\equiv X_{3,1}^1 X_{3,1}^{1 \dagger} + X_{3,1}^2 X_{3,1}^{2 \dagger} - X_{2,3}^\dagger X_{2,3} - X_{1,3}^\dagger X_{1,3}=-\frac{1}{2 \pi} \sigma_{3} + \epsilon_3~,
\label{dterm}
\end{align}
can be satisfied by seting $\sigma_1= \pi m^2 \ast \mathrm{diag}(1_{N \times N},0)$, $\sigma_2=\sigma_3= \pi m^2 \ast 1_{N \times N}$ and turning on FI parameters as follows: $\epsilon_1=0$, $\epsilon_2=-\frac{m^2}{2} 1_{N \times N}$ and $\epsilon_3=\frac{m^2}{2} 1_{N \times N}$.
This corresponds to turning on a negative FI paramter for the moment map $\mu _{2}-\mu _{3}$. 
Finally, it can be seen that the third set of equations in \eqref{VMSeqns} are also satisfied. 

We next discuss how this VEV Higgses the gauge group. Since we are giving a VEV to $X_{1,2}$ we are interested in the following part of the action 
\begin{align}
S&=\int \dd^3x\, \left[k_1 \epsilon^{\mu\nu\rho} {\cal A}^1_{\mu}\partial_{\nu}{\cal A}^1_{\rho}+k_2\epsilon^{\mu\nu\rho}{\cal A}^2_{\mu}\partial_{\nu}{\cal A}^2_{\rho} - (D_{\mu} X_{1,2})^\dagger.(D_{\mu} X_{1,2})+\cdots\right]~.
\label{relevant}
\end{align}
Recall the definition of the covariant derivative:
\begin{equation}
D_{\mu} X_{1,2}=\partial_{\mu} X_{1,2}-\ii({\cal A}^1_{\mu} X_{1,2}- X_{1,2}{\cal A}^2_{\mu})~,
\end{equation}
wherein it is convenient to define 
\bea
{\cal A}^1=
\left(
\begin{array}{ll}
{\cal B}^1_{N \times N} & {\cal A}^1_{OD,N \times 1} \\
{\cal A}^{1 \dagger}_{OD,1 \times N} & {\cal C}^1
\end{array}
\right) \ ,
\eea
whence,
\begin{equation}
|(D_{\mu} X_{1,2})|^2 \supset m^2\left\vert
\left(
\begin{array}{l}
{\cal B}^1_{N \times N} \\
{\cal A}^{1 \dagger}_{OD,1 \times N} 
\end{array}
\right)-
\left(
\begin{array}{l}
{\cal A}^2_{N \times N} \\
0 
\end{array}
\right)
\right\vert ^2= 
m^2|{\cal B}^1-{\cal A}^2|^2+m^2|{\cal A}^1_{OD}|^2 \ .
\end{equation}
Substituting this result into \eqref{relevant} we can rewrite the relevant part of the Lagrangian as
\begin{align}
\nonumber
S
&= \int \dd^3x\, \Big[ k_1 \epsilon^{\mu\nu\rho}( {\cal B}^1_{\mu}\partial_{\nu}{\cal B}^1_{\rho}+{\cal A}^1_{OD\mu}\partial_{\nu}{\cal A}^{1 \dagger}_{OD\rho}+{\cal A}^{1 \dagger}_{OD\mu}\partial_{\nu}{\cal A}^1_{OD\rho}+{\cal C}^1_{\mu}\partial_{\nu}{\cal C}^1_{\rho})+k_2\epsilon^{\mu\nu\rho}{\cal A}^2_{\mu}\partial_{\nu}{\cal A}^2_{\rho}  \\ 
& \qquad - m^2|{\cal B}^1-{\cal A}^2|^2-m^2|{\cal A}_{OD}^1|^2+\cdots\Big]~.
\label{relevant2}
\end{align}
We see that ${\cal A}_{OD}^1$ has become massive. 
In the IR we can therefore consider this field as constant, $\partial{\cal A}^1_{OD}=0$. 
Solving the equations of motion we see that ${\cal A}_{OD}^1 \varpropto \frac{1}{m^2}$, and therefore terms that contain $A_{OD}^1$ can be deleted from the Lagrangian in the low-energy limit. 
Next, let us integrate out the second massive combination of gauge fields, namely, ${\cal B}^1-{\cal A}^2$. After deleting ${\cal A}_{OD}^1$ from \eqref{relevant2} and defining 
\begin{equation}
{\cal A}^{\pm}=\frac{1}{2}({\cal B}^1\pm {\cal A}^2) \ ,
\label{newgauge}
\end{equation}
together with $k_{\pm}=k_1 \pm k_2$, \eqref{relevant2} reduces to 
\begin{eqnarray}
\nn
S&=\int \dd^3x\, \Big[ k_+\epsilon^{\mu\nu\rho} {\cal A}^+_{\mu}\partial_{\nu}{\cal A}^+_{\rho}+k_+\epsilon^{\mu\nu\rho}{\cal A}^-_{\mu}\partial_{\nu}{\cal A}^-_{\rho} +2k_-\epsilon^{\mu\nu\rho} {\cal A}^-_{\mu}\partial_{\nu}{\cal A}^+_{\rho}\\
& \qquad 
+k_1\epsilon^{\mu\nu\rho} {\cal C}^1_{\mu}\partial_{\nu}{\cal C}^1_{\rho}
-4m^2({\cal A}^-_{\mu})^2+\cdots\ \Big]~.
\end{eqnarray}
Again, we see that by solving the equations of motion ${\cal A}^- \varpropto \frac{1}{m^2}$, and therefore terms which contain ${\cal A}^-$ can be deleted from the Lagrangian in the low-energy limit. 

\begin{figure}[H]
\includegraphics[trim=0mm 210mm 0mm 0mm, clip, width=6in]{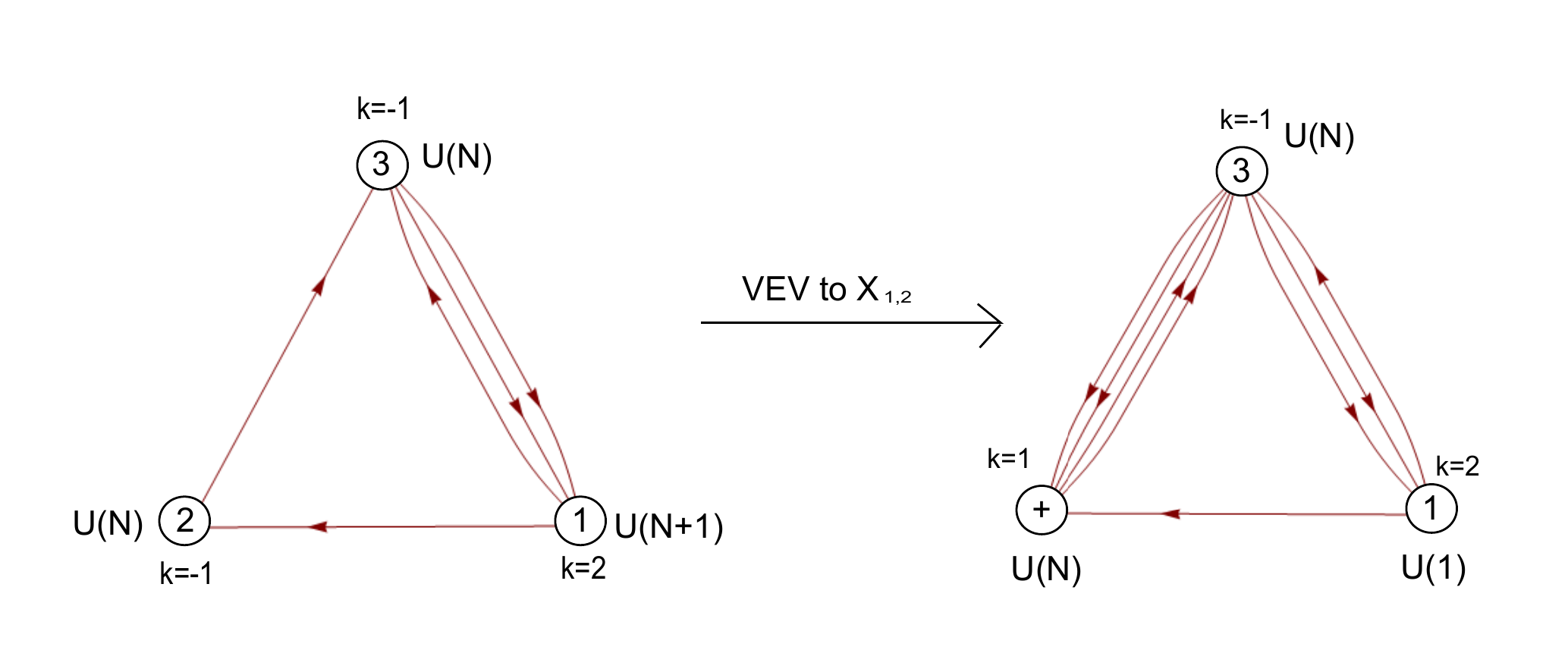}
\caption{{\sf The intermediate step of giving $X_{1,2}$ a VEV in Higgsing the theory $C(Y^{1,2}(\mathbb{CP}^2))_{Ib}$ with gauge group $U(N) \times U(N) \times U(N+1)$ going into
$U(N) \times U(N) \times U(1)$.
}}
\label{f:Vev_X12}
\end{figure}

Now that we have completed the discussion of the Higgsing of the gauge groups we would like to see how the action in the IR is modified. 
First, let us examine the covariant derivative term around the new vacuum, in which we define
$X_{1,2}=\left(
\begin{array}{l}
m*1_{N \times N}+X \\
\tilde{X}_{1,+} 
\end{array}
\right)$.
It is easy to see that 
\begin{align}
\nonumber
|D_{\mu} X_{1,2}|^2 &\supset \left\vert 
\left(
\begin{array}{l}
{\cal B}^1 (m*1_{N \times N}+X)+{\cal A}^1_{OD}\cdot\tilde{X}_{1,+} \\
{\cal A}^{1 \dagger}_{OD} (m*1_{N \times N}+X)+ {\cal C}^1\cdot\tilde{X}_{1,+}
\end{array}
\right)-
\left(
\begin{array}{l}
(m*1_{N \times N}+X) {\cal A}^2 \\
\tilde{X}_{1+}\cdot{\cal A}^2 
\end{array}
\right)\right\vert^2 \\ \nonumber
&= |{\cal B}^1 (m*1_{N \times N}+X)+{\cal A}^1_{OD}\cdot\tilde{X}_{1,+} - (m*1_{N \times N}+X) {\cal A}^2|^2 \\
& \qquad 
+ |{\cal A}^{1 \dagger}_{OD} (m*1_{N \times N}+X)+ {\cal C}^1\cdot\tilde{X}_{1,+}-\tilde{X}_{1,+}\cdot{\cal A}^2|^2 \ ,
\end{align}
which in the low-energy limit becomes $|{\cal C}^1\cdot\tilde{X}_{1,+}-\tilde{X}_{1,+}\cdot{\cal A}^2|^2$. 
Moreover, in the low-energy limit ${\cal A}^2={\cal A}^+ -{\cal A}^- \simeq {\cal A}^+$, and therefore
\begin{eqnarray}
(D_{\mu} X_{1,2})^\dagger\cdot(D_{\mu} X_{1,2}) &=& (\partial_{\mu} \tilde{X}_{1,+})^\dagger\cdot(\partial_{\mu} \tilde{X}_{1,+}) + |C^1\cdot\tilde{X}_{1,+}-\tilde{X}_{1,+}\cdot A^+|^2 \nn \\ 
&=&(D_{\mu} \tilde{X}_{1,+})^\dagger\cdot (D_{\mu} \tilde{X}_{1,+}) \ .
\label{newCD}
\end{eqnarray}
It is thus clear that the remaining degree of freedom in the field $X_{1,2}$ transforms in the $(N_1,\bar{N}_+)$ bifundamenal representation of the gauge group, where the subscript $+$ denotes the new node $2$ after the Higgsing, as shown in Figure \ref{f:Vev_X12}. 
Similarly, since in the IR ${\cal B}^{1}={\cal A}^+ + {\cal A}^- \simeq {\cal A}^+$, we see that the bifundamental charged under ${\cal A}^2$ or ${\cal B}^1$ now carries charge under ${\cal A}^+$. Therefore, the fields after Higgsing can be written as follows:
\bea
\nn
&&X_{2,3;N \times N} \rightarrow X^1_{+,3;N \times N} \ , \ 
X_{1,3;(N+1) \times N} \rightarrow 
\left(
\begin{array}{l}
X^2_{+,3;N \times N} \\
\tilde{X}_{1,3;1 \times N} 
\end{array}
\right) \\ 
&&X^i_{3,1;N \times (N+1)} \rightarrow 
\left(
\begin{array}{ll}
X^i_{3,+;N \times N} & \tilde{X}^i_{3,1;N \times 1} 
\end{array} 
\right)
\eea
We draw this intermediate step in Figure \ref{f:Vev_X12}.
This is not the quiver of the final theory as there are other fields which become massive, as we now explain. 
To see which of the fields become massive we need to examine the relevant terms in the action; these are of the form presented in \eqref{S-massive}.
In the notations of the intermediate quiver $\sigma_1=0$ and $\sigma_3 =\pi m^2 1_{N \times N}$, while from \eqref{newgauge} it follows that $\sigma_+ =\pi m^2 1_{N \times N}$. 
Therefore all the bifundamental fields between node 1 and the other nodes in the quiver in Figure~\ref{f:Vev_X12} become massive. 
To illustrate this point let us investigate an example by focusing on the term that contains $\tilde{X}_{1,+}$, namely
\begin{align}
(\sigma _1 \tilde{X}_{1,+} - \tilde{X}_{1,+} \sigma_+)(\sigma _1 \tilde{X}_{1,+} - \tilde{X}_{1,+} \sigma_+)^\dagger = \pi^2m^4|\tilde{X}_{1,+}|^2 \ .
\end{align}
From this we see that $\tilde{X}_{1,+}$ acquires a mass. 
After integrating this out the superpotential reduces to
\bea
\nonumber
W&=&X_ {3, 1}^{1} X_ {1, 2} X_ {2, 3} X_ {3, 1}^{2} X_ {1, 3} -
X_ {3, 1}^{2} X_ {1, 2} X_ {2, 3} X_ {3, 1}^{1} X_ {1, 3} \\ \nn \\ \nn
&=&
\left(
\begin{array}{ll}
X^1_{3,+} & 0 
\end{array}
\right)\cdot
\left(
\begin{array}{l}
1_{N \times N} \\
0
\end{array}
\right)\cdot
X^1_{+,3}\cdot
\left(
\begin{array}{ll}
X^2_{3,+} & 0 
\end{array}
\right)\cdot
\left(
\begin{array}{l}
X^2_{+,3} \\
0 
\end{array}
\right) \nonumber \\
&& - \left(
\begin{array}{ll}
X^2_{3,+} & 0 
\end{array}
\right)\cdot
\left(
\begin{array}{l}
1_{N \times N} \\
0
\end{array}
\right)\cdot
X^1_{+,3}\cdot
\left(
\begin{array}{ll}
X^1_{3,+} & 0 
\end{array}
\right)\cdot
\left(
\begin{array}{l}
X^2_{+,3} \\
0 
\end{array}
\right)\nn \\
&=& X^1_{3,+} X^1_{+,3} X^2_{3,+} X^2_{+,3}- X^2_{3,+} X^1_{+,3} X^1_{3,+} X^2_{+,3}~.
\eea
The resulting theory is (b) in Figure~\ref{f:phase0}. 
We see that this theory is just the $U(N)_1 \times U(N)_{-1}$ ABJM theory together with a decoupled $U(1)$ gauge group. 
This theory is dual to AdS$_4 \times S^7$. 

Finally, we will derive theory (c) in Figure~\ref{f:phase0}. The analysis proceeds very much analogously to that above.
We begin by giving the following VEV: $X_{2,3}=m*1_{N \times N}$.
As before, with this VEV it is still possible to satisfy the VMS equations. 
The F-term can be satisfied by setting all other spacetime fields to have zero VEV. 
The D-term \eqref{dterm} is satisfied by seting $\sigma_1=\sigma_2=\sigma_3=0$, $\epsilon_1=0$, $\epsilon_2=m^2*1_{N \times N}$ and $\epsilon_3=-m^2*1_{N \times N}$; this corresponds to turning on a positive FI parameter for the moment map $\mu_2-\mu_3$. 
It is easy to see that \eqref{SigmaEqu} are also satisfied since all the $\sigma$ are set to zero. 
Since we gave a diagonal VEV to the bifundamental field which is charged under $(N_2,\bar{N_3})$, it follows that the $U(N)_2 \times U(N)_3$ group is Higgsed to the diagonal $U(N)$, with $k_2+k_3$ as CS level. 
The superpotential precisely reduces to that of the ABJM theory. 
Since all $\sigma$s are set to zero, there are no additional massive fields which should be integrated out, and we obtain precisely theory (c) in Figure~\ref{f:phase0}. 
This is a theory which was conjectured in \cite{Aharony:2008gk} to be dual to AdS$_4 \times S^7/\IZ_2$ with one unit of torsion $G$-flux, as reflected by the ranks of the gauge groups. 

As a final remark, note that we could not find any other way to Higgs theory (a) in Figure~\ref{f:phase0} to obtain a theory which is dual to AdS$_4 \times S^7/\IZ_2$ with no torsion $G$-flux. 
Once more, this is in line with with the calculation in the gravity dual as presented in Table \ref{HiggsTable}. Moreover, one can repeat the same analysis for the $U(N)_{-1} \times U(N)_{-1} \times U(N+2)_{2}$ theory. It is easy to see that we will still obtain the same theory (b), but with an isolated node which corresponds to a $U(2)_2$ theory, having, of course, the same VMS.
Theory (c) will change to $U(N)_{-2} \times U(N+2)_{2}$, which according to \cite{Aharony:2008gk} is equivalent to $U(N)_{-2} \times U(N)_{2}$ and therefore to a background with no $G$-flux. 
Thus we see that the Higgsing behaviour of $U(N)_{-1} \times U(N)_{-1} \times U(N+2)_{2}$ and $U(N)_{-1} \times U(N)_{-1} \times U(N)_{2}$ are the same, and this is an indication that these theories are indeed dual to each other, as we suggested earlier.

%=====================================================
\subsection{$C(Y^{1,2}(\mathbb{CP}^2))_{II}$ : Some puzzles}\label{sec:extra}
%=====================================================
So far we have shown that there is a correspondence between gravity calculations and field theory results by examining the VMS and Higgsing behaviour of $C(Y^{1,2}(\mathbb{CP}^2))_{I}$, both with and without torsion $G$-flux. However, similiar examination of $C(Y^{1,2}(\mathbb{CP}^2))_{II}$, which was derived earlier by Higgsing $(\IC^4/\IZ_2^3)_{II}$, does not fit with these expectations. 
This raises some puzzles which should be further investigated in order to decide if this phase really describes the worldvolume theory on M2-branes at the singularity described by toric diagram (16). We note that this theory has a Type IIA derivation in \cite{Aganagic:2009zk}.

We begin by showing that by Higgsing $C(Y^{1,2}(\mathbb{CP}^2))_{II}$ it is {\em not} possible to obtain any theory corresponding to $\IC^4/\IZ_2$, with any $G$-flux configuration. This is related to the fact that the partial resolution $\hat{X}_+$ cannot be obtained as a VMS of the field theory, for any choice of FI parameters, and thus that the master space for this theory is ``too small''.
Next, we will show that the VMS of $C(Y^{1,2}(\mathbb{CP}^2))_{II}$ contains an \emph{extra} branch which does not appear in the VMS of $C(Y^{1,2}(\mathbb{CP}^2))_I$. It is not clear what the physical interpretation of this branch should be, in terms of M2-branes.
Finally, we will demonstrate that there is a mismatch between the mapping of toric divisors and baryonic operators, which does not occur in the case of $C(Y^{1,2}(\mathbb{CP}^2))_{I}$; again, this is closely related to the master space being too small.
\begin{figure}[H]
\includegraphics[trim=0mm 170mm 0mm 0mm, clip, width=6in]{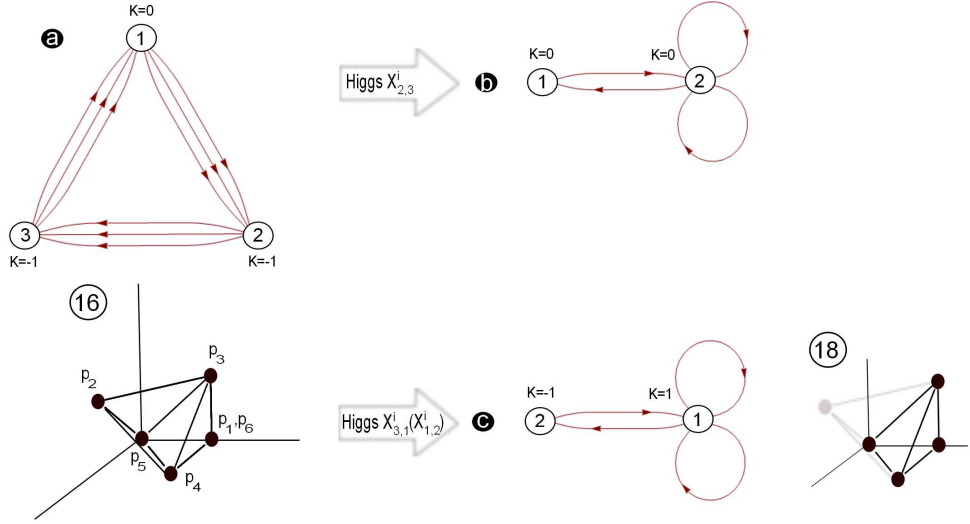}
\caption{{\sf Quiver diagrams for the Higgsed theories obtained from $C(Y^{1,2}(\mathbb{CP}^2))_{II}$, corresponding to toric diagram (16).}}
\label{f:phase1}
\end{figure}
The quiver and GLSM fields $p_{1,\ldots,6}$ corresponding to toric diagram (16), as well as its Higgsing to (18), are presented in Figure~\ref{f:phase1}. 
The superpotential $W$ of the theory and associated P-matrix relating spacetime and GLSM fields are: 
\begin{equation}
\begin{array}{rcl}
\begin{array}{rcl}
W&=&
-X_ {1, 2}^{1} X_ {2, 3}^{1} X_ {3, 1}^{3} + 
X_ {1, 2}^{3} X_ {2, 3}^{1} X_ {3, 1}^{1} - \\
&&
- X_ {1, 2}^{2} X_ {2, 3}^{2} X_ {3, 1}^{1} +  
+ X_ {1, 2}^{1} X_ {2, 3}^{2} X_ {3, 1}^{2} +\\ 
&&
+ X_ {1, 2}^{2} X_ {2, 3}^{3} X_ {3, 1}^{3} - 
X_ {1, 2}^{3} X_ {2, 3}^{3} X_ {3, 1}^{2} \ ;
\end{array}
&
\quad
&
P=
{\scriptsize
\left(
\begin{array}{l|llllll}
& p_1 & p_2 & p_3 & p_4 & p_5 & p_6 \\
\hline
X_{1,2}^1 & 1 & 0 & 1 & 0 & 0 & 0 \\
X_{1,2}^2 & 1 & 0 & 0 & 1 & 0 & 0 \\
X_{1,2}^3 & 1 & 0 & 0 & 0 & 1 & 0 \\
X_{2,3}^1 & 0 & 1 & 0 & 1 & 0 & 0 \\
X_{2,3}^2 & 0 & 1 & 0 & 0 & 1 & 0 \\
X_{2,3}^3 & 0 & 1 & 1 & 0 & 0 & 0 \\
X_{3,1}^1 & 0 & 0 & 1 & 0 & 0 & 1 \\
X_{3,1}^2 & 0 & 0 & 0 & 1 & 0 & 1 \\
X_{3,1}^3 & 0 & 0 & 0 & 0 & 1 & 1 
\end{array}
\right)
}.
\end{array}
\label{pmatrix}
\end{equation}

We see from Figure \ref{f:phase1} that by deleting nodes $\{p_3, p_6\}$ from the toric diagram we can arrive at toric diagram (18).
Now, if we were to be able to Higgs to diagram (19), we would have to delete GLSM fields $\{p_1,p_6\}$.
However, from the P-matrix in \eqref{pmatrix}, there are no spacetime fields which correspond to this deletion and hence (19) cannot be reached by Higgsing. 
This can also be seen from the GLSM picture. Because of the F-terms, the GLSM coming from the field theory is non-minimal and has multiplicities in $p$-fields. 
The GLSM equations (from the $Q_t$-matrix) are derived\footnote{Alternatively, these were derived by hand in \cite{Martelli:2008si}.} easily with the aid of the forward algorithm:
\bea
\nn\label{larry}
|p_6|^2+|p_1|^2+ |p_2|^2-|p_3|^2-|p_4|^2-|p_5|^2&=&0 \ , \\
 2|p_1|^2+|p_2|^2-|p_3|^2-|p_4|^2-|p_5|^2&=&\zeta \ .
\eea
The second line, containing only five GLSM fields, is precisely the \emph{minimal} GLSM description of $C(Y^{1,2}(\mathbb{CP}^2))$ discussed in Subsection \ref{sec:gravity}, {\it cf.} equation (\ref{momentum}). One can take the FI parameter $\zeta$ to be either positive or negative, and in the absence of the first line and the $p_6$ field this gives the (partial) resolutions $\hat{X}_+$ and $\hat{X}_-$ discussed earlier in Subsection \ref{sec:gravity}. However, in the field theory the effect of the F-terms is to add a sixth GLSM field $p_6$, together with the first line in (\ref{larry}) (for which the FI parameter is zero, since these are F-terms). The moduli space described by (\ref{larry}) for negative \emph{and} positive $\zeta$ is then $\hat{X}_-$. Indeed, combining the two equations in (\ref{larry}) gives $|p_6|^2 = |p_1|^2-\zeta$. For $\zeta<0$ one can simply eliminate $p_6$ using this equation, and the moduli space reduces to the second equation, which describes $\hat{X}_-$. On the other hand, for $\zeta>0$ one instead eliminates $p_1$ using $|p_1|^2 = |p_6|^2+\zeta$, which gives 
$2|p_6|^2 + |p_2|^2 - |p_3|^2-|p_4|^2-|p_5|^2= -\zeta$, which is again $\hat{X}_-$. 
Thus it is not possible to realize the partial resolution $\hat{X}_+$ as a branch of the field theory moduli space, which may be phrased as saying that the master space is ``too small'' -- it doesn't contain all partial resolutions of the cone geometry. On the other hand, the singularity $\IC^4/\IZ_2$ described by toric diagram (19) is the near-horizon (tangent cone) limit of the singular point in $\hat{X}_+$. This of course explains why it is not possible to Higgs $C(Y^{1,2}(\mathbb{CP}^2))_{II}$ to a theory for toric diagram (19), and is the reason why theory (19) was missing in the list in Figure~\ref{f:phaseIIIhiggs}.

Another puzzle arises when we examine the VMS of $C(Y^{1,2}(\mathbb{CP}^2))_{II}$, whose explicit equations are:
\bea
\nn \partial_{X_{i,j}} W &=& 0 \ , \\
\nn \mu_i &:=& -\sum\limits_{j=1}^3 X_{j,i}^{\dagger} {X_{j,i}}+ 
\sum\limits_{k=1}^3  {X_{i,k}} X_{i,k}^{\dagger} = 
\frac{k_i\sigma_i}{2\pi} \ , \\
\sigma_i X_{i,j} - X_{i,j} \sigma_j &=& 0 \ .
\label{DF1} 
\eea
The last equation in \eqref{DF1} can be written in the following way:
\begin{equation}\label{SEqn}
X_{1,2}^i (\sigma_1 - \sigma_2) =
X_{2,3}^i (\sigma_2 - \sigma_3) =
X_{3,1}^i (\sigma_3 - \sigma_1) = 0 \ .
\end{equation}
Solving these equations by taking $\sigma_1=\sigma_2=\sigma_3$ corresponds to the complex dimension four moduli space which is precisely $C(Y^{1,2}(\mathbb{CP}^2))$.
We can see this from the forward algorithm;
the $G_t$ matrix which describes the toric diagram of this branch is
\bea
G_t=
\left(
\begin{array}{llllll}
p_1&p_2&p_3&p_4&p_5&p_6 \\
\hline
 1 & 1 & 1 & 1 & 1 & 1 \\
 0 & 1 & 0 & 1 & 0 & 0 \\
 1 & 0 & 1 & 1 & 0 & 1 \\
 0 & 1 & 1 & 0 & 0 & 0
\end{array}
\right)~,
\eea
which is indeed toric diagram number (16).
Notice that $p_1$ and $p_6$ correspond to the same point in the toric diagram. 

However, we now show that there is an extra branch in the VMS. 
Since $k_1=0$ it is possible to solve the vacuum equations with $\sigma_2=\sigma_3=\sigma$ and arbitrary $\sigma_1$. 
Notice that these are the only solutions that satisfy the D-terms since after summing the D-term equations we have $0=\sigma_2-\sigma_3$. 
Moreover, we have to set $X_{3,1}^i=X_{1,2}^i=0$ in order to solve \eqref{SEqn}. 
The remaining D-term equation is  then
\begin{align}
|X_{2,3}^1|^2 + |X_{2,3}^2|^2  + |X_{2,3}^3|^2 =\frac{1}{2 \pi} \sigma \ .
\end{align}
The VEVs of these bifundamental fields simply describe $\IC^3$.
Since the F-term vanishes and there are no gauge transformation left by which we need to quotient, we see that this branch in the VMS is $\IC^* \times \IC^3$. 
Here $\IC^*=\IR\times S^1$, where $\IR$ corresponds to the the VEV of $\sigma_1$, which is unconstrained, while the $S^1$ is parametrized by the VEV of the periodic scalar dual to the gauge field $\mathcal{A}_1$.
In conclusion, then, we have two branches in the VMS, $C(Y^{1,2}(\mathbb{CP}^2))$ and $\IC^* \times \IC^3$, which intersect where $\sigma_1=\sigma$.

The last issue we wish to discuss is the mapping between baryonic operators in the field theory and M5-branes wrapped on toric divisors in the geometry. 
In the minimal GLSM presentation of the Calabi-Yau four-fold singularity, one can realise the four-fold as a K\"ahler quotient of $\IC^D$ by $U(1)^{D-4}$, where $D$ is the number of external vertices. Let $z_{a=1,\ldots,D}$, be complex coordinates on $\IC^D$. 
Then setting $z_a=0$ gives a complex codimension one submanifold of the Calabi-Yau four-fold, invariant under the torus $U(1)^D$ action on $\IC^D$, and which is a cone. We have $D$ such submanifolds, one for each external point in the toric diagram, and these are called the \textit{toric divisors}.

Now, on each submanifold $z_a=0$ one can wrap an M5-brane. 
For the $\IR^{1,2} \times X_4$ solution this is Euclidean (an instanton in fact). As is well-known \cite{Mikhailov:2000ya}, one can ``Wick rotate'' such a configuration to an M5-brane wrapping the five-manifold in $Y_7$ cut out by $z_a=0$, which in the near-horizon AdS$_4 \times Y_7$ solution leads to a BPS particle in AdS$_4$. 
AdS/CFT then implies there should be a chiral primary operator dual to this BPS particle. 
In the field theory realization of the Calabi-Yau four-fold these M5-branes wrapping toric five-cycles correspond to the external $p$-fields. 
Recall that via the P-matrix the GLSM $p$-fields are expressed in terms of the original spacetime fields $X_{i,j}$ and that this gives an explicit mapping between M5-branes wrapping divisors and baryonic operators.

In the case at hand, it is immediate from the P-matrix \eqref{pmatrix} that it is \emph{not} possible to find a baryonic operator which corresponds to single irreducible toric divisor: as one can see, the spacetime fields correspond to \emph{pairs} of divisors. For example, there is no obvious candidate operator in the field theory that is dual to an M5-brane wrapped on the toric divisor $p_2=0$, while a baryonic operator constructed from $X^{1}_{2,3}$ is a candidate dual to the reducible divisor $\{p_2=0\}\cup \{p_4=0\}$. 
Therefore it remains to explain whether there are BPS states in the field theory associated to M5-branes wrapped on single divisors. 
This seems to be a general problem for theories with parents, as discussed in \cite{Hanany:2008fj}. 
Notice that, on the contrary, for $C(Y^{1,2}(\mathbb{CP}^2))_{I}$ the P-matrix is the identity matrix, and therefore there is a 1-1 correspondence between toric divisors and spacetime fields. 

As a related comment, notice that it is {\it a priori} possible for a torsion $G$-flux to prevent one from wrapping an M5-brane on a cycle, due to the global anomaly discussed by Witten in \cite{Witten:1999vg}. Following the notations of \cite{Witten:1999vg} (the relevant equation is (5.10)), we see that the restriction of the $G$-flux to the M5-brane worldvolume $V$, the left hand side of (5.10), should be equal to a certain torsion class in $H^4_{\mathrm{tor}}(V,\IZ)$, defined on the right hand side of (5.10). In our case $V = \IR \times \Sigma$, where $\Sigma$ is a compact five-manifold. Then $H^4(V,\IZ) \cong H^4(\Sigma,\IZ) \cong H_1(\Sigma,\IZ)$ (using Poincar\'e duality), which is always a finite group for toric geometries, and indeed typically non-trivial. So a (torsion) $G$-flux in spacetime maps to an element of this finite group. Witten's global anomaly formula says this has to be the particular element defined in (5.10). It would be interesting to try to compute this in explicit examples.

We have seen that the attempt to interpret $C(Y^{1,2}(\mathbb{CP}^2))_{II}$ as a large $N$ theory on M2-branes raises some puzzles. A more careful examination of the string theory origin of this theory \cite{Aganagic:2009zk} might suggest a way out. In \cite{Aganagic:2009zk} $C(Y^{1,2}(\mathbb{CP}^2))_{II}$ was constructed from a Type IIA parent theory, namely the theory on $N$ D2-branes at the $\IC^3/\IZ_3$ singularity, by turning on RR two-form and four-form fluxes. While the two-form flux is lifted to pure geometry in M-theory, it is not clear how the four-form flux should be lifted. Turning on the RR two-form flux also requires one to fibre the Calabi-Yau three-fold ($\IC^3/\IZ_3$ and its resolution to $\mathcal{O}(-3)\rightarrow\mathbb{CP}^2$) over the real line $\IR$, and in passing from negative to positive $\IR$ one passes through the singular geometry $\IC^3/\IZ_3$. The fluxes change when crossing this singular point. 
This Type IIA construction suggests a non-trivial (non-flat) $G$-flux on the Calabi-Yau four-fold, and indeed a primitive (2,2) flux would preserve supersymmetry but break the conformal invariance\footnote{We thank Mina Aganagic for pointing this out to us.}. There have been recent suggestions that turning on such $G$-flux on a Calabi-Yau four-fold cone is dual to turning on a supersymmetric mass term \cite{Martelli:2009ga,Lambert:2009qw}. This might be an indication that the QCS theory engineered in Type IIA, at large $N$, does not flow to a SCFT dual to AdS$_4 \times Y^{1,2}(\mathbb{CP}^2)$. If this speculation turns out to be correct,  further investigation will be needed in order to determine what is the IR limit of this theory, and whether it has a supergravity dual.

%%%%%%%%%%%%%%%%%%%%%%%%%%%%%%
\subsection{Multiplicities}\label{sec:multiplicity}
We conclude this section with some parting remarks about multiplicities in the toric diagram. As reviewed in Subsection \ref{sec:forward}, expressing the spacetime fields $X_{i,j}$ in terms of the GLSM fields $p$, which parametrize the toric diagram, leads to a huge redundancy in the latter. This was noticed when the forward algorithm was first introduced \cite{Feng:2000mi}. So far, in all brane worldvolume theories which have appeared in the literature, whether for D3-branes or M2-branes, the VMS has the property that the toric diagram always has no multiplicity for external points in the toric diagram; that is, we have multiplicity 1 for the $p$-fields describing the external toric points. Moreover, it is an empirical observation that the multiplicities of points which are not \emph{extremal}\footnote{Notice such points exist only for non-isolated singularities.} external, but rather live on the interiors of external edges, tend to be binomial coefficients.

We have already mentioned that the external multiplicities for $C(Y^{1,2}(\mathbb{CP}^2))_{I}$ are all equal to 1, while for the apparently problematic theory $C(Y^{1,2}(\mathbb{CP}^2))_{II}$ there is an external point with multiplicity 2. Going back to our list of theories, we see that for the non-chiral $\IC^2/\IZ_n\times \IC^2/\IZ_m$ theories all the external multiplicities in the toric diagrams are equal to 1. 
For Phases $I$ and $II$ of $\IC^4/\IZ_2^3$, not all the Higgsed theories can be further Higgsed to have external multiplicities 1. 
For the $\IC^4$ theories which we present in Appendix \ref{sec:plethora} the result is interesting: the two standard theories known in the literature, and which we introduced in Subsection \ref{s:C4}, namely $(\IC^4)_I$ (the ABJM theory with $k=1$) and $(\IC^4)_{II}$, both have external multiplicities 1. 
However, all the other models in Figures \ref{f:c4plethoraA} and \ref{f:c4plethoraB} have more than two nodes, and all have external multiplicities greater than 1. 
For the $\IC^2/\IZ_2 \times \IC^2$ theories, all those with number of nodes in the quiver exceeding three (except theories (f) and (i)) in Figure \ref{f:c2z2plethora} have external multiplicities exceeding 1. 
Theories (f) and (i) have internal multiplicities which are not binomial coefficients, and
which are greater than the internal multiplicities of the other theories with three nodes.

%
%%%%%%%%%%%%%%%%%%%%%%%%%%%%%%%%%%%%%%%%%%%%%%%%%%%%%%%%%%%%%%
%=====================================================
\section{Summary and prospects}
%=====================================================

In this paper we have studied Higgsing and un-Higgsing in $(2+1)$-dimensional ${\cal N}=2$ quiver-Chern-Simons theories, in the context of M2-branes probing toric Calabi-Yau four-fold singularities.
Whereas the $(3+1)$-dimensional analogue of D3-branes probing Calabi-Yau three-folds is well understood, the story for $(2+1)$-dimensional M2-brane theories is considerably more complicated. 

From the outset there are two complications that do not arise in the D3-brane case: (1) In order to take an (Abelian) orbifold projection of a QCS theory, it is necessary that the order of the group divides the CS levels of the parent theory, subsequently leading to an additional discrete quotient in the VMS of the projected theory. The upshot is that there is currently no general method for constructing a QCS theory whose VMS is a given orbifold $\IC^4/\Gamma$. This is currently an important outstanding problem. 
(2) In AdS$_4 \times Y_7$, as opposed to AdS$_5 \times Y_5$ backgrounds, it is possible to turn on different torsion $G$-flux, classified by $H^4_{\mathrm{tor}}(Y_7,\IZ)$, and each such background corresponds to a physically distinct SCFT. 

Inspired by systematic studies for the D3-brane case, we set out to investigate toric Abelian orbifold singularities. In the D3-brane case one can construct all such orbifolds via a Douglas-Moore projection of $\mathcal{N}=4$ SYM theory, and moreover the D3-brane theory for any toric Calabi-Yau three-fold singularity may be obtained via Higgsing such a theory -- mathematically, this is related to the fact that the master space of an Abelian orbifold theory contains all toric partial resolutions of the singularity \cite{CrawIshii}. As a warm-up we started with the simple geometry $\IC^2/\IZ_n \times \IC^2/\IZ_n$, whose field theory may be obtained via a $\IZ_n$ projection of the ABJM theory. By Higgsing this theory\footnote{Since $\IC^2/\IZ_n \times \IC^2/\IZ_n$ is not an isolated singularity it is not clear to us how to classify $G$-flux on this background.} one can indeed produce QCS theories for all toric sub-diagrams, hence demonstrating a similarly straightforward behaviour to the D3-brane case.

We then proceeded to study another example, namely the orbifold $\IC^4/\IZ_2^3$. This has 18 inequivalent toric sub-diagrams, corresponding to 18 toric partial resolutions, including, of course, the flat space $\IC^4$ ({\it cf}.~Figure~\ref{f:c4z2cube}). 
Since a direct Douglas-Moore projection of the ABJM theory does not produce this theory, we constructed instead two different phases using other methods. First, we devised a {\it general un-Higgsing algorithm for toric quiver-Chern-Simons theories}, in analogy to \cite{Feng:2002fv}.
This algorithm, readily computerizable, at each step obeys the tiling condition \eqref{euler}, which is expected to be a physical requirement. (Indeed, in all examples studied it seems that violating this condition results in a QCS theory with VMS having the wrong dimension.) 
Sequential un-Higgsing starting from the ABJM theory for $\IC^4$ then leads to the first phase, $(\IC^4/\IZ_2^3)_{I}$. During this process, several sets of dual theories were obtained and observed to be connected by simple a CS level transformation \eqref{rules}. This leads to a general duality rule between theories with single-flavour nodes. Second, we used the method of lifting from Type II parents, discussed intensively in the literature, to derive a second phase from a descendent of the so-called Pseudo del Pezzo 5 theory. Phase $I$ can be Higgsed to theories for all 18 toric partial resolutions, whereas Phase $II$ is \emph{missing} a theory for $\IC^4/\IZ_2$, which is toric diagram (19). Moreover, we noticed the interesting phenomenon that $(\IC^4/\IZ_2^3)_I$ can also be Higgsed to another two theories with the same VMS but with fewer gauge nodes. This raises the issue of the M2-brane interpretation of these dual phases, especially since the latter two phases derived from Phase I can no longer be Higgsed to all toric sub-diagrams. 

In addition to the theories just mentioned, we have found a plenitude of other examples of this phenomenon. Specifically, via Higgsing we have produced a plethora of new QCS theories whose VMSs are $\IC^4$ (17 in addition to the two standard ones) and $\IC^2 \times \IC^2 / \IZ_2$ (13 theories) -- 
these are presented in Appendix \ref{sec:plethora}.

The ``missing'' toric sub-diagram (19) in the Higgsing of $(\IC^4/\IZ_2^3)_{II}$ is related to the fact that the Higgsed theory $C(Y^{1,2}(\mathbb{CP}^2))_{II}$ for toric diagram (16) cannot itself be Higgsed to diagram (19). We therefore focused attention on the singularity $C(Y^{1,2}(\mathbb{CP}^2))$, which has two toric (partial) resolutions, and studied two Phases $I$ and $II$ in detail, with Phase $I$ obtained via Higgsing $(\IC^4/\IZ_2^3)_I$. These display significantly different properties. A key ingredient on our analysis was a careful consideration of the effect of \emph{torsion $G$-flux} in the dual AdS backgrounds, which we described for general four-fold singularities. In the case at hand,
the AdS solution AdS$_4\times Y^{1,2}(\mathbb{CP}^2)$ has two choices of torsion $G$-flux, labelled by $H^4(Y^{1,2}(\mathbb{CP}^2),\IZ)\cong \IZ_2$. 
We conjectured that $C(Y^{1,2}(\mathbb{CP}^2))_I$, with equal gauge group ranks, corresponds to the background with no $G$-flux: the theory Higgses to both (partial) resolutions with no $G$-flux, and this is consistent with the supergravity analysis. 
Furthermore, we have shown that it is possible to deform the ranks of the $C(Y^{1,2}(\mathbb{CP}^2))_I$ phase so that it corresponds to a background with $G$-flux, and that this is again consistent with the Higgsings expected from the supergravity side. 
This analysis can in principle be used for other theories, and should serve as a further constraint on M2-branes theories at Calabi-Yau four-fold singularities.
In general, we expect from the existence of a supergravity dual, and also from the examples studied, that {\it theories with no $G$-flux should be Higgsable to all toric sub-diagrams}.

The $C(Y^{1,2}(\mathbb{CP}^2))_{II}$ phase, derived via Higgsing $(\IC^4/\IZ_2^3)_{II}$, presents us with some puzzles, however.
We have shown that it is not possible to Higgs this theory to its sub-diagram $\IC^4/\IZ_2$, with any configuration of $G$-flux. This does not seem to agree with the expectations of the large $N$ dual supergravity computation. Another potential problem with this theory is the extra branch in the VMS; this does not appear in Phase $I$, and is difficult to explain from the dual supergravity solution. (Of course, it is possible that quantum effects could lift this branch, a speculation that deserves further study.) Furthermore, there is an apparent mismatch between toric divisors and baryonic operators, again not present in Phase $I$ (although we note that baryonic operators are still rather poorly understood for QCS theories). However, it is also possible that these puzzles will be resolved somehow with a better understanding of the correspondence, perhaps along the lines suggested at the end of section \ref{sec:extra}. These issues, together with the proposals above, should provide ample future directions in investigating M2-brane quiver-Chern-Simons theories.

%%%%%%%%%%%
\subsection*{Acknowledgments}
\noindent 
We thank E.~Lerman, D.~Martelli and  M.~Aganagic for discussions. N.~B.~would like to thank the ISEF for their support; this work was completed with the support of a University of Oxford Clarendon Fund Scholarship.
Y.-H.~H. gratefully acknowledges the STFC for an Advanced Fellowship and the FitzJames Fellowship of Merton College, Oxford. 
J.~F.~S.~is funded by a Royal Society University Research Fellowship.

%=====================================================
\appendix
%=====================================================
%=====================================================
\section{Orbifold projections \label{sec:orbifold}}
%=====================================================
In this appendix we demonstrate the orbifold projection procedure with two examples. 
We begin with a $\IZ_3 \times \IZ_2$ projection of theory $(\IC^4)_I$ (the ABJM theory with CS level $k=1$). We take the orbifold group action generated by
\bea\label{z3z2}
\IZ_{3} &:& 
(x^{1},x^{2},x^{3},x^{4})\longrightarrow (\me^{\frac{2\pi \ii}{3}}x^{1},\me^{\frac{-2\pi \ii}{3}}x^{2},x^{3},x^{4}) \ , \nn \\
\IZ_{2} &:& 
(x^{1},x^{2},x^{3},x^{4})\longrightarrow (-x^{1},x^{2},-x^{3},x^{4}) \ ,
\eea
where $(x^1,x^2,x^3,x^4)$ are the coordinates of $\IC^4$. We next promote the theory $(\IC^4)_I$ to have $N=k=6$, {\it i.e.} our starting point will be a $U(6)_{6}\times U(6)_{-6}$ quiver theory. We then consider the $\IZ_{3}\times \IZ_{2}$ orbifold projection of this theory corresponding to (\ref{z3z2}). 
More precisely, the fields of the ABJM theory described in Subsection \ref{s:C4} are required to obey the conditions:
\bea
\nn
&&X_{1,2}^{1}=-\me^{\frac{2\pi \ii}{3}}\Omega_{1}\Omega_{2}X_{1,2}^1\Omega_{2}^{\dagger}\Omega_{1}^{\dagger} \ , \quad
X_{1,2}^2=\me^{\frac{-2\pi \ii}{3}}\Omega_{1}\Omega_{2}X_{1,2}^2\Omega_{2}^{\dagger}\Omega_{1}^{\dagger} \ ,
\\ 
&&X_{2,1}^1=-\Omega_{1}\Omega_{2} X_{2,1}^1\Omega_{2}^{\dagger}\Omega_{1}^{\dagger} \ , \quad
X_{2,1}^2=\Omega_{1}\Omega_{2} X_{2,1}^2\Omega_{2}^{\dagger}\Omega_{1}^{\dagger} \ , \\
\nn
&&\mathcal{A}^{1}=\Omega_{1}\Omega_{2}\mathcal{A}^{1}\Omega_{2}^{\dagger}\Omega_{1}^{\dagger} \ , \quad 
\mathcal{A}^{2}=\Omega_{1}\Omega_{2}\mathcal{A}^{2}\Omega_{2}^{\dagger}\Omega_{1}^{\dagger} \ ,
\eea
where $\mathcal{A}^1$, $\mathcal{A}^2$ are the gauge fields for the two nodes, and we have defined the diagonal matrices
\begin{equation}
\Omega_{1}:={\rm diag}(1,-1,1,-1,1,-1) \ , \qquad
\Omega_{2}={\rm diag}(\me^{\frac{2\pi \ii}{3}},\me^{\frac{2\pi \ii}{3}},\me^{\frac{-2\pi \ii}{3}},\me^{\frac{-2\pi \ii}{3}},1,1)~.
\end{equation}

On solving, we find that the invariant gauge fields are of the form
\bea
\nn
&&\mathcal{A}^{1}=
{\tiny
\left(
\begin{array}{llllll}
 v_1 & 0 & 0 & 0 & 0 & 0 \\
 0 & v_{2} & 0 & 0 & 0 & 0 \\
 0 & 0 & v_{3} & 0 & 0 & 0 \\
 0 & 0 & 0 & v_{4} & 0 & 0 \\
 0 & 0 & 0 & 0 & v_{5} & 0 \\
 0 & 0 & 0 & 0 & 0 & v_{6} 
\end{array}
\right)  
}
\ , \quad
\mathcal{A}^{2}=
{\tiny
\left(
\begin{array}{llllll}
 v_{7} & 0 & 0 & 0 & 0 & 0 \\
 0 & v_{8} & 0 & 0 & 0 & 0 \\
 0 & 0 & v_{9} & 0 & 0 & 0 \\
 0 & 0 & 0 & v_{10} & 0 & 0 \\
 0 & 0 & 0 & 0 & v_{11} & 0 \\
 0 & 0 & 0 & 0 & 0 & v_{12}
\end{array}
\right)
} \ .
\eea
The diagonal nature of these matrices implies that the daughter gauge group is $U(1)^6 \times U(1)^6$.
The invariant bifundamental fields are of the form:
\bea
\nn
&&X_{1,2}^1=
{\tiny \left(
\begin{array}{llllll}
 0 & 0 & 0 & X_{1,10} & 0 & 0 \\
 0 & 0 & X_{2,9} & 0 & 0 & 0 \\
 0 & 0 & 0 & 0 & 0 & X_{3,12} \\
 0 & 0 & 0 & 0 & X_{4,11} & 0 \\
 0 & X_{5,8} & 0 & 0 & 0 & 0 \\
 X_{6,7} & 0 & 0 & 0 & 0 & 0
\end{array}
\right)}
\ , \quad
X_{1,2}^2=
{\tiny
\left(
\begin{array}{llllll}
 0 & 0 & 0 & 0 & X_{1,11} & 0 \\
 0 & 0 & 0 & 0 & 0 & X_{2,12} \\
 X_{3,7} & 0 & 0 & 0 & 0 & 0 \\
 0 & X_{4,8} & 0 & 0 & 0 & 0 \\
 0 & 0 & X_{5,9} & 0 & 0 & 0 \\
 0 & 0 & 0 & X_{6,10} & 0 & 0
\end{array}
\right) \ ,
}
\\
\nn
&&X_{2,1}^1=
{\tiny
\left(
\begin{array}{llllll}
 0 & X_{7,2} & 0 & 0 & 0 & 0 \\
 X_{8,1} & 0 & 0 & 0 & 0 & 0 \\
 0 & 0 & 0 & X_{9,4} & 0 & 0 \\
 0 & 0 & X_{10,3} & 0 & 0 & 0 \\
 0 & 0 & 0 & 0 & 0 & X_{11,6} \\
 0 & 0 & 0 & 0 & X_{12,5} & 0
\end{array}
\right)
}
\ , \quad
X_{2,1}^2=
{\tiny
\left(
\begin{array}{llllll}
 X_{7,1} & 0 & 0 & 0 & 0 & 0 \\
 0 & X_{8,2} & 0 & 0 & 0 & 0 \\
 0 & 0 & X_{9,3} & 0 & 0 & 0 \\
 0 & 0 & 0 & X_{10,4} & 0 & 0 \\
 0 & 0 & 0 & 0 & X_{11,5} & 0 \\
 0 & 0 & 0 & 0 & 0 & X_{12,6}
\end{array}
\right)
} \ .
\eea
The $C$ matrix encoding the Chern-Simons levels is then
\begin{equation}
C=
{\scriptsize
\left(
\begin{array}{llllllllllll}
 1 & 1 & 1 & 1 & 1 & 1 & 1 & 1 & 1 & 1 & 1 & 1 \\
 1 & 1 & 1 & 1 & 1 & 1 & $-1$ & $-1$ & $-1$ & $-1$ & $-1$ & $-1$
\end{array}
\right)} \ ,
\end{equation}
and the resulting superpotential is
\bea
\nn
W &=&-X_{2, 9} X_{9, 3} X_{3, 7} X_{7, 2} + 
 X_{2, 9} X_{9, 4} X_{4, 8} X_{8, 2}  + 
 X_{1, 10} X_{10, 3} X_{3, 7} X_{7, 1}  \\ \nn
&&- X_{1, 10} X_{10, 4} X_{4, 8} X_{8, 1} + 
 X_{1, 11} X_{11, 5} X_{5, 8} X_{8, 1} - 
 X_{4, 11} X_{11, 5} X_{5, 9} X_{9, 4} \\ \nn 
&&- X_{1, 11} X_{11, 6} X_{6, 7} X_{7, 1} + 
 X_{4, 11} X_{11, 6} X_{6, 10} X_{10, 4} - 
 X_{2, 12} X_{12, 5} X_{5, 8} X_{8, 2} \\ 
&&+ X_{3, 12} X_{12, 5} X_{5, 9} X_{9, 3} + 
 X_{2, 12} X_{12, 6} X_{6, 7} X_{7, 2} - 
 X_{3, 12} X_{12, 6} X_{6, 10} X_{10, 3}  \ .
\label{WorbC4-I}
\eea
Using the forward algorithm we calculate 
\bea
\nn
\tiny
G_t=\left(
\begin{array}{llllllllllllllllllllllllllllllllllllllll}
 2 & 3 & 1 & 2 & 3 & 4 & 2 & 3 & 1 & 2 & 1 & 2 & 1 & 2 & 1 & 2 & 0 & 1
   & 2 & 1 & 1 & 2 & 1 & 0 & 2 & 1 & 0 & 2 & 1 & 1 & 1 & 0 & 2 & 1 & 1
   & 1 & 0 & 0 & 0 & 1 \\
 3 & 2 & 4 & 3 & 2 & 1 & 3 & 2 & 2 & 1 & 2 & 1 & 2 & 1 & 2 & 1 & 1 & 0
   & 1 & 2 & 0 & 1 & 2 & 1 & 1 & 2 & 1 & 1 & 2 & 0 & 0 & 1 & 0 & 1 & 1
   & 1 & 0 & 0 & 2 & 1 \\
 0 & 1 & $-1$ & 0 & 1 & 2 & 0 & 1 & 0 & 1 & 0 & 1 & 0 & 1 & 0 & 1 & 0 &
   1 & 1 & 0 & 1 & 1 & 0 & 0 & 1 & 0 & 0 & 1 & 0 & 1 & 1 & 0 & 1 & 0 &
   0 & 0 & 0 & 0 & 0 & 1 \\
 1 & 1 & 1 & 1 & 1 & 1 & 1 & 1 & 1 & 1 & 1 & 1 & 1 & 1 & 1 & 1 & 1 & 1
   & 1 & 1 & 1 & 1 & 1 & 1 & 1 & 1 & 1 & 1 & 1 & 1 & 1 & 1 & 1 & 1 & 1
   & 1 & 1 & 1 & 1 & 1  
\end{array}\right.~
\\ \tiny \left.~
\begin{array}{llllllllllllllllllllllllllllllllllllllll}
  1 & 1 & 0 & 0 & 1 & 2 & 0 & 1 & 1 & 1 & 0 & 1 & 0 & 0 & 1 & 2 & 1 & 0 & $-1$ & 0 & 1 & 1 & 0 & 1 & 2 & 0 & 0 & 1 & 2 & 1 & 0 & 0 & 1 & 2 & 0 & 1 & 0 & 0 & 0 & 1 \\
  1 & 1 & 0 & 0 & 2 & 1 & 1 & 0 & 1 & 1 & 0 & 1 & 2 & 0 & 2 & 1 & 0 & 1 & 0 & 0 & 1 & 1 & 0 & 1 & 0 & 0 & $-1$ & 2 & 1 & 0 & 1 & 0 & 2 & 1 & 1 & 0 & 0 & 0 & 1 & 0 \\
  0 & 1 & 1 & 1 & 0 & 1 & 0 & 1 & 1 & 1 & 1 & 1 & 0 & 1 & 0 & 1 & 1 & 0 & 1 & 1 & 0 & 0 & 0 & 0 & 1 & 0 & 0 & 0 & 1 & 1 & 0 & 0 & 0 & 1 & 0 & 1 & 0 & 1 & 0 & 1 \\
  1& 1 & 1 & 1 & 1 & 1 & 1 & 1 & 1 & 1 & 1 & 1 & 1 & 1 & 1 & 1 & 1 & 1 & 1 & 1 & 1 & 1 & 1 & 1 & 1 & 1 & 1 & 1 & 1 & 1 & 1 & 1 & 1 & 1 & 1 & 1 & 1 & 1 & 1 & 1
\end{array} \right)~.
\eea
Using the Delzant construction we find that the VMS of this theory is $\IC^4/\IZ_6 \times \IZ_3 \times \IZ_2$, where the generators of the three groups acts on the coordinates of $\IC^4$ as follows: 
\bea
\IZ_6: (\me^{\frac{2\pi \ii}{6}}, \me^{\frac{2\pi \ii}{6}}, \me^{\frac{-2\pi \ii}{6}}, \me^{\frac{-2\pi \ii}{6}})~, \quad \IZ_3 : (\me^{\frac{2\pi \ii}{3}}, \me^{\frac{-2\pi \ii}{3}}, 1,1)~, \quad \IZ_2:(-1,1,-1,1)~.
\eea
Compare this with (\ref{z3z2}): one sees that, in addition to the original $\IZ_3\times \IZ_2$ action, there is also a quotient by $\IZ_6$ corresponding to the ABJM quotient with CS level $k=6$.

As another example, we begin with theory $(\IC^4)_{II}$ and consider instead a $\IZ_{2}\times \IZ_{2}$ projection. We first promote the theory to $N=4$ M2-branes, {\it i.e.} to a $U(4)_{4}\times U(4)_{-4}$ quiver theory. The $\IZ_{2}\times \IZ_{2}$ orbifold projection on the matter fields is then
\bea
\nn &
\phi_1^1=-\Omega_{1}\phi_1^1\Omega_{1}^{\dagger}=-\Omega_{2}\phi_1^1\Omega_{2}^{\dagger}  \ , \quad
\phi_1^2=-\Omega_{1}\phi_1^2\Omega_{1}^{\dagger}=\Omega_{2}\phi_1^2\Omega_{2}^{\dagger} \ ,
\\
&
X_{2,1}=\Omega_{1}X_{2,1}\Omega_{1}^{\dagger}=-\Omega_{2}X_{2,1}\Omega_{2}^{\dagger}  \ , \quad
X_{1,2}=\Omega_{1}X_{1,2}\Omega_{1}^{\dagger}=\Omega_{2}X_{1,2}\Omega_{2}^{\dagger}  \ ,
\\
\nn &
\mathcal{A}^{1}=\Omega_{1}\mathcal{A}^{1}\Omega_{1}^{\dagger}=\Omega_{2}\mathcal{A}^{1}\Omega_{2}^{\dagger}  \ , \quad
\mathcal{A}^{2}=\Omega_{1}\mathcal{A}^{2}\Omega_{1}^{\dagger}=\Omega_{2}\mathcal{A}^{2}\Omega_{2}^{\dagger} \ ,
\eea
where again $\mathcal{A}^1$, $\mathcal{A}^2$ are the gauge fields for the two nodes and
\begin{equation}
\Omega_{1}={\rm diag}(1,1,-1,-1) \ , \qquad
\Omega_{2}={\rm diag}(1,-1,1,-1) \ .
\end{equation}

The invariant gauge fields are then
\begin{equation}
\mathcal{A}^{1} = {\rm diag}(v_1,v_3,v_5,v_7) \ , \quad
\mathcal{A}^{2} = {\rm diag}(v_2,v_4,v_6,v_8) \ ,
\end{equation}
signifying that the resulting gauge group is $U(1)^4 \times U(1)^4$. The invariant bifundamental fields
and adjoints are subsequently:
\bea
&&
\phi_1^1=
{\scriptsize
\left(
\begin{array}{llll}
 0 & 0 & 0 & X_{1,7} \\
 0 & 0 & X_{3,5} & 0 \\
 0 & X_{5,3} & 0 & 0 \\
 X_{7,1} & 0 & 0 & 0 
\end{array}
\right)
}
\ ,  \quad
\phi_1^2=
{\scriptsize
\left(
\begin{array}{llll}
 0 & 0 & X_{1,5} & 0 \\
 0 & 0 & 0 & X_{3,7} \\
 X_{5,1} & 0 & 0 & 0 \\
 0 & X_{7,3} & 0 & 0 
\end{array}
\right) \ ,
}
\\
&&
X_{2,1}=
{\scriptsize
\left(
\begin{array}{llll}
 0 & X_{2,3} & 0 & 0 \\
 X_{4,1} & 0 & 0 & 0 \\
 0 & 0 & 0 & X_{6,7} \\
 0 & 0 & X_{8,5} & 0 
\end{array}
\right)
}
\ , \quad
X_{1,2}=
{\scriptsize
\left(
\begin{array}{llll}
 X_{1,2} & 0 & 0 & 0 \\
 0 & X_{3,4} & 0 & 0 \\
 0 & 0 & X_{5,6} & 0 \\
 0 & 0 & 0 & X_{7,8}
\end{array}
\right)
} \ .
\eea
The resulting $C$ matrix for the Chern-Simons levels is 
\begin{equation}
C=
\left(
\begin{array}{llllllll}
 1 & 1 & 1 & 1 & 1 & 1 & 1 & 1 \\
 1 & $-1$ & 1 & $-1$ & 1 & $-1$ & 1 & $-1$
\end{array}
\right) \ ,
\end{equation}
while the superpotential is
\bea
\nn
W&=&X_{1, 2} X_{2, 3} X_{3, 5} X_{5, 1} - 
 X_{1, 5} X_{5, 3} X_{3, 4} X_{4, 1} - 
 X_{1, 2} X_{2, 3} X_{3, 7} X_{7, 1} \\ \nn
&&+ X_{1, 5} X_{5, 6} X_{6, 7} X_{7, 1} + 
 X_{1, 7} X_{7, 3} X_{3, 4} X_{4, 1} - 
 X_{3, 5} X_{5, 6} X_{6, 7} X_{7, 3}  \\
&&- X_{1, 7}  X_{7, 8} X_{8, 5} X_{5, 1} + 
 X_{3, 7} X_{7, 8} X_{8, 5} X_{5, 3} ~.
\label{WorbC4-II}
\eea
Using the forward algorithm we calculate 
\bea
{\tiny
G_t=
\left(
\begin{array}{llllllllllllllllllllllllllllllllllll}
 1 & 1 & 1 & 1 & 1 & 1 & 1 & 1 & 1 & 1 & 1 & 1 & 1 & 1 & 1 & 1
   & 1 & 1 & 1 & 1 & 1 & 1 & 1 & 1 & 1 & 1 & 1 & 1 & 1 &
   1 & 1 & 1 & 1 & 1 & 1 & 1 \\
 0 & 2 & 1 & 2 & 1 & 1 & 0 & 1 & 0 & 0 & 0 & 0 & 4 & 3 & 3 & 2 & 3 & 2
   & 2 & 1 & 2 & 1 & 2 & 1 & 3 & 2 & 2 & 1 & 2 & 1 & 1 & 0 & 1 & 0 & 1
   & 0 \\
 1 & 0 & 0 & 1 & 1 & 0 & 0 & 1 & 1 & 2 & 0 & 1 & 0 & 0 & 0 & 0 & 0 & 0
   & 0 & 0 & 1 & 1 & 0 & 0 & 0 & 0 & 0 & 0 & 0 & 0 & 0 & 0 & 1 & 1 & 0
   & 0 \\
 $-1$ & 0 & 0 & 0 & 0 & 0 & 0 & 0 & 0 & $-1$ & $-1$ & $-1$ & 1 & 1 & 1 & 1 & 1
   & 1 & 1 & 1 & 0 & 0 & 0 & 0 & 1 & 1 & 1 & 1 & 1 & 1 & 1 & 1 & 0 & 0
   & 0 & 0
\end{array}
\right)} ~.
\eea
Using the Delzant construction we find that the VMS of this theory is $\IC^4/\IZ_4 \times \IZ_2 \times \IZ_2$, where the generators of the groups acts on the coordinates of $\IC^4$ as follows:
\bea
\IZ_4:(1,\me^{\frac{-2\pi \ii}{4}},\me^{\frac{2\pi \ii}{4}},1)~, \quad \IZ_2:(-1,1,1,-1)~, \quad \IZ_2 : (1,-1,1,-1)~.
\eea
Notice the additional $\IZ_4$ quotient along the CS level direction of the original theory.

%=====================================================
\section{A plethora of models for $\IC^4$ and $\IC^2/\IZ_2 \times \IC^2$}\label{sec:plethora}
\begin{figure}[H]
\includegraphics[trim=-30mm 13mm 0mm 12mm, clip, width=4.3in]{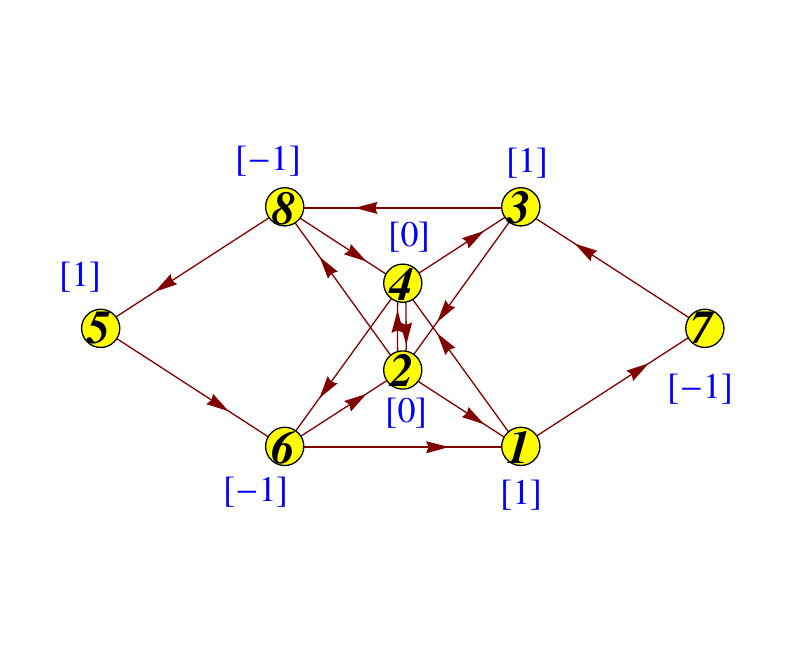}
\caption{{\sf Quiver diagram for the third phase of $\IC^4/\IZ_2^3$. The Chern-Simons levels are labelled in (blue) square brackets.}}
\label{f:oldphase}
\end{figure}

%=====================================================
In addition to the phases of $\IC^4/\IZ_2^3$ studied in the main text, we have also studied the Higgsing behaviour of yet a third phase of $\IC^4/\IZ_2^3$, which we present here. 
The quiver is shown in Figure~\ref{f:oldphase} and the superpotential is
\bea
\nn
W&=&-X_{1, 4} X_{4, 2} X_{2, 1} - X_{2, 4} X_{4, 3} X_{3, 2} + X_{2, 4} X_{4, 6} X_{6, 2}  \\ \nn
 &&+ X_{2, 8} X_{8, 4} X_{4, 2} + X_{1, 7} X_{7, 3} X_{3, 2} X_{2, 1} - X_{2, 8} X_{8, 5} X_{5, 6} X_{6, 2}  \\
 &&- X_{1, 7} X_{7, 3} X_{3, 8} X_{8, 4} X_{4, 6} X_{6, 1} + X_{1, 4} X_{4, 3} X_{3, 8} X_{8, 5} X_{5, 6} X_{6, 1} \ .
\eea

By Higgsing this theory it is \emph{not} possible to obtain QCS theories which correspond to toric diagrams (10), (16) and (19) in Figure~\ref{f:c4z2cube}. 
Moreover, we have noticed that there are typically many theories produced which share the same VMS. 
Indeed, for the two simplest and perhaps most-studied four-folds in the context of QCS theories, namely $\IC^4$ and $\IC^4/\IZ_2 \times \IC^2$, we have found a plethora of toric phases; many of these are new to the literature. 
For the purposes of completeness and of illustration, we list all possible Higgsings of the third phase of $\IC^4/\IZ_2^3$ which lead to these two particular moduli spaces. 

\begin{figure}[h!b!t!]
\includegraphics[trim=0mm 0mm 0mm 0mm, clip, width=5.5in]{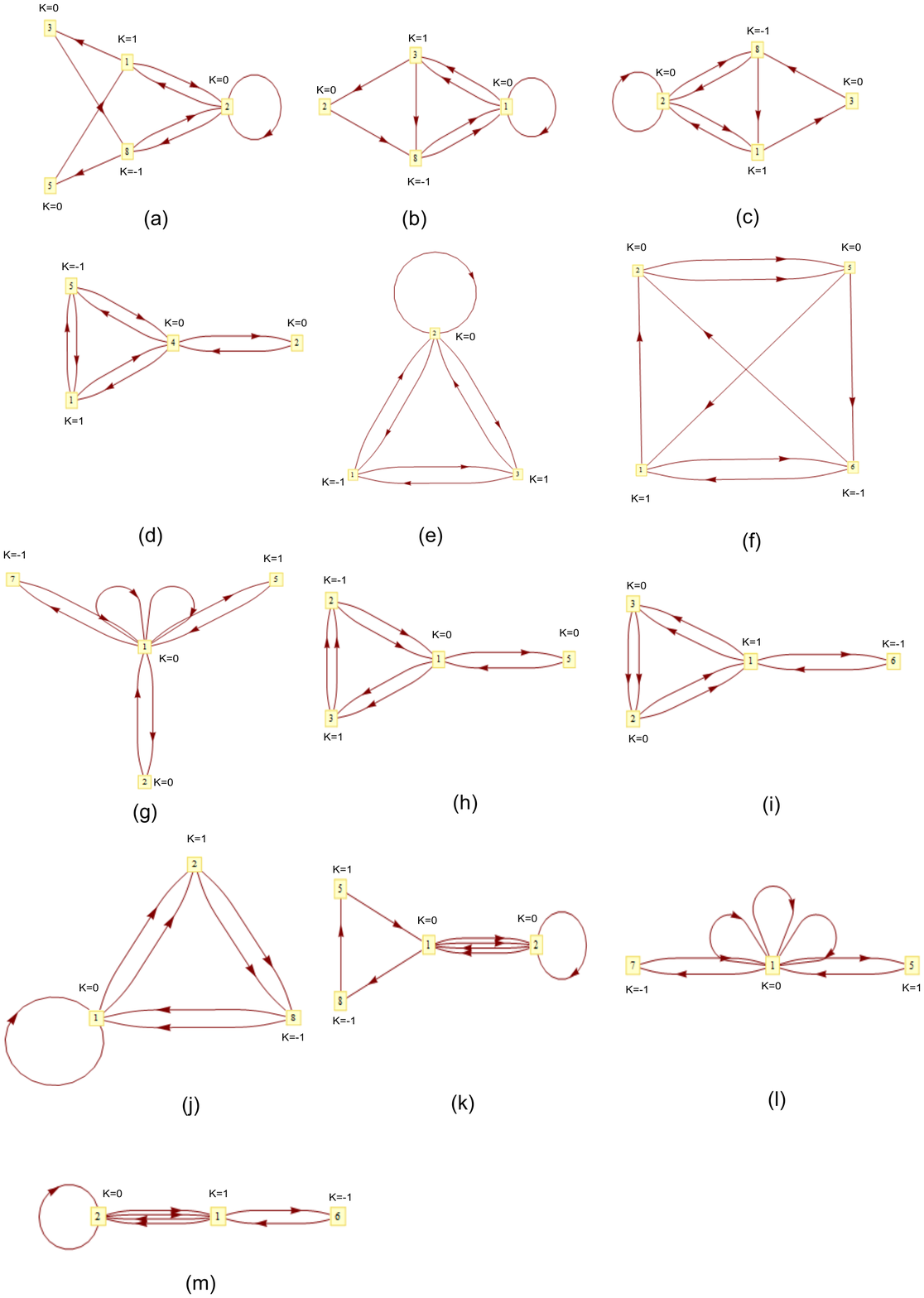}
\caption{{\sf Quiver diagrams for various phases of $\IC^2/\IZ_2 \times \IC^2$.}}
\label{f:c2z2plethora}
\end{figure}

\begin{figure}[H]
\includegraphics[trim=0mm 20mm 0mm 13mm, clip, width=6in]{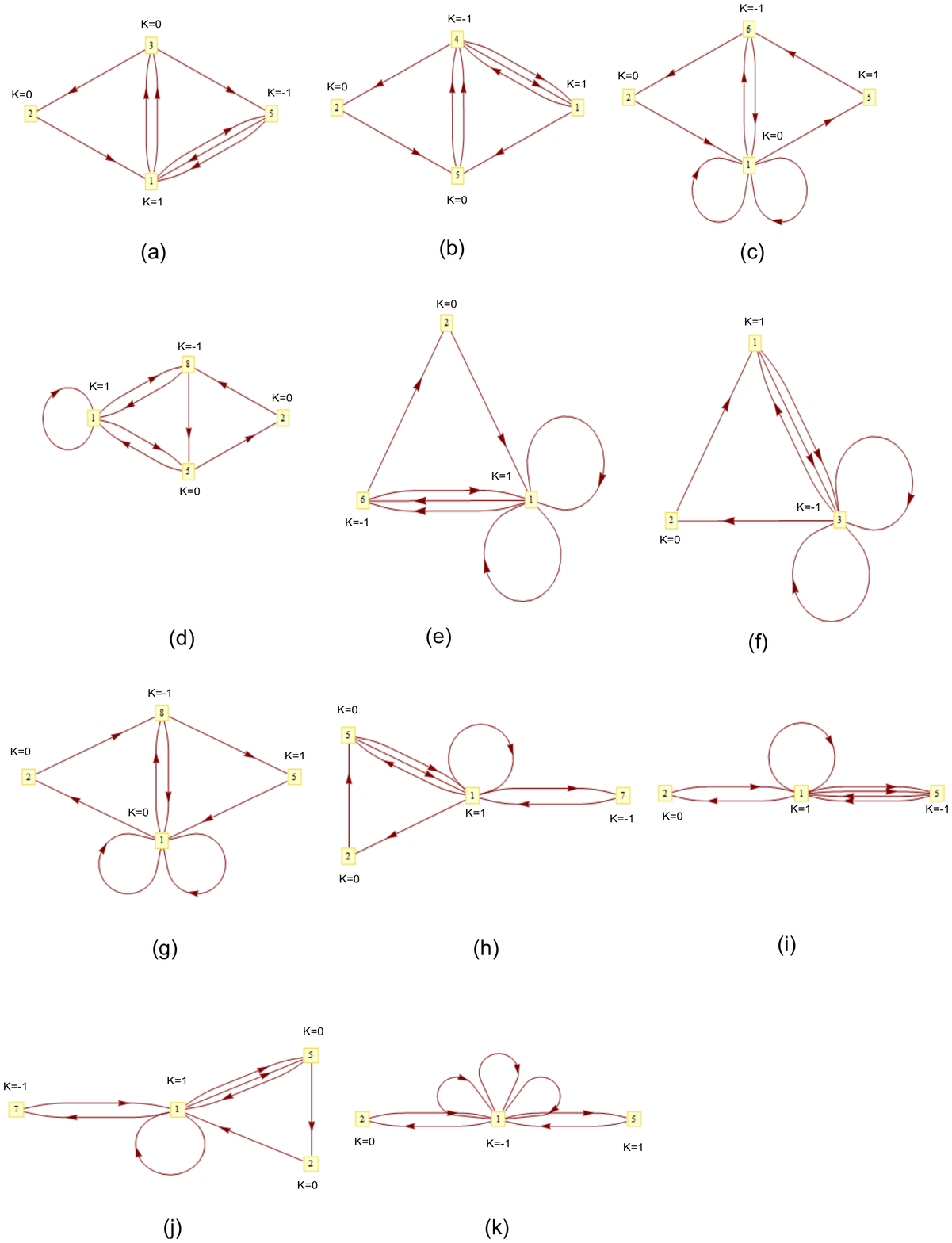}
\caption{{\sf Quiver diagrams for phases of $\IC^4$ with non-vanishing superpotential.}}
\label{f:c4plethoraA}
\end{figure}

\begin{figure}[H]
\includegraphics[trim=0mm 100mm 0mm 0mm, clip, width=5in]{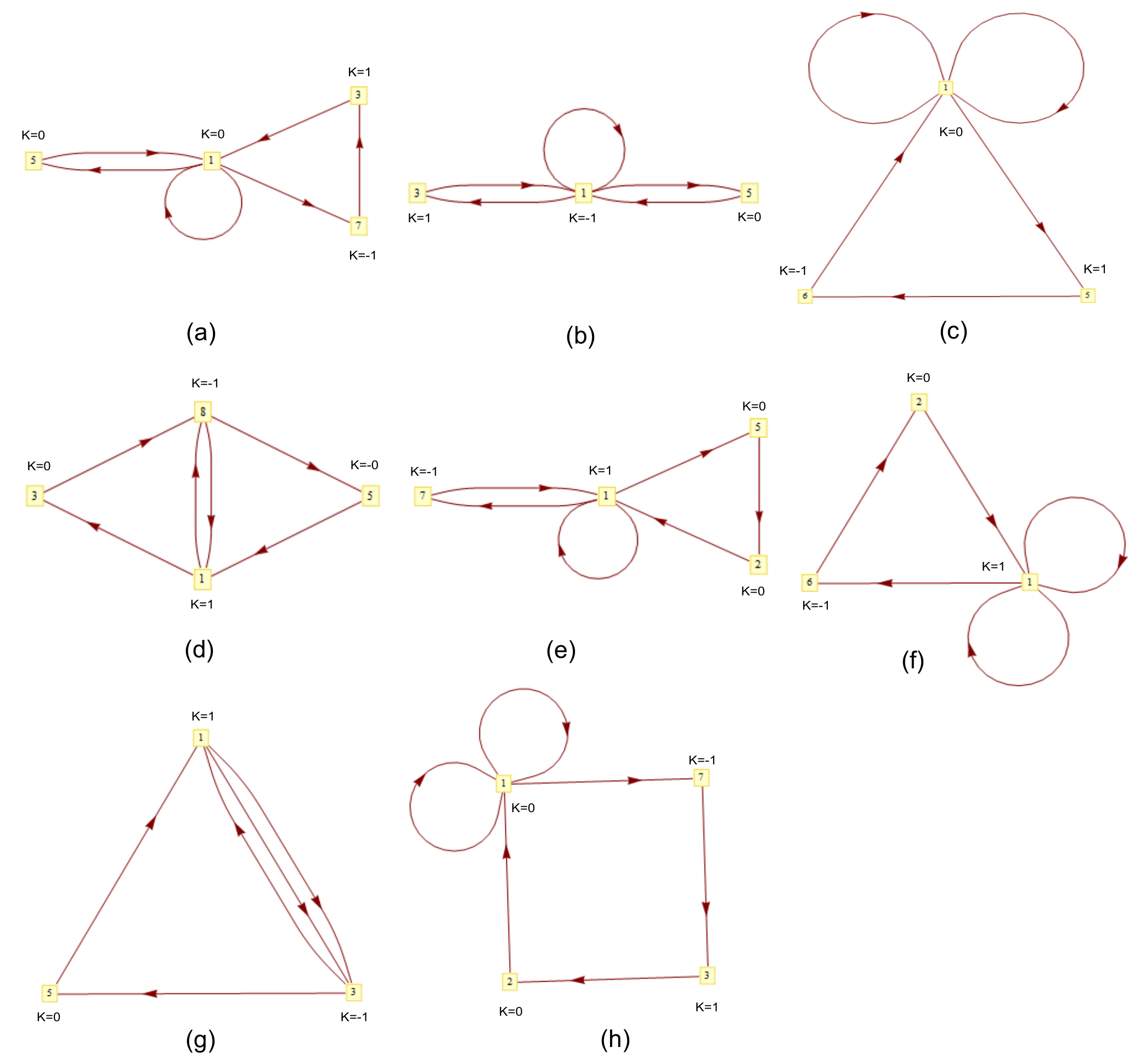}
\caption{{\sf Quiver diagrams for phases of $\IC^4$ with vanishing (Abelian) superpotential.}}
\label{f:c4plethoraB}
\end{figure}

For $\IC^4$ we find a total of 17 QCS theories, in addition to $(\IC^4)_I$ and $(\IC^4)_{II}$. The quivers for these are presented in Figures~\ref{f:c4plethoraA} and \ref{f:c4plethoraB}, while the toric matrices $G_t$ and superpotentials are given in Tables~\ref{t:c4plethoraA} and \ref{t:c4plethoraB}.
For the orbifold $\IC^2/\IZ_2 \times \IC^2$ we find a total of 13 QCS theories. The quivers for these are presented in Figure~\ref{f:c2z2plethora}, while the toric matrices $G_t$ and superpotentials are given in Table~\ref{t:c2z2plethora}.

\begin{table}[H]\begin{center}
\begin{tabular}{|c|c||c|c|} \hline
& Toric diagram $G_t$ & & Toric diagram $G_t$\\ \hline \hline
(a) &
{\tiny
$\left(
\begin{array}{llllll}
 0 & 0 & 0 & 0 & 0 & 1 \\
 0 & 1 & 0 & 0 & 1 & 0 \\
 0 & 0 & 1 & 1 & 0 & 0 \\
 1 & 0 & 0 & 0 & 0 & 0
\end{array}
\right)
$}
&
(e) &
{\tiny
$\left(
\begin{array}{llllll}
 0 & 0 & 0 & 0 & 0 & 1 \\
 0 & 1 & 0 & 1 & 1 & 0 \\
 0 & 0 & 1 & 0 & 0 & 0 \\
 1 & 0 & 0 & 0 & 0 & 0
\end{array}
\right)
$}
\\ \hline
(b) &
{\tiny
$\left(
\begin{array}{lllll}
 0 & 0 & 1 & 0 & 1 \\
 0 & 0 & 0 & 1 & 0 \\
 0 & 1 & 0 & 0 & 0 \\
 1 & 0 & 0 & 0 & 0
\end{array}
\right)
$} &
(f) &
{\tiny
$\left(
\begin{array}{lllll}
 0 & 0 & 0 & 1 & 1 \\
 0 & 0 & 1 & 0 & 0 \\
 0 & 1 & 0 & 0 & 0 \\
 1 & 0 & 0 & 0 & 0
\end{array}
\right)
$}
\\ \hline
(c) &
{\tiny
$\left(
\begin{array}{lllll}
 0 & 0 & 1 & 0 & 1 \\
 0 & 0 & 0 & 1 & 0 \\
 0 & 1 & 0 & 0 & 0 \\
 1 & 0 & 0 & 0 & 0
\end{array}
\right)
$} &
(g) &
{\tiny
$\left(
\begin{array}{lllll}
 0 & 0 & 0 & 1 & 1 \\
 0 & 0 & 1 & 0 & 0 \\
 0 & 1 & 0 & 0 & 0 \\
 1 & 0 & 0 & 0 & 0
\end{array}
\right)
$}
\\ \hline
(d) &
{\tiny
$\left(
\begin{array}{llllll}
 0 & 0 & 0 & 1 & 0 & 1 \\
 0 & 0 & 0 & 0 & 1 & 0 \\
 1 & 0 & 1 & 0 & 0 & 0 \\
 0 & 1 & 0 & 0 & 0 & 0
\end{array}
\right)
$}&
(h) &
{\tiny
$\left(
\begin{array}{llllll}
 0 & 0 & 0 & 0 & 0 & 1 \\
 0 & 0 & 1 & 1 & 1 & 0 \\
 0 & 1 & 0 & 0 & 0 & 0 \\
 1 & 0 & 0 & 0 & 0 & 0
\end{array}
\right)
$}
\\ \hline
\end{tabular}
\caption{{\sf The toric diagrams for phases of $\IC^4$ with vanishing (Abelian) superpotential.}}
\label{t:c4plethoraB}
\end{center}
\end{table}
%}

We have therefore found a host of new theories.
The authors of \cite{Aharony:2008ug} conjectured the ABJM theory to be the worldvolume theory on an M2-brane in flat spacetime, and one may similarly wonder whether any of our $\IC^4$ theories are really M2-brane worldvolume theories. 
A necessary condition would be that the manifest $\mathcal{N}=2$ supersymmetry of the theory is in fact enhanced to $\mathcal{N}=8$. 
In fact even for the ABJM theory $(\IC^4)_I$ this amount of symmetry is not manifest either; it is believed that the additional supersymmetries are described by certain monopole operators \cite{Aharony:2008ug}, which are currently rather poorly understood. It seems difficult, therefore, to address this question directly and deserves further study. As another hint, recall that in the D3-brane case the number of nodes in the quiver is precisely the Euler number of a (any) Calabi-Yau resolution of the singularity. In \cite{Imamura:2009ph} it was argued that for non-chiral QCS theories on M2-branes probing Calabi-Yau four-fold singularities, the number of nodes is instead $2+\mathrm{rank}\, H_2(Y_7)$.
If correct, this implies that the number of nodes for $Y_7=S^7$ should be 2. 
Clearly, this is true for models $(\IC^4)_I$ and $(\IC^4)_{II}$, but the two non-chiral theories (i) in Figures~\ref{f:c4plethoraA} and (b) in \ref{f:c4plethoraB} fail to satisfy this condition. This suggests that there might be problems in interpreting these as M2-branes theories.

\TABLE[h!b!p!]{
\caption{{\sf The toric diagrams and superpotentials for phases of $\IC^4$ with non-vanishing superpotential.}}
\begin{tabular}{|c|c|c|} \hline
& Toric diagram $G_t$ & Superpotential $W$ \\ \hline \hline
(a) &
{\tiny
$\left(
\begin{array}{lllllll}
 0 & 0 & 0 & 0 & 0 & 0 & 1 \\ [-1.3mm]
 0 & 0 & 0 & 0 & 1 & 1 & 0 \\ [-1.3mm]
 1 & 0 & 1 & 1 & 0 & 0 & 0 \\ [-1.3mm]
 0 & 1 & 0 & 0 & 0 & 0 & 0
\end{array}
\right)
$}
&
{\scriptsize
$\begin{array}{l}
-X_{1, 3}^{2} X_{3, 2} X_{2, 1} + 
 X_{1, 3}^{2} X_{3, 5} X_{5, 1}^{1} + \\
 X_{1, 3}^{1} X_{3, 2} X_{2, 1} X_{1, 5} X_{5, 1}^{2} - 
 X_{1, 3}^{1} X_{3, 5} X_{5, 1}^{1} X_{1, 5} X_{5, 1}^{2}
\end{array}$
}
\\ [-1mm] \hline
(b) &
{\tiny
$\left(
\begin{array}{lllllll}
 1 & 0 & 1 & 0 & 0 & 0 & 1 \\ [-1.3mm]
 0 & 0 & 0 & 0 & 0 & 1 & 0 \\ [-1.3mm]
 0 & 0 & 0 & 0 & 1 & 0 & 0 \\ [-1.3mm]
 0 & 1 & 0 & 1 & 0 & 0 & 0
\end{array}
\right)
$}
&
{\scriptsize
$\begin{array}{l}
X_{1, 4} X_{4, 1}^{2} X_{1, 5} X_{5, 4}^{1} X_{4, 1}^{1} - 
 X_{1, 4} X_{4, 2} X_{2, 5} X_{5, 4}^{1} X_{4, 1}^{2} - \\
 X_{1, 5} X_{5, 4}^{2} X_{4, 1}^{1} +
 X_{2, 5} X_{5, 4}^{2} X_{4, 2} 
\end{array}$
}
\\  [-1mm]  \hline
(c) &
{\tiny
$\left(
\begin{array}{lllllll}
 0 & 1 & 0 & 0 & 1 & 0 & 1 \\ [-1.3mm]
 0 & 0 & 1 & 0 & 0 & 1 & 0 \\ [-1.3mm]
 0 & 0 & 0 & 1 & 0 & 0 & 0 \\ [-1.3mm]
 1 & 0 & 0 & 0 & 0 & 0 & 0
\end{array}
\right)
$}
&
{\scriptsize
$\begin{array}{l}
-X_{1, 1}^{1} X_{1, 6} X_{6, 1} + 
 X_{1, 1}^{1} X_{1, 1}^{2} X_{1, 5} X_{5, 6} X_{6, 1} + \\
 X_{1, 6} X_{6, 2} X_{2, 1} -
 X_{1, 1}^{2} X_{1, 5} X_{5, 6} X_{6, 2} X_{2, 1} 
\end{array}$
}
\\  [-1mm] \hline
(d) &
{\tiny
$\left(
\begin{array}{lllllll}
 0 & 0 & 0 & 0 & 0 & 0 & 1 \\ [-1.3mm]
 0 & 1 & 0 & 1 & 0 & 1 & 0 \\ [-1.3mm]
 0 & 0 & 0 & 0 & 1 & 0 & 0 \\ [-1.3mm]
 1 & 0 & 1 & 0 & 0 & 0 & 0
\end{array}
\right)
$}
&
{\scriptsize
$\begin{array}{l}
-X_{1, 1} X_{1, 5} X_{5, 1} X_{1, 8} X_{8, 1} + 
 X_{1, 1} X_{1, 5} X_{5, 2} X_{2, 8} X_{8, 1} + \\
 X_{1, 8} X_{8, 5} X_{5, 1} - 
 X_{2, 8} X_{8, 5} X_{5, 2} 
\end{array}$
}
\\  [-1mm] \hline
(e) &
{\tiny
$\left(
\begin{array}{llllll}
 0 & 0 & 0 & 1 & 0 & 1 \\ [-1.3mm]
 0 & 0 & 1 & 0 & 1 & 0 \\ [-1.3mm]
 0 & 1 & 0 & 0 & 0 & 0 \\ [-1.3mm]
 1 & 0 & 0 & 0 & 0 & 0
\end{array}
\right)
$}
&
{\scriptsize
$\begin{array}{l}
X_{1, 1}^{1} X_{1, 1}^{2} X_{1, 6}^{2} X_{6, 1} - 
 X_{1, 1}^{1} X_{1, 6}^{1} X_{6, 1} - \\
 X_{1, 1}^{2} X_{1, 6}^{2} X_{6, 2} X_{2, 1} + 
 X_{1, 6}^{1} X_{6, 2} X_{2, 1} 
\end{array}$
}
\\  [-1mm] \hline
(f) &
{\tiny
$\left(
\begin{array}{llllll}
 0 & 0 & 0 & 0 & 0 & 1 \\ [-1.3mm]
 0 & 0 & 0 & 1 & 1 & 0 \\ [-1.3mm]
 0 & 1 & 1 & 0 & 0 & 0 \\ [-1.3mm]
 1 & 0 & 0 & 0 & 0 & 0
\end{array}
\right)
$}
&
{\scriptsize
$\begin{array}{l}
-X_{1, 3}^{2} X_{3, 2} X_{2, 1} + 
 X_{1, 3}^{2} X_{3, 3}^{1} X_{3, 1} + \\
 X_{1, 3}^{1} X_{3, 3}^{2} X_{3, 2} X_{2, 1} - 
 X_{1, 3}^{1} X_{3, 3}^{1} X_{3, 3}^{2} X_{3, 1} 
\end{array}$
}
\\  [-1mm] \hline
(g) &
{\tiny
$\left(
\begin{array}{lllllll}
 0 & 0 & 0 & 0 & 0 & 0 & 1 \\ [-1.3mm]
 0 & 1 & 0 & 1 & 0 & 1 & 0 \\ [-1.3mm]
 0 & 0 & 0 & 0 & 1 & 0 & 0 \\ [-1.3mm]
 1 & 0 & 1 & 0 & 0 & 0 & 0
\end{array}
\right)
$}
&
{\scriptsize
$\begin{array}{l}
-X_{1, 1}^{1} X_{1, 8} X_{8, 1} + 
 X_{1, 2} X_{2, 8} X_{8, 1} + \\
 X_{1, 1}^{1} X_{1, 1}^{2} X_{1, 8} X_{8, 5} X_{5, 1} - 
 X_{1, 1}^{2} X_{1, 2} X_{2, 8} X_{8, 5} X_{5, 1} 
\end{array}$
}
\\  [-1mm] \hline
(h) &
{\tiny
$\left(
\begin{array}{lllllll}
 0 & 0 & 0 & 0 & 0 & 0 & 1 \\ [-1.3mm]
 0 & 1 & 0 & 1 & 0 & 1 & 0 \\ [-1.3mm]
 0 & 0 & 0 & 0 & 1 & 0 & 0 \\ [-1.3mm]
 1 & 0 & 1 & 0 & 0 & 0 & 0
\end{array}
\right)
$}
&
{\scriptsize
$\begin{array}{l}
  X_{1, 1} X_{1, 5} X_{5, 1}^{1} - 
  X_{1, 2} X_{2, 5} X_{5, 1}^{1} - \\
  X_{1, 1} X_{1, 5} X_{5, 1}^{2} X_{1, 7} X_{7, 1} + 
  X_{1, 2} X_{2, 5} X_{5, 1}^{2} X_{1, 7} X_{7, 1}
\end{array}$
}
\\  [-1mm] \hline
(i) &
{\tiny
$\left(
\begin{array}{llllll}
 0 & 0 & 0 & 0 & 0 & 1 \\ [-1.3mm]
 0 & 0 & 0 & 1 & 1 & 0 \\ [-1.3mm]
 1 & 0 & 1 & 0 & 0 & 0 \\ [-1.3mm]
 0 & 1 & 0 & 0 & 0 & 0
\end{array}
\right)
$}
&
{\scriptsize
$\begin{array}{l}
-X_{1, 1} X_{1, 2} X_{2, 1} + 
 X_{1, 1} X_{1, 5}^{2} X_{5, 1}^{1} + \\
 X_{1, 2} X_{2, 1} X_{1, 5}^{1} X_{5, 1}^{2} - 
 X_{1, 5}^{2} X_{5, 1}^{1} X_{1, 5}^{1} X_{5, 1}^{2}
\end{array}$
}
\\  [-1mm] \hline
(j) &
{\tiny
$\left(
\begin{array}{lllllll}
 0 & 0 & 0 & 0 & 0 & 0 & 1 \\ [-1.3mm]
 1 & 0 & 0 & 1 & 1 & 1 & 0 \\ [-1.3mm]
 0 & 0 & 1 & 0 & 0 & 0 & 0 \\ [-1.3mm]
 0 & 1 & 0 & 0 & 0 & 0 & 0
\end{array}
\right)
$}
&
{\scriptsize
$\begin{array}{l}
X_{1,1} X_{1,5}^{2} X_{5,1} -
X_{1,5}^{2} X_{5,2} X_{2,1} -\\
X_{1,1} X_{1,5}^{1} X_{5,1} X_{1,7} X_{7,1} +
X_{1,5}^{1} X_{5,2} X_{2,1} X_{1,7} X_{7,1} 
\end{array}$
}
\\  [-1mm] \hline
(k) &
{\tiny
$\left(
\begin{array}{llllll}
 0 & 0 & 0 & 0 & 0 & 1 \\ [-1.3mm]
 0 & 1 & 0 & 0 & 1 & 0 \\ [-1.3mm]
 1 & 0 & 0 & 1 & 0 & 0 \\ [-1.3mm]
 0 & 0 & 1 & 0 & 0 & 0
\end{array}
\right)
$}
&
{\scriptsize
$\begin{array}{l}
-X_{1, 1}^{3} X_{1, 1}^{2} X_{1, 1}^{1} + 
 X_{1, 1}^{1} X_{1, 2} X_{2, 1} + \\
 X_{1, 1}^{3} X_{1, 1}^{2} X_{1, 5} X_{5, 1} - 
 X_{1, 2} X_{2, 1} X_{1, 5} X_{5, 1}
\end{array}$
}
\\ \hline
\end{tabular}
\label{t:c4plethoraA}
%\end{table}
}

%%%%%%%%%%%%%%%%%%%%%%%%---------------

\begin{table}[h!b!t!]\begin{center}
\begin{tabular}{|c|c|c|} \hline
& Toric diagram $G_t$ & Superpotential $W$ \\ \hline \hline
(a) &
{\tiny
$\left(
\begin{array}{llllllll}
 0 & 0 & 0 & 0 & 1 & 0 & 0 & 1 \\ [-1mm]
 0 & $-1$ & 0 & 0 & 0 & 0 & 1 & 0 \\ [-1mm]
 0 & 2 & 1 & 0 & 0 & 1 & 0 & 0 \\ [-1mm]
 1 & 0 & 0 & 1 & 0 & 0 & 0 & 0
\end{array}
\right)
$}
&
{\scriptsize
$\begin{array}{l}
-X_{1, 2} X_{2, 2} X_{2, 1} + 
 X_{2, 2} X_{2, 8} X_{8, 2} + \\
 X_{1, 2} X_{2, 1} X_{1, 3} X_{3, 8} X_{8, 5} X_{5, 1}  - 
 X_{1, 3} X_{3, 8} X_{8, 2} X_{2, 8} X_{8,5} X_{5, 1}  
\end{array}$
}
\\ \hline
(b) &
{\tiny
$
\left(
\begin{array}{llllllll}
 1 & 0 & 0 & 0 & 0 & 2 & 0 & 1 \\ [-1mm]
 0 & 0 & 0 & 0 & 0 & $-1$ & 1 & 0 \\ [-1mm]
 0 & 0 & 0 & 1 & 1 & 0 & 0 & 0 \\ [-1mm]
 0 & 1 & 1 & 0 & 0 & 0 & 0 & 0
\end{array}
\right)
$}
&
{\scriptsize
$\begin{array}{l}
-X_{1, 3}^{2} X_{3, 2} X_{2, 8} X_{8, 1}^{1} + 
 X_{1, 1} X_{1, 3}^{2} X_{3, 8} X_{8, 1}^{1} + \\
 X_{1, 3}^{1} X_{3, 2} X_{2, 8} X_{8, 1}^{2} - 
 X_{1, 1} X_{1, 3}^{1} X_{3, 8} X_{8, 1}^{2}
\end{array}$
}
\\ \hline
(c) &
{\tiny
$
\left(
\begin{array}{lllllll}
 0 & $-1$ & 0 & 0 & 0 & 0 & 1 \\ [-1mm]
 0 & 2 & 1 & 0 & 0 & 1 & 0 \\ [-1mm]
 0 & 0 & 0 & 0 & 1 & 0 & 0 \\ [-1mm]
 1 & 0 & 0 & 1 & 0 & 0 & 0
\end{array}
\right)
$}
&
{\scriptsize
$\begin{array}{l}
-X_{1, 2} X_{2, 2} X_{2, 1} + 
 X_{1, 2} X_{2, 1} X_{1, 3} X_{3, 8} X_{8, 1} + \\
 X_{2, 2} X_{2, 8} X_{8, 2} -
 X_{1, 3} X_{3, 8} X_{8, 2} X_{2, 8} X_{8, 1} 
\end{array}$
}
\\ \hline
(d) &
{\tiny
$
\left(
\begin{array}{llllllll}
 0 & 0 & $-1$ & 0 & 0 & 0 & 0 & 1 \\ [-1mm]
 0 & 0 & 2 & 1 & 0 & 0 & 1 & 0 \\ [-1mm]
 0 & 0 & 0 & 0 & 1 & 1 & 0 & 0 \\ [-1mm]
 1 & 1 & 0 & 0 & 0 & 0 & 0 & 0
\end{array}
\right)
$}
&
{\scriptsize
$\begin{array}{l}
-X_{1, 4} X_{4, 2} X_{2, 4} X_{4, 1} + 
 X_{1, 4} X_{4, 1} X_{1, 5} X_{5, 1} + \\
 X_{2, 4} X_{4, 5} X_{5, 4} X_{4, 2} - 
 X_{1, 5} X_{5, 4} X_{4, 5} X_{5, 1} 
\end{array}$
}
\\ \hline
(e) &
{\tiny
$
\left(
\begin{array}{llllll}
 0 & $-1$ & 0 & 0 & 0 & 1 \\ [-1mm]
 0 & 2 & 1 & 0 & 1 & 0 \\ [-1mm]
 0 & 0 & 0 & 1 & 0 & 0 \\ [-1mm]
 1 & 0 & 0 & 0 & 0 & 0
\end{array}
\right)
$}
&
{\scriptsize
$\begin{array}{l}
 X_{1, 2} X_{2, 2} X_{2, 1} - 
 X_{1, 2} X_{2, 1} X_{1, 3} X_{3, 1} - \\
 X_{2, 2} X_{2, 3} X_{3, 2} + 
 X_{1, 3} X_{3, 2} X_{2, 3} X_{3, 1} 
\end{array}$
}
\\ \hline
(f) &
{\tiny
$
\left(
\begin{array}{lllllll}
 0 & 0 & $-1$ & 0 & 0 & 0 & 1 \\ [-1mm]
 0 & 1 & 2 & 1 & 0 & 1 & 0 \\ [-1mm]
 0 & 0 & 0 & 0 & 1 & 0 & 0 \\ [-1mm]
 1 & 0 & 0 & 0 & 0 & 0 & 0
\end{array}
\right)
$}
&
{\scriptsize
$\begin{array}{l}
X_{1, 2} X_{2, 5}^{2} X_{5, 1} - 
 X_{1, 2} X_{2, 5}^{1} X_{5, 1} X_{1, 6} X_{6, 1} - \\
 X_{2, 5}^{2} X_{5, 6} X_{6, 2} + 
 X_{1, 6} X_{6, 2} X_{2, 5}^{1} X_{5, 6} X_{6, 1}
\end{array}$
}
\\ \hline
(g) &
{\tiny
$
\left(
\begin{array}{llllllll}
 0 & 0 & 1 & 0 & 0 & 2 & 0 & 1 \\ [-1mm]
 0 & 0 & 0 & 0 & 0 & $-1$ & 1 & 0 \\ [-1mm]
 0 & 1 & 0 & 0 & 1 & 0 & 0 & 0 \\ [-1mm]
 1 & 0 & 0 & 1 & 0 & 0 & 0 & 0
\end{array}
\right)
$}
&
{\scriptsize
$\begin{array}{l}
X_{1, 1}^{1} X_{1, 1}^{2} X_{1, 5} X_{5, 1} - 
 X_{1, 2} X_{2, 1} X_{1, 5} X_{5, 1} - \\
 X_{1, 1}^{1} X_{1, 1}^{2} X_{1, 7} X_{7, 1} + 
 X_{1, 2} X_{2, 1} X_{1, 7} X_{7, 1}
\end{array}$
}
\\ \hline
(h) &
{\tiny
$
\left(
\begin{array}{lllllll}
 0 & 0 & 0 & 0 & 1 & 0 & 1 \\ [-1mm]
 0 & 0 & 0 & 0 & 0 & 1 & 0 \\ [-1mm]
 1 & 2 & 0 & 1 & 0 & 0 & 0 \\ [-1mm]
 0 & $-1$ & 1 & 0 & 0 & 0 & 0
\end{array}
\right)
$}
&
{\scriptsize
$\begin{array}{l}
-X_{1, 3}^{2} X_{3, 2}^{2} X_{2, 1}^{1} + 
 X_{1, 3}^{1} X_{3, 2}^{2} X_{2, 1}^{2} + \\
 X_{1, 3}^{2} X_{3, 2}^{1} X_{2, 1}^{1} X_{1, 5} X_{5, 1} - 
 X_{1, 3}^{1} X_{3, 2}^{1} X_{2, 1}^{2}  X_{1, 5} X_{5, 1}
\end{array}$
}
\\ \hline
(i) &
{\tiny
$
\left(
\begin{array}{lllllll}
 0 & 0 & 0 & 0 & 0 & 0 & 1 \\ [-1mm]
 1 & 2 & 0 & 1 & 0 & 1 & 0 \\ [-1mm]
 0 & 0 & 0 & 0 & 1 & 0 & 0 \\ [-1mm]
 0 & $-1$ & 1 & 0 & 0 & 0 & 0
\end{array}
\right)
$}
&
{\scriptsize
$\begin{array}{l}
-X_{1, 3}^{2} X_{3, 2}^{2} X_{2, 1}^{1} + 
 X_{1, 3}^{1} X_{3, 2}^{2} X_{2, 1}^{2} + \\
 X_{1, 3}^{2} X_{3, 2}^{1} X_{2, 1}^{1} X_{1, 6} X_{6, 1} - 
 X_{1, 3}^{1} X_{3, 2}^{1} X_{2, 1}^{2} X_{1, 6} X_{6, 1}
\end{array}$
}
\\ \hline
(j) &
{\tiny
$
\left(
\begin{array}{llllll}
 1 & 0 & 0 & 2 & 0 & 1 \\ [-1mm]
 0 & 0 & 0 & $-1$ & 1 & 0 \\ [-1mm]
 0 & 0 & 1 & 0 & 0 & 0 \\ [-1mm]
 0 & 1 & 0 & 0 & 0 & 0
\end{array}
\right)
$}
&
{\scriptsize
$\begin{array}{l}
X_{1, 1} X_{1, 2}^{2} X_{2, 8}^{1} X_{8, 1}^{1} - 
 X_{1, 2}^{2} X_{2, 8}^{2} X_{8, 1}^{1} - \\
 X_{1, 1} X_{1, 2}^{1} X_{2, 8}^{1} X_{8, 1}^{2} + 
 X_{1, 2}^{1} X_{2, 8}^{2} X_{8, 1}^{2}
\end{array}$
}
\\ \hline
(k) &
{\tiny
$
\left(
\begin{array}{lllllll}
 0 & 0 & 0 & 0 & 0 & 0 & 1 \\ [-1mm]
 0 & 0 & 0 & 0 & 1 & 1 & 0 \\ [-1mm]
 1 & 2 & 0 & 1 & 0 & 0 & 0 \\ [-1mm]
 0 & $-1$ & 1 & 0 & 0 & 0 & 0
\end{array}
\right)
$}
&
{\scriptsize
$\begin{array}{l}
X_{1, 2}^{2} X_{2, 2} X_{2, 1}^{1} - 
 X_{1, 2}^{1} X_{2, 2} X_{2, 1}^{2} - \\
 X_{1, 2}^{2} X_{2, 1}^{1} X_{1, 8}  X_{8, 5}  X_{5, 1} + 
 X_{1, 2}^{1} X_{2, 1}^{2} X_{1, 8}  X_{8, 5} X_{5, 1}
\end{array}$
}
\\ \hline
(l) &
{\tiny
$
\left(
\begin{array}{llllll}
 0 & 0 & 1 & 2 & 0 & 1 \\ [-1mm]
 0 & 0 & 0 & $-1$ & 1 & 0 \\ [-1mm]
 0 & 1 & 0 & 0 & 0 & 0 \\ [-1mm]
 1 & 0 & 0 & 0 & 0 & 0
\end{array}
\right)
$}
&
{\scriptsize
$\begin{array}{l}
-X_{1, 1}^{3} X_{1, 5} X_{5, 1} + 
 X_{1, 1}^{1} X_{1, 1}^{2} X_{1, 5} X_{5, 1} + \\
 X_{1, 1}^{3} X_{1, 7} X_{7, 1} - 
 X_{1, 1}^{1} X_{1, 1}^{2} X_{1, 7} X_{7, 1}
\end{array}$
}
\\ \hline
(m) &
{\tiny
$
\left(
\begin{array}{llllll}
 0 & 0 & 0 & 0 & 0 & 1 \\ [-1mm]
 0 & 0 & 0 & 0 & 1 & 0 \\ [-1mm]
 1 & 2 & 0 & 1 & 0 & 0 \\ [-1mm]
 0 & $-1$ & 1 & 0 & 0 & 0
\end{array}
\right)
$}
&
{\scriptsize
$\begin{array}{l}
-X_{1, 2}^{2} X_{2, 2} X_{2, 1}^{1} + 
 X_{1, 2}^{1} X_{2, 2} X_{2, 1}^{2} + \\
 X_{1, 2}^{2} X_{2, 1}^{1} X_{1, 6} X_{6, 1} - 
 X_{1, 2}^{1} X_{2, 1}^{2} X_{1, 6} X_{6, 1}
\end{array}$
}
\\ \hline
\end{tabular}
\caption{{\sf The toric diagrams and superpotentials for various phases of $\IC^2/\IZ_2 \times \IC^2$.}}
\label{t:c2z2plethora}
\end{center}
\end{table}

%%%%%%%%%%%%%%%%%%%%%%%%%%%%%%%%%%%%%%%%%%%%%%%%%%%%%%%%%%%%%%%%%%%%%%%%%%%%%%%%%%%%%%%%%%%%%%%%%%%%%%%%%%%%%%%%%%%%%%%%%%%%%%%%%%%%%%%%%%%%%%%%%%%%%%%
\bibliographystyle{JHEP}

\end{document}